\documentclass[usenatbib]{mn2e}
\usepackage{multirow}
\usepackage{rotating}
\usepackage{colortbl}
\usepackage{color}
\usepackage[fleqn]{amsmath}
\usepackage{amssymb}
\usepackage{amsfonts}
\usepackage{verbatim}
\usepackage{scalefnt}
\usepackage[percent]{overpic}
\usepackage[bookmarks,bookmarksnumbered,colorlinks=true, citecolor=blue, linkcolor=cyan]{hyperref}

\newcommand{\apjl}{ApJ}
\newcommand{\apjs}{ApJS}
\newcommand{\ao}{Appl.~Opt.}
\newcommand{\aap}{A\&A}

\newcommand{\beq}{\begin{equation}}
\newcommand{\eeq}{\end{equation}}

\definecolor{grey}{rgb}{0.5,0.6,0.7}
\def \simlt { \lower .75ex \hbox{$\sim$} \llap{\raise .27ex \hbox{$<$}} }
\definecolor{purple}{rgb}{0.65,0.15,0.9}
\definecolor{darkorange}{rgb}{0.8,0.3,0}
\definecolor{olive}{rgb}{0.4,0.6,0.25}
\definecolor{darkgreen}{rgb}{0,0.7,0}
\definecolor{darkred}{rgb}{0.5,0,0}

\title[Linking galaxy structural properties and star formation activity to BH activity]{Linking galaxy structural properties and star formation activity to black hole activity with IllustrisTNG}
\author[Habouzit et al.]{M\'{e}lanie Habouzit$^{1}$\thanks{E-mail: mhabouzit@flatironinstitute.org},
Shy Genel$^{1,}$$^{2}$,
Rachel S. Somerville$^{1,}$$^{3}$,
Dale Kocevski$^{4}$,
\newauthor
Michaela Hirschmann$^{5}$,
Avishai Dekel$^{6}$,
Ena Choi$^{2}$,
Dylan Nelson$^{7}$,
\newauthor
Annalisa Pillepich$^{8}$,
Paul Torrey$^{9}$,
Lars Hernquist$^{10}$,
Mark Vogelsberger$^{11}$,
\newauthor
Rainer Weinberger$^{10}$,
Volker Springel$^{7}$
\\
$^1$ Center for Computational Astrophysics, Flatiron Institute, 162 5th Avenue, New York, NY 10010, USA\\
$^2$ Columbia Astrophysics Laboratory, Columbia University, 550 West 120th Street, New York, NY 10027, USA\\
$^3$ Department of Physics and Astronomy, Rutgers University, 136 Frelinghuysen Rd, Piscataway, NY 08854, USA\\
$^4$ Department of Physics and Astronomy, Colby College, Waterville, ME 04901, USA\\
$^5$ Sorbonne Universit\'es, UPMC-CNRS, UMR7095, Institut d'Astrophysique de Paris, F-75014 Paris, France\\
$^6$ Center for Astrophysics and Planetary Science, Racah Institute of Physics, The Hebrew University, Jerusalem 91904, Israel\\
$^7$ Max-Planck-Institut f{\"u}r Astrophysik, Karl-Schwarzschild-Strasse 1, 85740 Garching bei M{\"u}nchen, Germany\\
$^8$ Max-Planck-Institut f{\"u}r Astronomie, K{\"o}nigstuhl 17, 69117 Heidelberg, Germany\\
$^{9}$ Department of Astronomy, University of Florida, Bryant Space Sciences Center, Gainesville, FL 32611, USA\\
$^{10}$ Harvard-Smithsonian Center for Astrophysics, 60 Garden Street, Cambridge, MA 02138, USA\\
$^{11}$ Department of Physics, Kavli Institute for Astrophysics and Space Research, MIT, Cambridge, MA 02139, USA\\
}               

\begin{document}
\maketitle

\begin{abstract}
We study the connection between active galactic nuclei (AGN) and their host galaxies through cosmic time in the large-scale cosmological IllustrisTNG simulations. 
We first compare BH properties, i.e. the hard X-ray BH luminosity function, AGN galaxy occupation fraction, and distribution of Eddington ratios, to available observational constraints. The simulations produce a population of BHs in good agreement with observations, but we note an excess of faint AGN in hard X-ray ($L_{\rm x}\sim 10^{43-44}\, \rm erg/s$), and a lower number of bright AGN ($L_{\rm x}>10^{44} \, \rm erg/s$), a conclusion that varies quantitatively but not qualitatively with BH luminosity estimation method. 
The lower Eddington ratios of the $10^{9}\, \rm M_{\odot}$ BHs compared to observations suggest that AGN feedback may be too efficient in this regime.
We study galaxy star formation activity and structural properties, and design sample-dependent criteria to identify different galaxy types (star-forming/quiescent, extended/compact) that we apply both to the simulations and observations from the {\sc candels} fields.
We analyze how the simulated and observed galaxies populate the specific star formation rate - stellar mass surface density diagram. A large fraction of the $z=0$ $M_{\star}\geqslant 10^{11}\, \rm M_{\odot}$ quiescent galaxies first experienced a compaction phase (i.e. reduction of galaxy size) while still forming stars, and then a quenching event.
We measure the dependence of AGN fraction on galaxies' locations in this diagram. 
After correcting the simulations with a redshift and AGN luminosity-dependent model for AGN obscuration, we find good qualitative and quantitative agreement with observations. The AGN fraction is the highest among compact star-forming galaxies ($16-20\%$ at $z\sim1.5-2$), and the lowest among compact quiescent galaxies ($6-10\%$ at $z\sim1.5-2$).

\end{abstract}



\section{Introduction}
\label{sec:intro}

Our current understanding of galaxy buildup is closely tied to the star-forming and structural properties of galaxies (e.g., galaxy sizes). These properties are regulated by the baryonic content of galaxies, such as gas and its angular momentum, and star formation regulation mechanisms such as stellar and black hole (BH) feedback.
There is growing evidence for correlations between BH properties and those of their host galaxies.
Indeed, we observe BHs with millions of solar masses and greater in the center of most local galaxies, for various types of galaxies \citep{2012NatCo...3E1304G}.
We also observe BHs in their most powerful form, the so-called quasars, in the early epochs of the Universe up to $z=7$ \citep{Mortlock2011,2018ApJ...856L..25B}, when the Universe was less than 1 Gyr old, suggesting that galaxies could host BHs since the Universe was in its infancy. Accreting BHs, known as active galactic nuclei (AGN), are observed from high redshift down to the local Universe, and are able to drive feedback on their host galaxies, which is believed to be key in shaping the massive end of the galaxy stellar mass function \citep{DiMatteo2005,Croton2006,Silk2013}. Finally, several empirical scaling relations between BH mass and properties of host galaxies have been identified, such as galaxy mass, luminosity, velocity dispersion \citep[][and references therein]{2013ARA&A..51..511K}. This suggests the existence of co-evolution mechanisms between BHs and galaxies from high redshift until today. Therefore the build up of the galaxy population and the global cosmic evolution of the Universe need to be studied jointly with the cosmic evolution of BHs.

Large-scale cosmological hydrodynamical simulations offer the advantage of self-consistently modeling the formation and evolution of galaxies, as well as the evolution of BHs \citep{2014MNRAS.444.1453D,2014MNRAS.445..175G,2014MNRAS.442.2304H,2014MNRAS.444.1518V,2015MNRAS.446..521S,2015MNRAS.452..575S}. A recent suite of papers have presented the new IllustrisTNG simulations \citep{2017arXiv170703396M,2017arXiv170703401N,2017arXiv170703397S,2017arXiv170703395N,2017arXiv170302970P}, which currently consist of two large-scale cosmological magneto-hydrodynamical simulations of side length $\sim100$ cMpc and $\sim300$ cMpc. The simulations are able to capture a wide range of BH and galaxy properties and their evolution with time. The population of BHs at $z\sim 0$ is consistent with current observational constraints \citep{2017arXiv171004659W}, i.e. the bolometric AGN luminosity function \citep{Hop_bol_2007,2014ApJ...786..104U,2015ApJ...802..102L}, and BH mass-galaxy stellar mass relation \citep{2016ApJ...817...21S}. Compared to the previous Illustris simulations \citep{2014MNRAS.444.1518V,2014Natur.509..177V,2014MNRAS.445..175G}, the stellar and gas properties of galaxies are better reproduced in IllustrisTNG \citep{2017arXiv170703406P,2017arXiv170302970P}. The galaxy stellar sizes are of particular interest for the present study, and were too large by a factor of a few for galaxies with $M_{\star}\leqslant 10^{10.7}\, \rm M_{\odot}$ in Illustris \citep{2015MNRAS.454.1886S}. 
The discrepancy between the mass-size relation of the galaxies in the new IlustrisTNG (100 cMpc box) and observations is less than $0.2$ dex across the full mass range \citep{2017arXiv170705327G}, which is generally within the observational uncertainties. In particular, global observational trends are well reproduced, i.e. sizes of quenched galaxies are smaller than star-forming galaxies, and galaxy sizes increase with time, all of which represent uncalibrated outcomes of the simulations. 
These differences between star-forming galaxies and their quiescent counterparts have been found in many observational studies \citep[][and references therein]{2010ApJ...713..738W,2012ApJ...746..162N,2014ApJ...788...28V,2015ApJS..219...15S,2017ApJ...840...47B}.

Understanding the buildup of the massive quiescent population of galaxies, and the responsible mechanisms, is one of the biggest challenges of galaxy formation today. A small fraction of quiescent galaxies has been observed at very early times at redshifts about $\sim 4-3$ \citep{2012ApJ...759L..44G,2013ApJ...768...92S,2014ApJ...783L..14S,2015ApJ...808L..29S,2016ApJ...830...51S}. It is generally believed that quiescent galaxies start forming at high redshift when the Universe is no older than 2-3 Gyr, and that they dominate the number density of massive $M_{\star}\geqslant 10^{11}\, \rm M_{\odot}$ galaxies by $z\sim 2$ \citep{2009A&A...501...15F,2010ApJ...709..644I,2011ApJ...739...24B,2013ApJ...777...18M}. 
The properties of quiescent galaxies\footnote{The terminology quiescent galaxies is used to refer to galaxies with a low star formation rate (SFR) or SFR below a given detection limit, and many different specific definitions have been adopted in the literature. We will explicitly define later in the text the definition of quiescent and SF galaxies used in this paper.} 
are very different across cosmic time: while the quiescent galaxies that we observe today are extended, their high redshift counterparts ({\it compact quiescent galaxies} in the following) have sizes about 3 to 5 times smaller.

At $z\geqslant 2$, many quiescent galaxies are spheroid dominated\footnote{Galaxy morphologies at high redshift are most commonly measured via the projected light profile, frequently characterized via fitting with a S\'{e}rsic profile, which is characterized by the S\'{e}rsic index $n_s$. Galaxies with $n_s \geqslant 2$ are typically considered ``spheroid dominated''.},
and also exhibit several structural properties different from star-forming (SF) galaxies observed at the same redshift. The high redshift SF galaxies tend to be more clumpy, irregular, and disk-dominated \citep{2004ApJ...603...74E,2008ApJ...687...59G,2015ApJ...800...39G}. An intermediate population of galaxies at $z\sim3-2$ was recently discovered, with some properties shared with high redshift SF galaxies, and some others similar to quiescent galaxies. This abundant population of massive SF galaxies \citep{2011ApJ...742...96W,2013ApJ...765..104B,2014ApJ...791...52B,2013ApJ...766...15P,2013ApJ...768...92S,2014ApJ...780....1W,2014Natur.513..394N}, with both smaller sizes and spheroid-like morphologies (centrally concentrated light profiles) comparable to those of quiescent galaxies, are described as {\it compact SF galaxies}.

One can place these different categories of galaxies on a specific star formation rate - stellar mass surface density diagram (sSFR-$\Sigma$ diagram hereafter), where the projected stellar mass surface density is used as an indicator of galaxy compactness. The evolution of galaxies in this diagram is likely to be a function of galaxy mass.
This diagram was presented for galaxies in the {\sc candels} survey with $M_{\star}\geqslant 10^{10}\, \rm M_{\odot}$ and $z \sim 0.5$--3 by \citet{2013ApJ...765..104B}. This study \citep[see also][]{2014ApJ...791...52B,2017ApJ...840...47B,2015ApJ...813...23V} found that galaxies at $2.6 < z < 3$ mainly lie in the  {\it extended SF} and {\it compact SF} quadrants of the diagram. At $2.2 < z < 2.6$, a significant population of galaxies appears in the {\it compact quiescent} quadrant. The fraction of galaxies in the {\it compact SF} quadrant declines steadily over cosmic time from $z\sim 3$ to $z\sim 0.5$, while the fraction of {\it compact quiescent} galaxies increases from $z\sim 3$ to $z\sim 1$, perhaps implying that the former are evolving {\it into} the latter.
Qualitatively similar behavior is seen in the VELA hydrodynamic zoom-in simulations and in semi-analytic models {2017ApJ...846..112K}.
%
In the Illustris simulation, the formation of massive compact galaxies at $z=2$ (most of which were quiescent galaxies) was either triggered by gas-rich major mergers inducing central starbursts at $z=4-2$, or took place at very early times with galaxies remaining compact until $z=2$ \citep{2015MNRAS.449..361W}. Later on the compact quiescent galaxies are believed to eventually become more extended and massive galaxies, primary due to gas-poor (dry) minor mergers that do not initiate new star formation episodes \citep{Naab2009,Hopkins2010,2010ApJ...725.2312O,2014MNRAS.444..942P}. In Illustris, $z=2$ massive galaxies with $1-3\times 10^{11}\, \rm M_{\odot}$ either acquire an ex-situ envelope, remain almost identical with very little mass or size growth, are merged into more massive galaxies, or more marginally can be completely mixed by major mergers \citep{2016MNRAS.456.1030W}.
It has been suggested that the 4 quadrants of the sSFR-$\Sigma$ diagram could actually show the different epochs of galaxy evolution from extended SF galaxies at high redshift to compact and extended quiescent galaxies at low-redshift. This would suggest that galaxies experienced two crucial phases, a {\it compaction event} (while the galaxy is forming stars) that converts extended SF galaxies to compact SF galaxies, and a {\it quenching phase} that turn those compact SF galaxies into compact quiescent galaxies.

There has been much debate in the literature about the physical processes that drive a galaxy's evolution through this diagram \citep[see e.g.][]{2015MNRAS.450.2327Z, 2016MNRAS.458..242T}, and the role of AGN feedback.
One scenario to form compact SF galaxies requires the prior formation of a gas-rich disk, subject to violent instabilities, i.e. a highly perturbed disk. Compacting a star-forming disk would require the disk to lose angular momentum and energy, which is unlikely for a stellar disk. Large gaseous inflows toward the center of the disk are therefore needed, and consist of a dissipative mechanism called a {\it wet process}. The following compaction of the disk is thought to be possibly due to the following processes or a combination of them: gas-rich mergers \citep{1991ApJ...370L..65B,1996ApJ...471..115B,2006ApJS..163...50H}, low angular momentum recycled gas, compression due to tidal torques, counter-rotating streams, or gas inflows driven by the violent instabilities within the disk \citep[see][and references therein]{2009ApJ...703..785D,2013MNRAS.435..999D}. 

The compaction of the disk, and the intense inflows will likely also trigger accretion into a central BH, and will also certainly provide enough inflowing gas to form new stars. Consequently, these processes could set the end of the star formation within the galaxy, by SN or AGN feedback driven powerful outflows, which could clear the galaxy of its gas content and/or prevent gas from outside the galaxy from accreting, or simply due to gas exhaustion after the burst of star formation following the large gas inflows. Again, the relative importance of these mechanisms is subject to debate in the literature. 
The depletion of gas by high star formation rates, and the ability of SN feedback to quench galaxies have been studied in a suite of zoom-in cosmological galaxy simulations (without AGN feedback modeling) in \citet{2015MNRAS.450.2327Z}. In these simulations, the quenching is due to gas consumption, and SN feedback preventing the inflow of gas from the halos. 

\citet{2016MNRAS.463.3948D} have analyzed the role of AGN feedback in shaping galaxy properties, by using two large-scale cosmological simulations, with and without AGN feedback, and have shown that AGN feedback is needed to fully and persistently quench galaxies. \citet{2018arXiv180902143C} investigate both SN and AGN feedback in a new suite of zoom-in cosmological galaxy simulations, focussing on massive galaxies with $M_{\star}>10^{10.9}\, \rm M_{\odot}$.
They find that even if SN feedback plays an important role in decreasing the star formation rate of galaxies (e.g., decreasing the ratio ${\rm SFR}/M_{\star}$ by about one order of magnitude), it is not able to fully quench galaxies.
Conversely, the galaxies in the simulations with AGN feedback can see their SFR decreased by several orders of magnitude. As we will see in the following, the implementation of AGN feedback (i.e. the kinetic feedback mode, as explained in the following) in IllustrisTNG provides a sufficient amount of energy to the gas to overcome cooling losses \citep{2017arXiv171004659W}, preventing massive galaxies from forming stars efficiently.

Observationally, correlations between structural properties of galaxies and BH activity have been observed.
\citet{2014ApJ...791...52B} find in the GOODS-S field that compact star-forming sources at $z\sim 2$ are more likely to host X-ray bright AGN compared to more extended counterparts. \citet{2013ApJ...763L...6T} also show that clumpy and compact galaxies at $z\sim 0.11$ harbor a high fraction of AGN in the SDSS survey.
\citet{2017ApJ...846..112K} confirm that $\sim 40\%$ of $M_{\star}>10^{10}\, \rm M_{\odot}$ compact SF galaxies observed in four fields of {\sc candels} with $3>z>1.4$ host an AGN. This high fraction of AGN in the compact SF quadrant is 3.2 times higher than the fraction harbored by extended SF galaxies, and 4.7 times higher than in the compact quiescent quadrant of the sSFR - compactness diagram. Galaxies are identified as AGN if the BH X-ray luminosity exceed $L_{\rm x}=10^{42}\, \rm erg \, s^{-1}$ in either the soft ($0.5-2 \rm keV$), the hard ($2-8 \rm keV$) or full band ($0.5-8 \rm keV$).
At lower redshift, the picture seems slightly different, but still suggests the presence of a strong connection between compactness and processes triggering BH accretion.
Indeed, \citet{2017MNRAS.466L.103C} study a sample of obscured AGN at $z\leqslant 1.5$, selected in the COSMOS field. The selection relies on a high infrared to X-ray luminosity ratio of $L_{\rm IR}/L_{\rm Xray, 2-10 keV}>20$. They find that the host galaxies of those obscured AGN are more compact than a control sample of SF galaxies with the same stellar mass $10^{10.5}<M_{\star}/\rm M_{\odot}<10^{11.5}$ and the same redshift. These obscured AGN could be hosted by star-forming galaxies that have undergone dynamical contraction. Indeed, it has been shown that the inflow causing compaction and BH feeding can also lead to a high column density of $N_{\rm H}\sim10^{23}\, \rm cm^{-2}$ \citep{2011ApJ...741L..33B}. Such a high column density could result in a population of highly obscured AGN hosted by compact star-forming galaxies. Therefore there is growing evidence in observations for a connection between the compactness of galaxies before they eventually quench and the probability for these galaxies to host an AGN. The primary goal of the present paper is to measure the dependence of the AGN fraction on galaxy star formation activity and structural properties, i.e. SF/quiescent and extended/compact galaxies in the IllustrisTNG simulations.

The structure of the paper is as follows. In Section 2, we introduce the IllustrisTNG suite of simulations, and the different sub-grid physics models that are relevant for our analysis. In Section 3, we introduce the observations from the {\sc candels} fields that we will use in the paper to compare with the simulations. We examine the BH population in Section 4. We focus our attention on the accretion properties of the BH population. We present the BH hard X-ray luminosity function, the BH and AGN occupation fraction in galaxies, the fraction of bright AGN in massive galaxies, the distribution of Eddington ratios, and address a comparison with available observational data when possible. After this first step of validating the evolution of the BH population and related effects in IllustrisTNG, we study the time evolution of galaxy properties both in the simulation and in the {\sc candels} observations. In Section 5, we identify simulated and observed galaxies according to their star-forming and size properties, which allows us to populate the specific star formation - compactness diagram. In Section 6, we measure the dependence of the AGN fraction on a galaxy's location in this diagram, conduct a close comparison with the {\sc candels} observations, and investigate the mechanisms that quench galaxies and transform their structural properties. Finally, Section 7 provides examples of time evolution of two individual galaxies, which are representative of the population of massive simulated galaxies. We summarize and discuss our results in Section 8.

\section{Simulation parameters}
\subsection{{\sc IllustrisTNG} model and initial conditions}
\label{subsec:illustris}
In this paper, we analyse the two large-scale magneto-hydrodynamical cosmological simulations IllustrisTNG with side lengths of $\sim100$ cMpc (referred as TNG100 hereafter) and $\sim300$ cMpc (TNG300). The simulations were run using the moving mesh code {\sc AREPO} \citep{2010MNRAS.401..791S}, which solves the Euler equations with a second-order accurate \citep{2016MNRAS.455.1134P} finite-volume Godunov scheme on a quasi-Lagrangian moving Voronoi mesh. Poisson's equation is solved using a tree-particle-mesh algorithm. 
The initial conditions are obtained with the code {\sc N-GENIC} \citep{2005MNRAS.364.1105S}, for $z=127$. 
The two TNG simulations use a standard $\Lambda$ cold dark matter cosmology, with parameters compatible with those of the Planck results \citep{Planck15}
: dark energy density $\Omega_{\rm \Lambda}=0.691$, total matter density $\Omega_{\rm m}=0.31$, baryon density $\Omega_{\rm b}=0.0486$, Hubble constant $h=0.677$, amplitude of the matter power spectrum $\sigma_{8}=0.8159$, and spectral index $n_{\rm s}=0.97$.  
The number of dark matter particles is $1820^3$ and $2500^{3}$ respectively for TNG100 and TNG300, which corresponds to particle mass of $m_{\rm DM}=7.5\times 10^{6}\, \rm M_{\odot}$ and $m_{\rm DM}=59 \times 10^{6}\, \rm M_{\odot}$. The number of gas particles in the initial conditions is  $1820^3$ and $2500^{3}$ for TNG100 and TNG300 respectively, corresponding to baryonic mass resolution of $m_{\rm gas}=1.4 \times 10^{6}\, \rm{M_{\odot}}$ and $m_{\rm gas}=11 \times 10^{6}\, \rm{M_{\odot}}$. The spatial resolution of TNG100 is two times better than  the resolution of TNG300. Similarly the mass resolution is $2^3$ better.\\

In this paper we primarily present the results for the simulation TNG300, discussing any differences with TNG100 when relevant.

\subsection{Physics of the simulations}
We here describe the subgrid physics employed in the simulations, i.e. primordial and metal-line gas cooling, star formation, stellar feedback, metal enrichment, supermassive BH formation, growth and feedback. 
Complete descriptions of the subgrid models can be found in \citet{2017arXiv170302970P,2017MNRAS.465.3291W,2017arXiv171004659W}. These models are based on those developed for the original Illustris simulation \citep{2013MNRAS.436.3031V}.\\

\noindent {\bf Radiative cooling and photoheating by UV background}\\
The simulations include radiative cooling \citep{2013MNRAS.436.3031V} from primordial species and a cooling contribution of the metals, and a uniform redshift-dependent ionizing UV background from stars and AGN, corrected for self-shielding in the dense ISM \citep{1992ApJ...399L.109K,2009ApJ...703.1416F,2013MNRAS.430.2427R}.\\

\noindent {\bf Star formation and SN feedback}\\
Star formation occurs in cold and dense gas cells above the gas density threshold of $n_{\rm H}=0.13 \,\rm cm^{-3}$. Star formation is modeled with a Kennicutt-Schmidt law, with a star formation efficiency of $\epsilon_{\star}=0.057$ per free-fall time.
Stellar particles evolve and return mass and metals through stellar feedback to the ISM.
Stellar feedback is modeled assuming a Chabrier initial mass function with a low-mass cut-off of $0.1 \, \rm M_{\odot}$, and a high-mass cut-off of $100 \, \rm M_{\odot}$.
The production and evolution of H, He C, N, O, Ne, Mg, Si, and Fe, are tracked in the simulations \citep[see][for more details]{2017arXiv170302970P,2018MNRAS.477.1206N}.
Mass is returned to the ISM for these different elements and from the three different stellar phases in the evolution, i.e. the asymptotic giant branch (AGB, $m_{\star}=1-8\, \rm M_{\odot}$) stars, core collapse SNII ($m_{\star}=8-100\, \rm M_{\odot}$) and Type Ia SN.\\

\noindent {\bf Black hole formation, growth and BH feedback}\\
BH formation in IllustrisTNG is halo-based \citep[see also][]{Sijacki2009,2012ApJ...745L..29D,2014MNRAS.442.2304H,2015MNRAS.452..575S}, i.e. all BH free halos with mass exceeding the threshold of $M_{\rm h,thres}=7.38 \times 10^{10}\rm \, M_{\odot}$ 
are seeded in their centre with a BH of fixed mass $M_{\rm seed}=8\times 10^{5} h^{-1} \rm M_{\odot}$. 

The accretion into the BHs $\dot{M}_{\rm BH}$ follows the Bondi formalism, and is capped at the Eddington limit $\dot{M}_{\rm Edd}$, i.e. $\dot{M}_{\rm BH} = \min(\dot{M}_{\rm Bondi},\dot{M}_{\rm Edd})=\min(4\pi G^{2} M_{\rm BH}^{2}\bar{\rho}/\bar{c^{3}_{\rm s}}, 4\pi G M_{\rm BH} m_{\rm p}/\epsilon_{\rm r}\sigma_{\rm T} c)$, with $c_{\rm s}=(c^{2}_{\rm s,therm}+(B^{2}/4\pi \rho))^{1/2}$, with $c_{\rm s}$ the kernel weighted ambient sound speed including here the magnetic signal propagation speed, and $\epsilon_{\rm r}$ the radiative efficiency. Unlike the previous Illustris simulations, there is no boost factor in IllustrisTNG. 
BHs can also gain mass by BH-BH mergers, where BHs merge once they are within the {\it feedback radius} of each other, i.e. the radius where thermal energy is released \citep{2013MNRAS.436.3031V}.
Finally, BH particles are forced to remain in the bottom of the potential well of the dark matter halos; this is done by repositioning the BH directly at the location of the bottom of the  potential wells every coarse time-step \citep[see][for more details]{2017arXiv171004659W}. \\

BHs are able to provide feedback to their host galaxy, and surroundings. As in several cosmological simulations \citep{Sijacki2007,2014MNRAS.444.1453D}, AGN feedback is here modeled as a combination of two modes \citep[but see][]{2015MNRAS.446..521S}. At high accretion rate, a pure thermal mode (often called {\it quasar mode}) is employed, whereas at low accretion rate, a pure kinetic mode (often called {\it radio mode}) is instead invoked. 
In the quasar regime, more likely at high redshift, BHs are accreting sufficiently rapidly to power luminous quasars. This regime consists of a thermal feedback that heats the surrounding gas content around BHs.
At lower redshift, most BHs are accreting at rates much lower than the Eddington limit, and the feedback is therefore modeled as radiatively inefficient.
In IllustrisTNG, the previous bubble model described in \citet{Sijacki2007} is replaced by a {\it BH-driven kinetic wind feedback model} \citep{2017MNRAS.465.3291W}. The injection of momentum is in random directions, changing at every event.
Transition between the two modes is determined by the accretion rate onto the BHs, and BH mass. The accretion rate threshold is set to be the minimum between 0.1 and a function of the BH mass:
\begin{eqnarray}
f_{\rm Edd,thres}=\min \left(f_{\rm Edd,thres,0} \left(\frac{M_{\rm BH}}{10^{8}\, \rm M_{\odot}} \right)^{\beta}, 0.1\right),
\end{eqnarray}
with $\beta=2$, and $f_{\rm Edd,thres,0}=2\times10^{-3}$. BHs with Eddington ratio $f_{\rm Edd}\equiv \dot{M}_{\rm BH}/\dot{M}_{\rm Edd}\geqslant f_{\rm Edd,thres}$ will fall in the thermal regime, while BHs with $f_{\rm Edd}< f_{\rm Edd,thres}$ are in the kinetic AGN feedback mode. 
The minimum value is set to enforce that any BH can transition to the high accretion BH feedback mode if there is enough gas supply to fuel it. 

The role of AGN feedback on the population of BHs and galaxies in the IllustrisTNG simulations has been analyzed in \citet{2017arXiv171004659W}, which shows that the quasar luminosity function is dominated by BHs in the thermal AGN feedback mode, and that the decrease of the star formation rate in massive galaxies happens after the kinetic AGN feedback has kicked in. This modeling of AGN feedback has also been shown to provide the correct metal distribution in clusters \citep{2018MNRAS.474.2073V}.

\subsection{Dark matter halo and galaxy catalogs}
Dark matter halos are identified with a friends-of-friends (FOF) algorithm, using a linking length of 0.2 times the mean particle spacing. Galaxies are identified as gravitationally bound systems with non-zero stellar content with the algorithm {\sc SUBFIND} \citep{2001MNRAS.328..726S} for each dark matter halo.
Here, we focus on analyzing galaxies with $M_{\star}\geqslant 10^{9}\, \rm M_{\odot}$. 
In the last section of the paper, we track the evolution of individual galaxies in time. To do so, we use the {\sc SubLink} galaxy merger tree of the simulation \citep{2015MNRAS.449...49R}, which is based on stellar particles and gas cells.

\section{Observations from {\sc candels}}
One of the main objectives of our study is to carry out a close comparison between galaxy star formation activity and structural properties in observations and in the simulations. To do so, we will use galaxy catalogs based on observations from four of the five fields of the {\sc candels} survey \citep[][for an overview of the suvey]{2011ApJS..197...35G}, i.e. GOODS-S, UDS, EGS, and GOODS-N. For this comparison, we follow \citet{2017ApJ...846..112K} and select a robust sample of relevant galaxies. We select galaxies with $M_{\star}\geqslant 10^{10}\, \rm M_{\odot}$. 
Stellar masses are taken from the public catalogs, and were derived by fitting the spectral energy distributions assuming a Chabrier initial mass function.
We first select galaxies in the $H$-band (hereafter $H_{160}$ band) with magnitude $M_{H_{160}}\leqslant 24.5$, where we have excluded bright point sources that have a SExtractor \citep{1996A&AS..117..393B} stellarity index $<0.9$. Galaxy sizes were measured \citep{2014ApJ...788...28V} using GALFIT \citep{2002AJ....124..266P} and the HST/WFC3 $H$-band images. We exclude galaxies with large uncertainty on their sizes derived by GALFIT, and therefore only keep galaxies with decent GALFIT fits ($\rm {\sc FGALFIT}=0$) and reliable photometry ($\rm {\sc PHOTFLAG}=0$). We also apply a structural k-correction $f_{\rm corr}$ to compute galaxy radius \citep{2014ApJ...788...28V}, which corrects galaxy sizes from observed $H_{160}$ to rest-frame r band:
\begin{eqnarray}
R=R_{\rm obs}\times f_{\rm corr},
\end{eqnarray}
\begin{eqnarray}
f_{\rm corr}=\left(\frac{1+z}{1+2.2}\right)^{\gamma},
\end{eqnarray}
with
\begin{eqnarray}
\gamma=-0.35 + 0.12\times z - 0.25\times \frac{M_{\star} }{10^{10} \rm M_{\odot}},
\end{eqnarray}
where $z$ is the best available redshift estimate from spectroscopic and photometric redshifts.
We use the sample of X-ray detections compiled by \citet{2017ApJ...846..112K}, which uses the publicly available {\it Chandra} point source catalogs for UDS, the 4 Ms and 2 Ms point source catalogs of \citet{2011ApJS..195...10X} for GOODS-S and of \citet{2016ApJS..224...15X} for GOODS-N, the 800 ks source catalog for EGS \citep{2015ApJS..220...10N}, and a new source catalog of $\sim 600$ ks depth for the UDS survey \citep{2018ApJS..236...48K}. This leaves us with a final sample of 3208 galaxies in the range $1.4\leqslant z \leqslant 3$, among which 313 are detected as AGN with $L_{\rm x}\geqslant10^{42}\, \rm erg\, s^{-1}$ ($\sim 10\%$ of the sample).  The galaxies with no X-ray detection with $L_{\rm x}\geqslant10^{42}\, \rm erg\, s^{-1}$ are considered not to be hosting an X-ray AGN in this study. In this observational sample, there is no correction for AGN obscuration.

\section{BH population in IllustrisTNG}

\begin{figure*}
\includegraphics[scale=0.68]{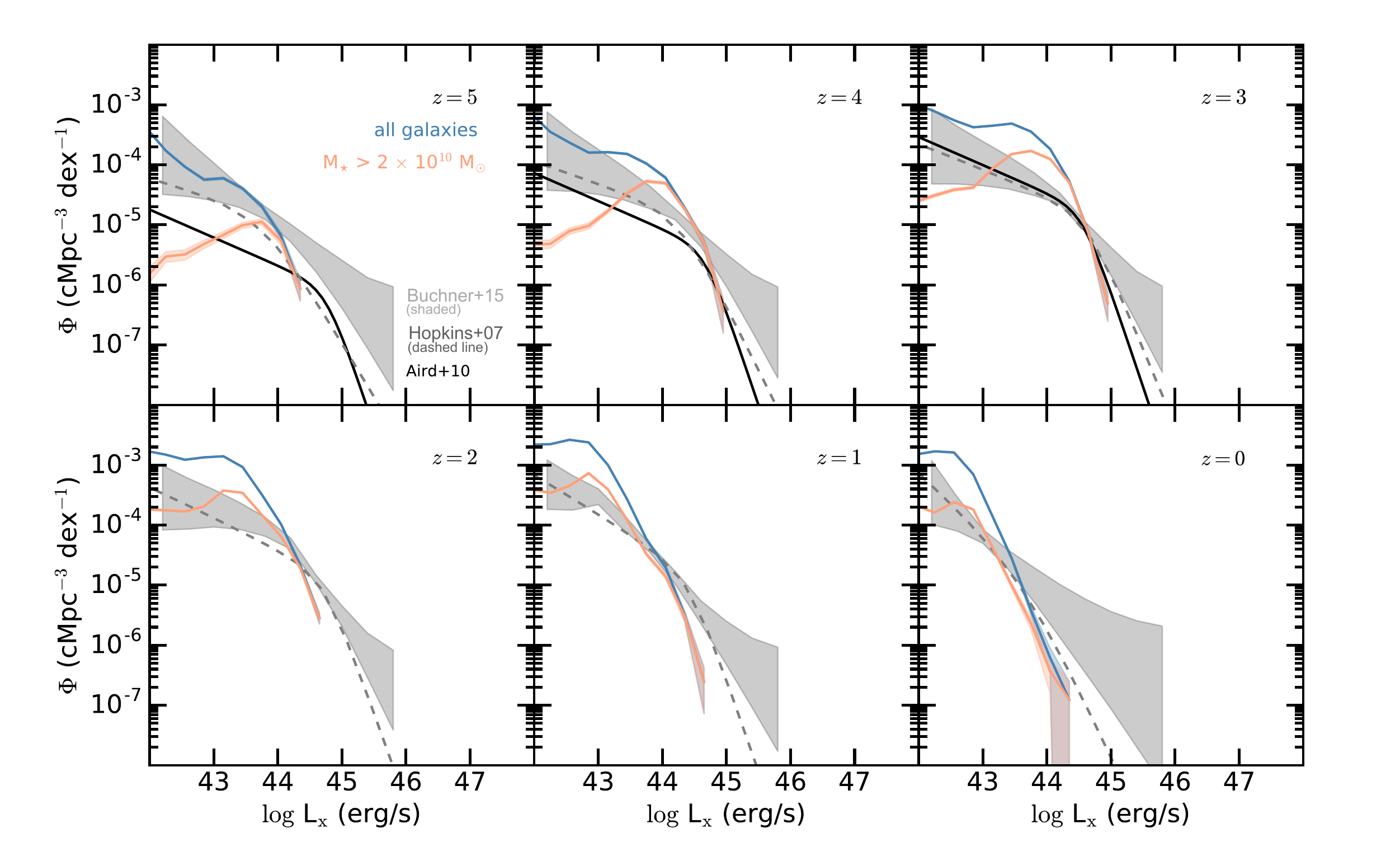}
\caption{Hard X-ray (2-10 keV) luminosity function of the BH population in TNG300 for $0\leqslant z\leqslant 5$, and comparison with observational constraints. We show the XLF of all BHs with blue solid lines, where the blue shaded regions represent Poissonian error bars, but are very tight because of the large population of BHs in the large simulated volume ($300^{3} \rm cMpc^{3}$). The orange solid lines represent a cut in galaxy mass, where we consider only the BHs in $M_{\star}>2\times 10^{10}\, \rm M_{\odot}$ galaxies to compute the X-ray luminosity function. The observational constraint from \citet{2010MNRAS.401.2531A} is shown as a black line, the one from \citet{Hop_bol_2007} as a dashed grey line, and the XLF from \citet{2015ApJ...802...89B} as a shaded grey area}
\label{fig:xlf}
\end{figure*}

We present in this section several diagnostics of the BH population in IllustrisTNG, i.e. the hard (2-10 keV) X-ray luminosity function, the occupation fraction of BH and AGN, the bright AGN fraction in massive galaxies as a function of time, and the Eddington ratio distribution of the BH population. 

\subsection{BH X-ray luminosity function}
\subsubsection{Estimating BH luminosities in the simulations}
The first method that has been widely used to compute the BH bolometric luminosity $L_{\rm bol}$ in cosmological simulations assumes a single expression:
\begin{eqnarray}
L_{\rm bol}=\frac{\epsilon_{\rm r}}{1-\epsilon_{\rm r}} \dot{M}_{\rm BH} c^{2}.
\end{eqnarray}
where $\epsilon_r$ is the radiative efficiency, $c$ the light speed, and $\dot{M}_{\rm BH}$ the accretion rate onto the BHs.

The second method assumes that $L_{\rm bol}$ depends on the Eddington ratio of the BH \citep{Churazov2005,2014MNRAS.442.2304H}, and therefore uses two different expressions for radiatively efficient and inefficient AGN. 
The radiative component of the total energy released by a BH is high for BHs with an Eddington ratio of $f_{\rm Edd}>0.1$, and therefore the bolometric luminosity is computed as in Eq. (5).
However, the radiative component is lower for BHs with a small Eddington ratio of $f_{\rm Edd}\leqslant 0.1$, and their bolometric luminosity can be written as:
\begin{eqnarray}
L_{\rm bol}=0.1 L_{\rm Edd} (10 f_{\rm Edd})^{2}=(10 f_{\rm Edd}) \epsilon_{\rm r} \dot{M}_{\rm BH} c^{2} .
\end{eqnarray}

In this paper we will use the single expression of $L_{\rm BH}$ to be able to compare our results with the majority of previous studies, but we compare and discuss the differences of the two methods in the following.

The hard X-ray luminosity is then computed by applying the bolometric correction, BC, of \citet{Hop_bol_2007}:
\begin{eqnarray}
\log_{10} L_{\rm 2-10\, keV,\odot}=\log_{10} L_{\rm bol,\odot} - \log_{10} \rm BC,
\end{eqnarray}
with
\begin{eqnarray}
{\rm BC} = 10.83 \left(\frac{L_{\rm bol,\odot}}{10^{10}\, \rm L_{\odot}} \right)^{0.28} + 6.08 \left(\frac{L_{\rm bol,\odot}}{10^{10}\, \rm L_{\odot}} \right)^{-0.020}.
\end{eqnarray}

An excess of faint AGN is found in the bolometric luminosity function of the simulations \citep[not shown here, but see][]{2017arXiv171004659W} when compared to observational constraints \citep{Hop_bol_2007,2015ApJ...802..102L}.
The radiative efficiency is not well constrained, and could be distributed over a large range of possible values. We choose here to adopt an intermediate value $\epsilon_r=0.1$, that we set the same for all BHs.

\subsubsection{Results}
In Fig.~\ref{fig:xlf}, we show the hard X-ray (i.e. in the band 2-10 keV, hereafter XLF) luminosity function of the BH population in IllustrisTNG (TNG300) for $0\leqslant z\leqslant 5$, and compare it with observational constraints. For all panels, we use a bin of 0.3 dex. This uses the first method described above with a single expression for the bolometric luminosity of BHs, while the same figure but using the second method \citep{Churazov2005} is shown in Fig.~\ref{fig:LF_churazov}.

We compare the simulated XLF to \citet{2015ApJ...802...89B}, which covers a large range of redshift and includes a correction for Compton thick AGN, i.e. AGN with large neutral hydrogen column density of $N_{\rm H}\geqslant 10^{24}\, \rm cm^{-2}$, and obscured AGN, i.e. with column density of $10^{22}<N_{\rm H}<10^{24}\, \rm cm^{-2}$.
We also include observational constraints by \citet{2010MNRAS.401.2531A} which include the Chandra Deep fields and AEGIS-X 200ks survey for $3<z<5$. There is good agreement between \citet{2015ApJ...802...89B} and \citet{2010MNRAS.401.2531A} at $z=3$. At high redshift, we note an absence of a drop in the space density reconstructed in \citet{2015ApJ...802...89B} compared to \citet{2010MNRAS.401.2531A}, potentially due to the large uncertainties in the redshift estimates in this redshift range \citep[as discussed in][]{2015ApJ...802...89B}. Finally we also include the hard X-ray luminosity function obtained by applying the bolometric correction to the best fit of the bolometric luminosity function derived in \citet{Hop_bol_2007} (grey dashed lines in Fig.~\ref{fig:xlf}).
Many more constraints on the X-ray BH luminosity function exist \citep[][and references therein]{2014ApJ...786..104U,2014MNRAS.445.3557V,2016MNRAS.463..348V,2015MNRAS.453.1946G,2010MNRAS.401.2531A,2015ApJ...804..104M,2012MNRAS.425..623L} but we only show two of them for clarity\footnote{Discrepancies between observational constraints on the XLF are most important at $z\geqslant 4$. For example, the XLFs from \citet{2014ApJ...786..104U, 2010MNRAS.401.2531A, 2014MNRAS.445.3557V,2016MNRAS.463..348V} show good agreement but have a normalization lower than \citet{2015ApJ...802...89B} for all $L_{\rm x}$ values, while the XLF derived by \citet{2015MNRAS.453.1946G} has a much lower normalization for the faint end ($L_{\rm x}<10^{44}\, \rm erg/s$). 
The XLF presented in \citet{2015A&A...578A..83G} has a faint end consistent with \citet{2015ApJ...802...89B}, while the bright end ($L_{\rm x}>10^{44}\, \rm erg/s$) is much lower. 
Discrepancies at lower redshift ($z\leqslant 3$) are smaller; all studies presented above, as well as the XLF from \citet{2015ApJ...804..104M, 2017arXiv170901926K}, are consistent with \citet{2015ApJ...802...89B}, with lower bright end normalizations ($L_{\rm x}>10^{44.5}\, \rm erg/s$) of the XLF for some of them.}.

The overall XLF is in relatively good agreement with available observational constraints. The normalization of the XLF (blue solid lines in Fig.~\ref{fig:xlf}) at high redshift $z=5$ is consistent with \citet{2015ApJ...802...89B}, and the steep decline at the bright end is in good agreement with most of the observational studies \citep{2014MNRAS.445.3557V, 2016MNRAS.463..348V, 2015MNRAS.453.1946G,2015A&A...578A..83G,2014ApJ...786..104U,2010MNRAS.401.2531A}. 
At lower redshifts $z<5$, the faint end of the XLF is higher in the simulation than for the observational constraints. Indeed, the number of $L_{\rm x}\leqslant 10^{44}\, \rm erg s^{-1}$ AGN increases with time down to $z=0$. The corresponding XLF values remain within less than an order of magnitude of the observed \citet{2015ApJ...802...89B} limits for $L_{\rm x}\leqslant 10^{44}\, \rm erg s^{-1}$.
The overproduction of faint AGN can be seen for $L_{\rm x}\leqslant 10^{42}-10^{43}\, \rm erg s^{-1}$ at $z=0$. 
The bright end of the XLF is consistent with some of the observational constraints \citep{Hop_bol_2007,2010MNRAS.401.2531A} at high redshift ($z\sim5-3$), and seems too steep compared to shallower constraints \citep{2015ApJ...802...89B}. There are fewer and fewer rapidly accreting BHs at the very bright end with time down to $z=0$, i.e. fewer AGN brighter than $L_{\rm x}\geqslant 10^{44}\, \rm erg s^{-1}$ for $z\leqslant 1$, which contrasts with observational constraints. This conclusion varies quantitatively but not qualitatively with BH luminosity estimation method, as discussed at the end of this section.

The overestimate of fainter AGN when including all resolved galaxies is a common problem in cosmological simulations \citep{2015MNRAS.452..575S,2016MNRAS.460.2979V, 2016MNRAS.462..190R}. The overproduction of faint AGN in the EAGLE simulation, which employed a thermal SN feedback, is slightly different than what we obtain for the IllustrisTNG simulations. The overproduction is present for $z>2$, but vanishes for lower redshfits. In the previous simulation Illustris \citep[whose analysis used $\epsilon_{\rm r}=0.05$; ][]{2015MNRAS.452..575S}, good agreement was found for the fainter end of the AGN luminosity function when only considering BHs more massive than $M_{\rm BH}\geqslant 5\times 10^{7}\, \rm M_{\odot}$, indicating that accretion onto low-mass BHs and/or BHs that were not yet in the self-regulated regime could have been overestimated in the simulation. 
Strong SN feedback, producing galaxies in good agreement with observations and empirical models, has been found to be able to regulate the growth of BHs \citep{2015MNRAS.452.1502D,2017MNRAS.468.3935H} because SN feedback is strong enough to heat or eject central cold gas and prevent BH fueling. When we enforce a cut in galaxy mass (orange lines in Fig.~\ref{fig:xlf}) to remove galaxies where BH growth is still subject to SN feedback, i.e. $M_{\star}>\rm few \, 10^{10}\, \rm M_{\odot}$ \citep{2015MNRAS.452.1502D}, we see that the number of faint AGN is greatly reduced. This suggests that poorly resolved feedback processes in low-mass galaxies may be responsible for the discrepancy. 

Regarding the possible smaller number of AGN at the bright end of the X-ray luminosity function compared to observational constraints, we suggest that this is due to the high efficiency of the kinetic mode of AGN feedback. In this mode the self-regulation of BHs is very efficient, and strongly suppresses their accretion rate/luminosity.
In the following section, we address this point again by comparing the AGN fraction in massive galaxies with observational constraints.

Finally, in the following we emphasize and discuss the uncertainties in computing the predicted X-ray luminosity function from the simulations. 
To be able to compare our results to others, in the main text of the paper we have derived the bolometric luminosity of BHs with a single expression. In Fig.~\ref{fig:LF_churazov}, we follow \citet{Churazov2005} and compute the bolometric luminosity assuming two different expressions for radiatively efficient and inefficient AGN. The main effect of this change on the XLF is to lower the number of faint AGN \citep{2014MNRAS.442.2304H} at low redshift $z\leqslant 1$. As the goal of the present paper is to quantify AGN fractions in galaxies and compare our findings to observations at $z>1$, we prefer to use the single expression approach described above. 
The radiative efficiency $\epsilon_r$ was set to $\epsilon_r=0.2$ self-consistently in the simulations, but we adopt $\epsilon_r=0.1$ for our post-processing analysis as this is the value commonly assumed in most studies. In Fig.~\ref{fig:xlf_eps} we show the same figure as Fig.~\ref{fig:xlf} when adopting $\epsilon_r=0.2$. The normalization of the XLF and therefore the excess of the number of faint AGN increases with increasing $\epsilon_r$. Using $\epsilon_r=0.2$ also leads to a bright end of the XLF in better agreement with the lower limits of \citet{2015ApJ...802...89B}. We computed the XLF for the two different simulation boxes TNG100/TNG300 to test resolution effects \citep[see also Appendix of ][]{2017arXiv171004659W}, and found that the main effect is that a higher resolution increases the normalization of the XLF, because the density around BHs is better resolved and higher than in lower resolution simulations, here by a factor of a few.
The bolometric corrections \citep{Hop_bol_2007,2007MNRAS.381.1235V,2009MNRAS.392.1124V,2012MNRAS.425..623L}, and the fraction of obscured and Compton-thick AGN \citep{2014ApJ...786..104U,2015ApJ...802...89B,2015MNRAS.451.1892A,2017MNRAS.469.3232G}, are also sources of uncertainty. Finally, we have not taken into account the short timescale variability that is not resolved by the simulations, which leads to an Eddington bias in the BH luminosity function.
Introducing a scatter in the AGN luminosities will move a few bright AGN to fainter bins, but will move a large number of faint AGN to brighter bins, which results in a flattening of the bright end of the luminosity function.
The impact of unresolved short time-scale variations has been studied in \citet{2016MNRAS.462..190R}, where the AGN luminosities are convolved with a log-normal distribution with quite large values of $\sigma=0.3-0.5$, leading to a slight increase of the number density of bright AGN (with $L_{\rm x}\geqslant 10^{44}\, \rm erg/s$).

\subsection{Time evolution of the AGN fraction in massive galaxies}

The very large simulated volume, i.e. 300$^{3}$ cMpc$^{3}$, allows us to probe the massive end of the galaxy population. There are 66, 877, 2164, and 4450 galaxies with $M_{\star}\geqslant 10^{11}\,\rm M_{\odot}$ in TNG300 at $z=4, 2,1,0$.
We investigate here the fraction of bright AGN ($L_{\rm bol}\geqslant 10^{44}\, \rm erg \, s^{-1}$) in massive galaxies with $M_{\star}\geqslant 10^{11} \, \rm M_{\odot}$. 

\begin{figure}
\centering
\includegraphics[scale=0.55]{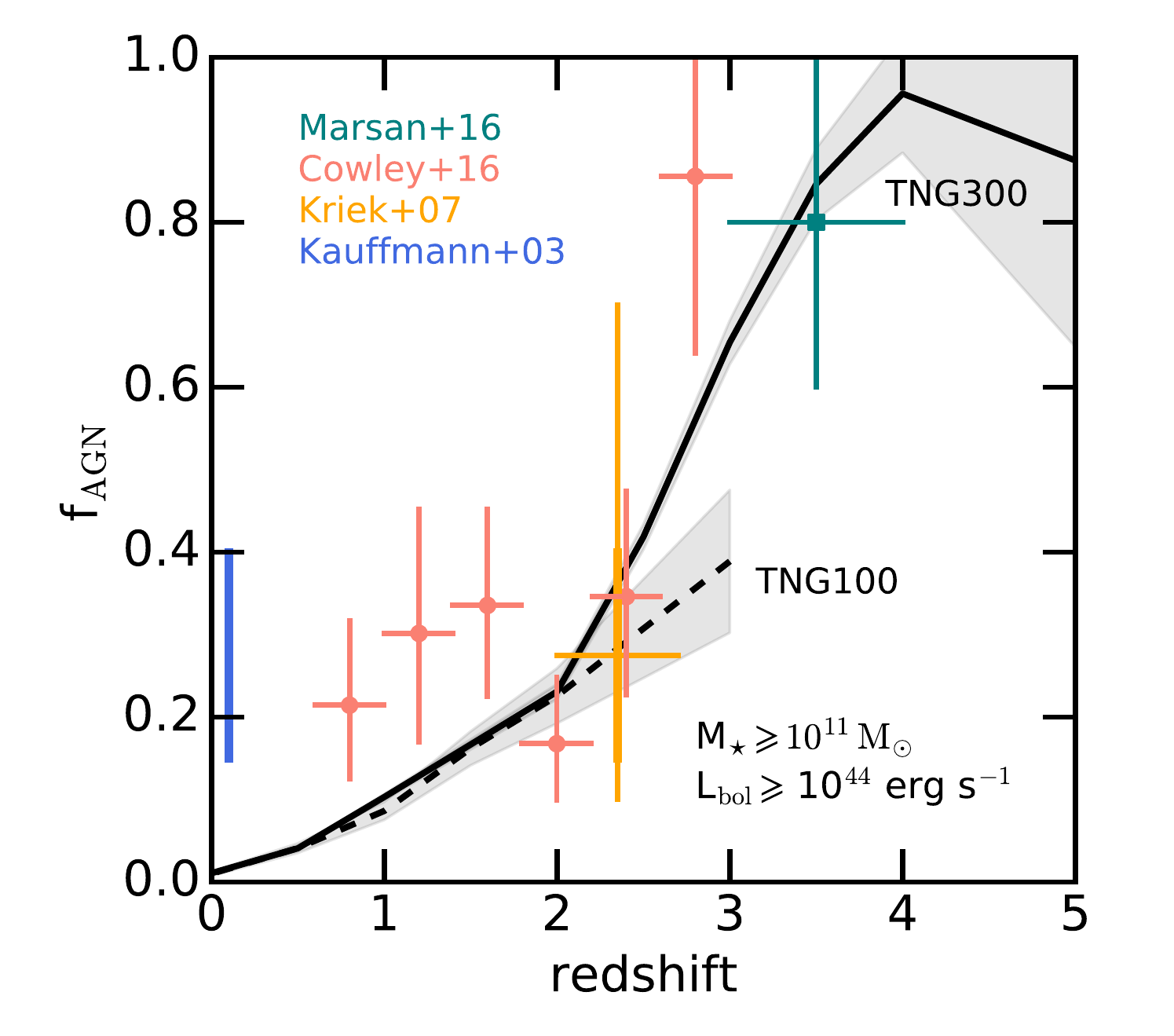}
\caption{Fraction of AGN (with $L_{\rm bol}\geqslant 10^{44}\, \rm erg/s$) in massive galaxies with $M_{\star}\geqslant 10^{11}\, \rm M_{\odot}$ in TNG300 (solid black line) and TNG100 (dashed black line), and comparison with observational constraints from \citet{2017ApJ...842...21M,2016MNRAS.457..629C,2007ApJ...669..776K}. We also show the constraint from \citet{Kauffmann2003} to guide the eye at $z\sim 0$. Grey shaded areas show the Poisson errors from the number of massive galaxies and the number of AGN.}
\label{fig:occupation_fraction_massivegal}
\end{figure}

Observationally, the time evolution of the fraction of very active AGN in massive galaxies is consistent with a step function from high to low redshift, where more than $80\%$ of massive galaxies host a very luminous AGN at $z\geqslant 3$, whereas only about $15-40\%$ of galaxies do so at $z<3$ \citep{2003cxo..prop.3999V,2004ApJ...613L...5R,Kauffmann2003,2006ApJ...640...92P,2006ApJ...647..128E}. From these observations, the transition happens at $z\sim 2.5$, when the Universe is $\sim 2.5$ Gyr old. 
At high redshift, we report on Fig.~\ref{fig:occupation_fraction_massivegal} the study of \citet{2017ApJ...842...21M} with a green dot, which consists of the spectroscopic analysis of 6 massive galaxies at $3<z<4$ with $M_{\star}\sim1.5-4\times 10^{11}\, \rm M_{\odot}$, and finds that $\geqslant 80 \%\pm 20\%$ of them host a very luminous AGN, with $L_{\rm bol}\sim 10^{44-46}\, \rm erg s^{-1}$. 
\citet{2016MNRAS.457..629C} study star formation properties in galaxies selected as radio, X-ray and IR AGN hosts using the zFOURGE dataset at $1<z<3$. They find a high AGN fraction of $\geqslant 80\%$ in massive galaxies binned in the redshift range $2.6<z<3.2$ (red dots in Fig.~\ref{fig:occupation_fraction_massivegal}).
For lower redshift, i.e. $z<3$, we compare the simulation with \citet{2007ApJ...669..776K} at $2<z<2.7$ (orange dot in Fig.~\ref{fig:occupation_fraction_massivegal}), which studied a sample of 11 $M_{\rm gal}\sim3\times 10^{11}\, \rm M_{\odot}$ K-selected galaxies with detected H$\alpha$ emission, and with $L_{0.5-8.0 \rm keV}$ in the range $(< 10^{42}, < 6\times 10^{43})$ erg/s, which corresponds to $L_{\rm bol}$ in ($<10^{44}, <10^{45})$ erg/s.
Only one of these AGN is detected as an X-ray AGN with $L_{0.5-8.0 \rm keV} =2.7\times 10^{42}\, \rm erg/s$, which roughly corresponds to $L_{\rm bol}=10^{44}\, \rm erg/s$, indicating that the other AGN may be highly obscured or accreting at lower sub-Eddington accretion rates. Among these galaxies four of them also present narrow-line emission from AGN, leading to an estimate of the AGN fraction of $20\%$, which includes large uncertainties.
Finally in the local Universe at $z\sim 0$ we use the observational constraint provided by the SDSS sample \citep{Kauffmann2003} to guide the eye, which indicates that less than $40\%$ of galaxies host an AGN, but this constraint includes all galaxies and is not based on X-ray detection. 

In Fig.~\ref{fig:occupation_fraction_massivegal}, we show the fraction of very active AGN with $L_{\rm BH, bol}\geqslant 10^{44}\, \rm erg \, s^{-1}$ among massive galaxies ($M_{\star}\geqslant 10^{11}\, \rm M_{\odot}$) with the solid black line for TNG300, and the dashed black line for TNG100. Observations are shown as colored symbols. We find good agreement with observations at high redshift $z\geqslant 2$, but the simulation does not seem to reproduce the observational plateau of $20\%$ of very active AGN in massive galaxies going down $z=0$.
We find similar behavior when using the bolometric estimation method assuming two different expressions for radiatively efficient and inefficient AGN; the $f_{\rm AGN}$ curve has a slightly lower amplitude.
We find good convergence between TNG300 and TNG100 for $z\leqslant 2$, but TNG100 hosts a lower fraction of very bright AGN in massive galaxies at higher redshift. This result is of course affected by the smaller number of massive galaxies present in TGN100 (due to the smaller simulated volume). In Fig.~\ref{fig:occupation_fraction_massivegal_variation}, we show and discuss the impact of small variations in the bolometric luminosity cut (left panel) and the galaxy stellar mass cut (right panel) on the AGN fraction in massive galaxies.

The measurement of the AGN fraction in massive galaxies is a difficult task, and suffers from many uncertainties. Many of the observed galaxies presented here are also star-forming, and the diagnostics of the presence of an AGN in these galaxies are particularly subject to contamination from the stellar emission (e.g. line emission in HII regions). Obscuration is also one of the major reasons that AGN are missed in observational samples, and only a few galaxies have mid-IR observations, which are less sensitive to obscuration. Galaxy selection bias can be an issue as well.
Finally the reliability of the occupation fraction estimate strongly depends on the size of the galaxy samples, which are very small in the existing observational studies at high redshift.

\subsection{Occupation fraction of AGN in galaxies}
The occupation fraction of BHs in low-mass galaxies is a crucial diagnostic of BH formation in the early Universe \citep{2012NatCo...3E1304G}. Indeed, the two main models for BH formation predict theoretically very different efficiencies of the mechanisms forming BH seeds. Light seed models, like the so-called {\it Pop III remnant model} \citep{MadauRees2001,VHM}, predict a high occupation fraction in low-mass galaxies, reaching unity for galaxies with total stellar mass of $M_{\star}\geqslant 10^{9}\, \rm M_{\odot}$ \citep[][for a physical seeding sub-grid model based on local properties and following theoretical prescriptions of light seed models in cosmological hydrodynamical simulations]{2017MNRAS.468.3935H}.
However, heavy seed models such as the direct collapse of primordial gas \citep{Loeb1994,2003ApJ...596...34B,2004MNRAS.354..292K,BVR2006,LN2006,Spaans2006,2008MNRAS.391.1961D,Wise2008,Regan09,2013MNRAS.433.1607L} predict a low occupation fraction. 
In a large simulated volume of $\rm (100\, cMpc)^{3}$ or more, the resolution is not sufficiently high to accurately resolve the crucial low-mass galaxies of $M_{\star}<10^{9}\, \rm M_{\odot}$ and to make reliable predictions on the occupation fraction, especially in the low-mass regime. 
However, the AGN occupation fraction can provide crucial information on the activity of the BH population as a function of host galaxy mass. We focus on this aspect in the following.

\begin{figure}
\centering
\includegraphics[scale=0.55]{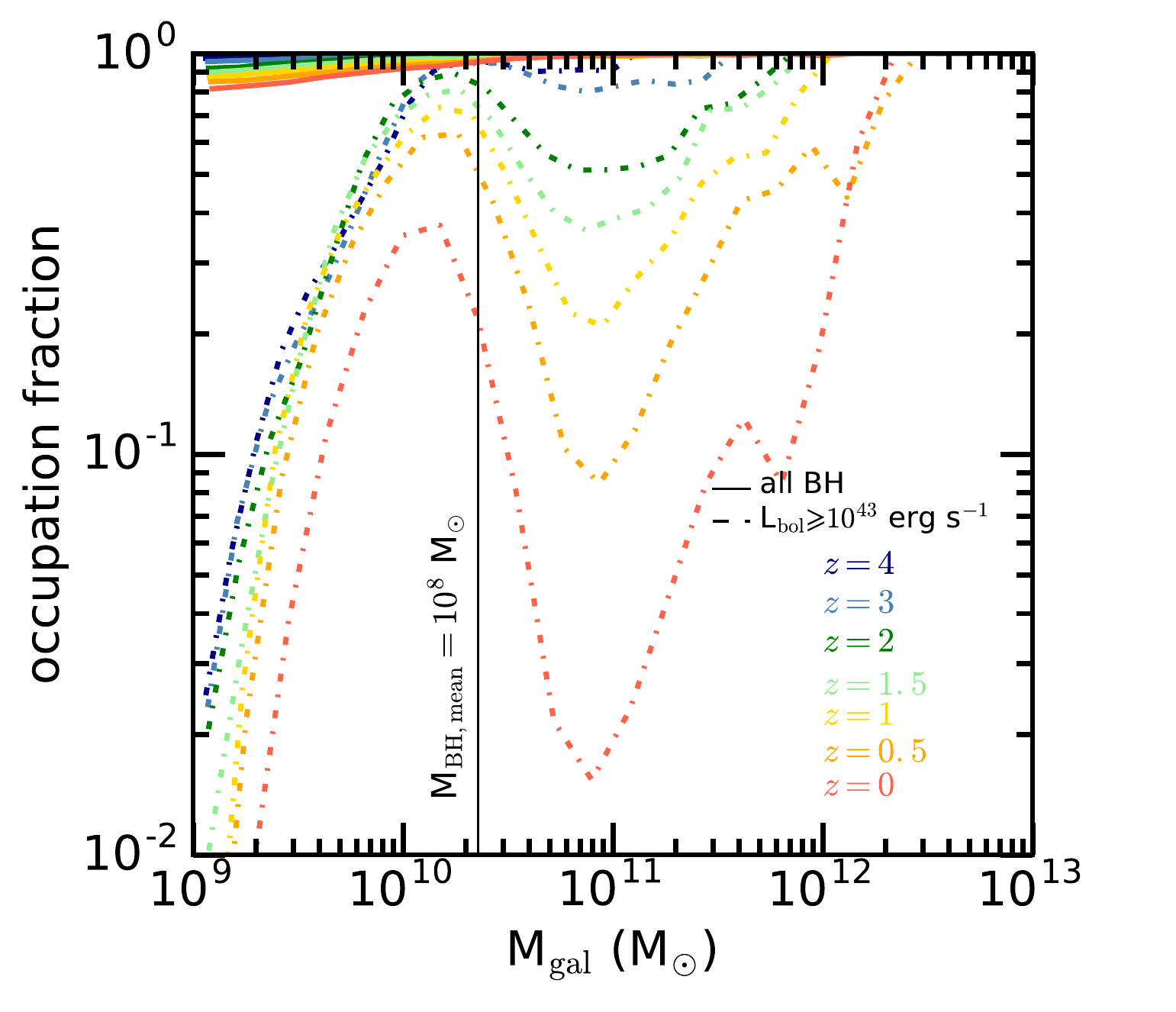}
\caption{BH occupation fraction (solid lines), i.e. the fraction of galaxies hosting a BH, and the AGN occupation fraction (dashed lines), i.e. the fraction of galaxies that experiences high BH activity, with $L_{\rm BH, bol}\geqslant 10^{43} \rm erg s^{-1}$, in TNG300. The BH occupation fraction is very close to unity, and gets closer for more massive galaxies, as expected in a hierarchical formation of structures. The AGN occupation fraction traces the global activity of BHs as a function of galaxy mass. The AGN occupation fraction increases with galaxy mass of $M_{\star}\leqslant 10^{10}\, \rm M_{\odot}$, and then suddenly drops, before increasing again for more massive galaxies of $M_{\star}>10^{11}\, \rm M_{\odot}$. This drop reflects the transition between the quasar AGN feedback mode to the more efficient kinetic mode at $M_{\rm BH}\sim 10^{8}\, \rm M_{\odot}$. }
\label{fig:bh_occ_fraction}
\end{figure}

In Fig.~\ref{fig:bh_occ_fraction} we show the BH occupation fraction as a function of galaxy total stellar mass, and redshift (solid lines, colors are labelled on the figure). Here, we show the occupation fraction of all galaxies, i.e. both central and satellite galaxies, but the behavior of the two populations is similar. As a consequence of the seeding model employed in the simulation (i.e. placement of a seed BH with fixed mass in halos above a mass threshold), the BH occupation fraction is very close to unity for $M_{\star}~10^{9}\, \rm M_{\odot}$, and gets closer for more massive galaxies. A BH occupation fraction close to unity for $M_{\star}\geqslant 10^{9}\, \rm M_{\odot}$ is what is expected from light seed BH formation models \citep{2017MNRAS.468.3935H}.

In Fig.~\ref{fig:bh_occ_fraction} we also present the AGN occupation fraction (dashed lines) as a function of the stellar mass of their host galaxies and redshift. Here, we define AGN as BHs with $L_{\rm bol}\geqslant 10^{43} \rm erg\, s^{-1}$. The choice of the AGN definition only affects the normalization of the occupation fraction, where higher luminosity thresholds decrease the normalization, because BHs with higher accretion rates are rarer. The function increases with galaxy mass for $M_{\star}\geqslant 10^{9}\, \rm M_{\odot}$, similar to the BH occupation fraction. However, the values reached by the AGN occupation fraction are lower than those of the BH occupation, as expected. This reflects the BH duty cycle: not all BHs are active and seen as AGN at the same time. The fraction then suddenly drops at around the characteristic galaxy mass $M_{\star}\sim10^{10}\, \rm M_{\odot}$, which corresponds approximately to the mean BH mass $M_{\rm BH}= 10^{8}\, \rm M_{\odot}$, that we report on the figure as a solid black vertical line. 
The Eddington ratio and the BH mass are used in the AGN feedback model of the simulation to switch between the quasar mode of the feedback (which is modeled as isotropic injection of thermal energy in the surroundings of the BH) to a pure kinetic mode \citep{2017MNRAS.465.3291W}. Indeed, BHs with $M_{\rm BH}=10^{7}\, \rm M_{\odot}$ and Eddington ratio below $f_{\rm Edd}\leqslant 2\times 10^{-5}$ (see Eq. (1)), which represents a minority of the entire population of BHs, enter in the kinetic AGN feedback mode. However, more massive BHs of $M_{\rm BH}=10^{8}\, \rm M_{\odot}$ enter the kinetic AGN feedback regime with $f_{\rm Edd}\leqslant 2\times 10^{-3}$, which represents a larger population of BHs. Therefore many more BHs of about $M_{\rm BH}=10^{8}\, \rm M_{\odot}$ have their activity regulated by their own kinetic feedback than the  $M_{\rm BH}=10^{7}\, \rm M_{\odot}$ BHs. We discuss this in more detail in the following subsection, and more details regarding the impact of AGN feedback can be found in \citet{2017arXiv171004659W}.
The drop in the AGN occupation fraction is here a direct illustration of the impact of the kinetic AGN feedback on the population of BHs and host galaxies. When the kinetic feedback is active the global activity of BHs is shut down, and this starts happening for an increasing number of BHs in galaxies with $M_{\star}\geqslant 10^{10}\, \rm M_{\odot}$. Galaxies keep growing in mass, and eventually some of them overcome the effect of the AGN feedback. The occupation fraction therefore increases again for high mass galaxies with $M_{\star}\geqslant 10^{11}\, \rm M_{\odot}$. As explained in Section 7, we have followed back in time these massive galaxies from $z=0$ toward higher redshift. From our analysis, we note that a fraction of these galaxies experience new episodes of accretion onto the BHs and simultaneously of SFR. These episodes are correlated with an increase of the inner gas mass (in the half-mass radius), and an increase of the ex-situ stellar mass. This indicates that galaxy-galaxy mergers at later times are able to increase the available gas mass to trigger both new SFR and fuel the central BHs between AGN feedback episodes.

\subsection{Eddington ratio}
\begin{figure}
\centering
\includegraphics[scale=0.52]{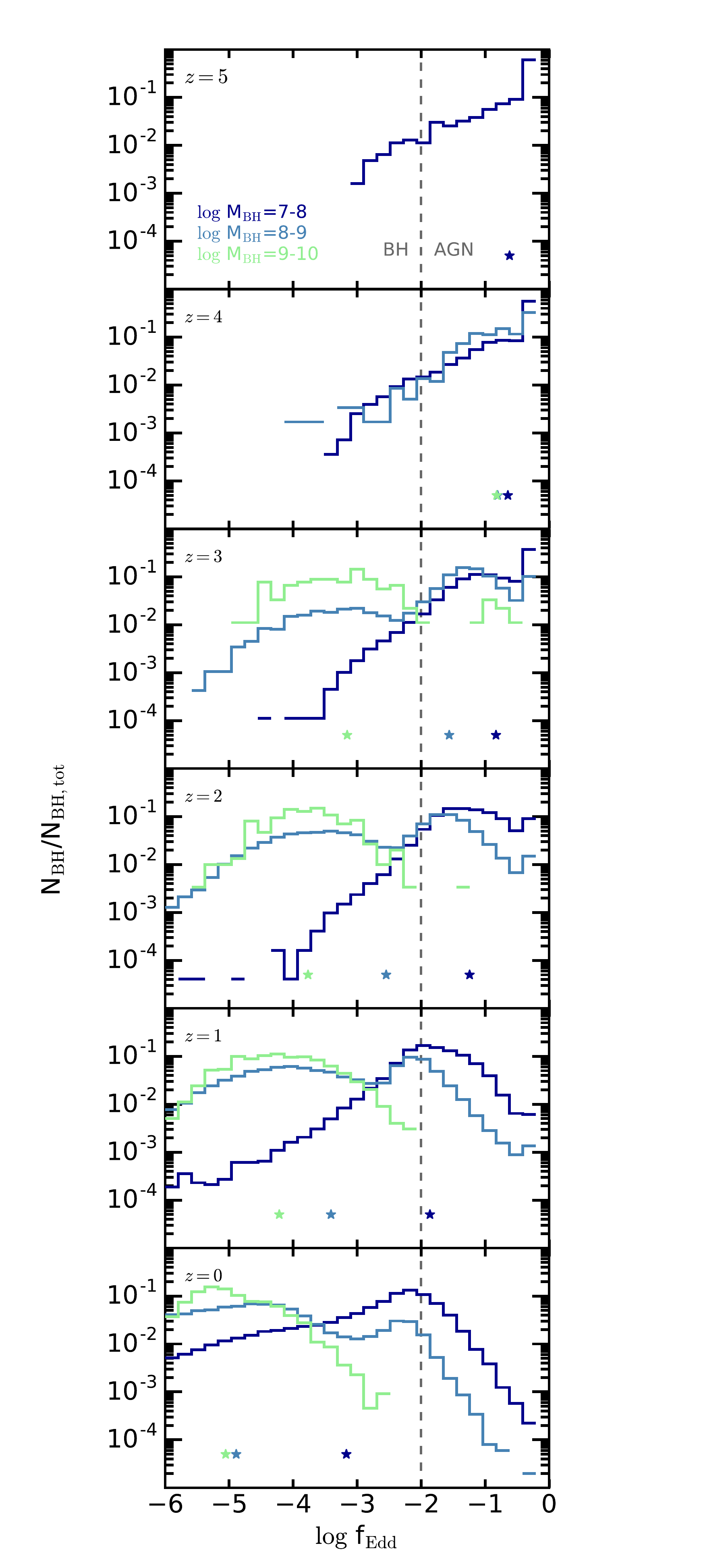}
\caption{Distributions of BH Eddington ratios in TNG300 at $z=5,4,3,2,1,0$, as indicated on the legend. Here, we assume $\epsilon_{\rm r}=0.1$. The distributions are normalized to the total number of BHs within the given BH mass range. To guide the eye, we draw the limit $\log_{10} f_{\rm Edd}=-2$ commonly used to distinguish between radiatively efficient and radiatively inefficient accretion (vertical grey dashed line). There is a clear evolution of the Eddington ratio with time, with BHs being mostly in the AGN regime at early times, and moving to the radiatively inefficient regime at later times. Star symbols represent the mean $\langle\log_{10} f_{\rm Edd}\rangle$ of the distributions.}
\label{fig:edd_ratio}
\end{figure}

In Fig.~\ref{fig:edd_ratio} we show the distribution of BH Eddington ratios at $z=5,4,3,2,1,0$ (top to bottom panels), again assuming $\epsilon_{\rm r}=0.1$. Each bin of the distributions is normalized to the total number of BHs within the given BH mass range, $\log_{10} M_{\rm BH}/\rm M_{\odot}=7-8,8-9,9-10$. To guide the eye, we draw the limit $\log_{10} f_{\rm Edd}=-2$ commonly used to distinguish between radiatively efficient and radiatively inefficient accreting BHs (vertical grey dashed line).

\begin{figure}
\centering
\includegraphics[scale=0.525]{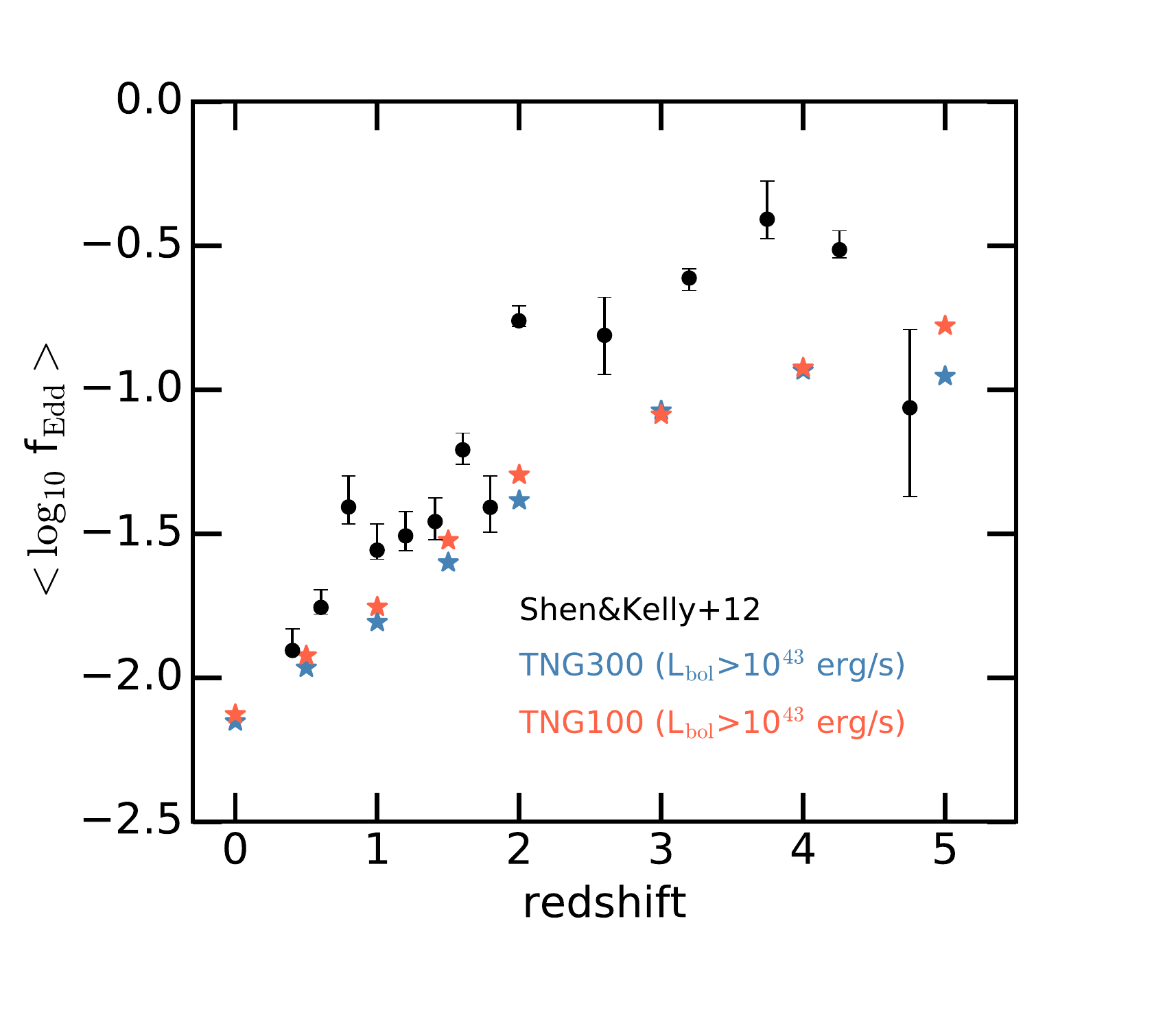}
\caption{Mean values of the Eddington ratio $<\log_{10} f_{\rm Edd}>$ of the BH population for $L_{\rm bol}>10^{43}\, \rm erg/s$ (blue star symbols for TNG300, and red star symbols for TNG100), and comparison with the observational estimates of \citet{2012ApJ...746..169S}. Good overall agreement is found between observations and simulations. The discrepancies are larger at $z>2$ where, however, observational estimates are more uncertain.} 
\label{fig:mean_fedd}
\end{figure}

There is a clear evolution of the Eddington ratio with time in Fig.~\ref{fig:edd_ratio}. This can be seen easily from the star symbols on the bottom of each panel, which indicate the mean $\langle \log_{10} f_{\rm Edd}\rangle$ of the distributions. 
The evolution of the mean $f_{\rm Edd}$ with redshift (shown in Fig.~\ref{fig:mean_fedd}) is in good agreement with the observational constraints of \citet{2012ApJ...746..169S} in the range $0 \leqslant z \leqslant 2$, and has a slightly lower normalization for higher redshifts but is still consistent with observations.

For the three BH mass ranges presented in Fig.~\ref{fig:edd_ratio},
the mean of the distribution moves from high to low accretion rate with time. At early times, almost all BHs are accreting close to the Eddington limit, but with time more and more BHs transition from the AGN regime (right of the dashed vertical line) to the radiatively inefficient BH regime (left of the vertical dashed line). Moreover at any redshift while the lower mass BHs (dark blue histogram) follow a distribution which peaks at $\log_{10} f_{\rm Edd}\geqslant -2$, the distribution of the most massive BHs (green histogram) peaks at very sub-Eddington ratios at all redshifts.

For the intermediate BH range $\log_{10} M_{\rm BH}/\rm M_{\odot}=8-9$, we observe a bimodality of the Eddington ratio distribution from $z=3$ to $z=0$, i.e. two peaks in the distribution: the first one close to the AGN regime limit (i.e. $\log_{10} f_{\rm Edd}\sim [-1,-2]$), and the second one at much lower Eddington ratio (i.e. $\log_{10} f_{\rm Edd}\sim [-3.5,-4.5]$). The bimodality is due to the transition between the thermal and kinetic modes of the AGN feedback model. 
For BHs of mass $\log_{10} M_{\rm BH}/\rm M_{\odot}=8$, the transition happens at $\log_{10} f_{\rm Edd}=-2.7$, for more massive BHs of $\log_{10} M_{\rm BH}/\rm M_{\odot}=9$, the transition happens for higher Eddington ratios of $\log_{10} f_{\rm Edd}=-1$.
Some of the BHs in this BH mass range, i.e. the most massive, already suffer from the efficient kinetic feedback. Their gas accretion and growth is regulated by their own kinetic feedback, which leads to a population with strongly sub-Eddington ratios. The modeling of the AGN feedback is constructed to favor this mode once it is reached in the first place \citep{2017MNRAS.465.3291W,2017arXiv171004659W}, and indeed the distribution is almost unchanged for more massive BHs (green histogram). Some of the BHs in the range $\log_{10} M_{\rm BH}/\rm M_{\odot}=8-9$ are not yet in this AGN kinetic mode, and therefore are able to accrete gas at rates close to $\log_{10} f_{\rm Edd}\geqslant -2$; their distribution peaks at similar Eddington ratios as the lower BH mass range $\log_{10} M_{\rm BH}/\rm M_{\odot}=7-8$. 

\begin{figure}
\centering
\includegraphics[scale=0.6]{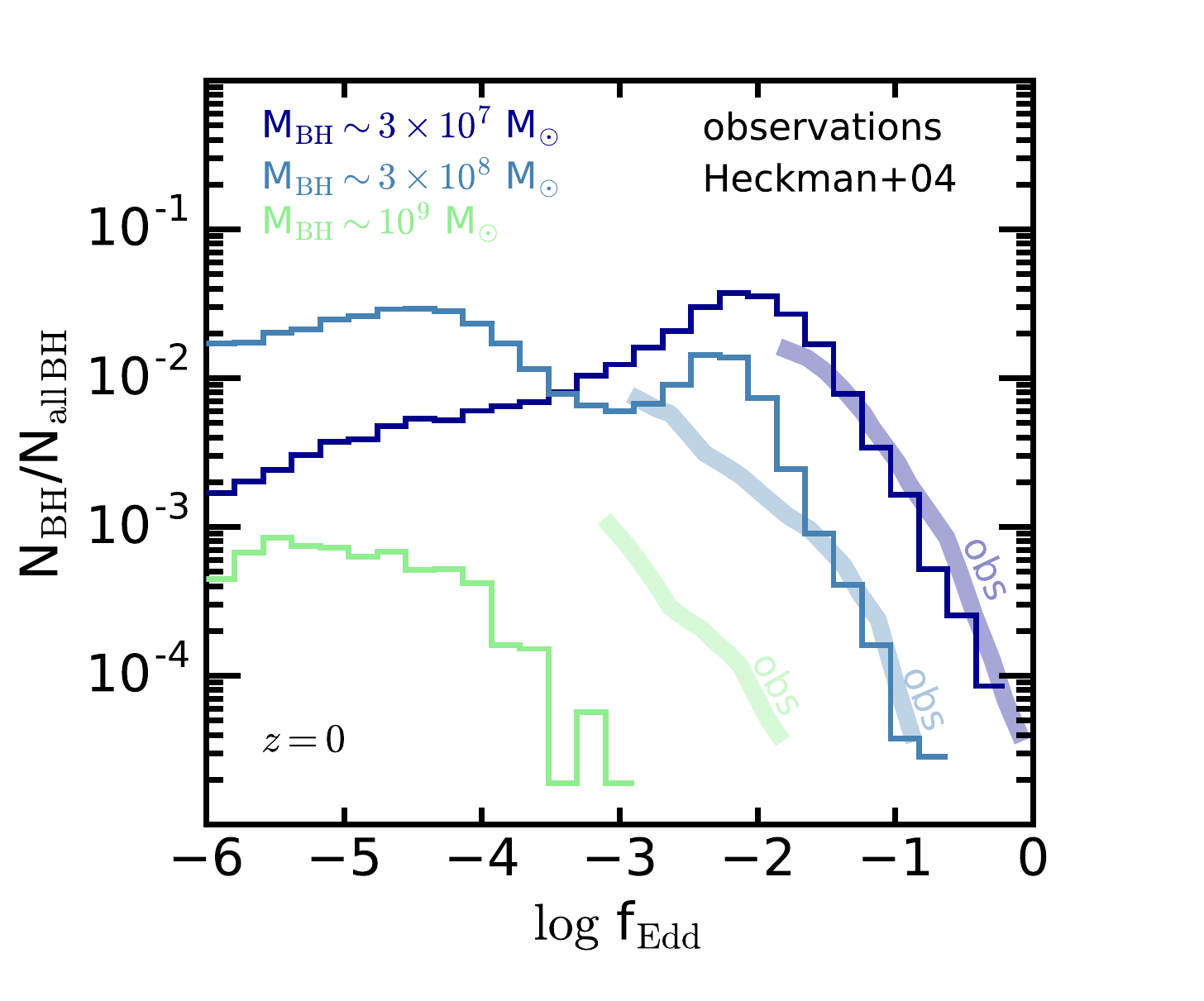}
\caption{Comparison between the distribution of BH Eddington ratios in TNG300 (thin lines) and observations from the SDSS sample at $z\sim 0$ \citep{2004ApJ...613..109H} (thick shaded lines). To compare with the observational data, each bin is normalized to the total number of BHs in the three mass bins.}
\label{fig:edd_ratio_compheckman}
\end{figure}

We now turn to comparing the Eddington ratios of the simulated BHs to the Eddington ratio distribution found in the SDSS sample at $z\sim0$. \citet{2004ApJ...613..109H} classify BHs in the SDSS sample according to the luminosity of the emission line [OIII]$\lambda$5007, which is the strongest emission line in the optical spectra of Type 2 AGN, and has the advantage of not being too contaminated by emission lines from star-forming HII regions. Their selection favors relatively luminous AGN. BH masses are computed with the empirical relation \citep{Tremaine02} between BH mass and bulge velocity dispersion, which are taken from the SDSS catalog. We report their Eddington distribution in Fig.~\ref{fig:edd_ratio_compheckman} as thick shaded lines, for the BH masses $M_{\rm BH}=3\times 10^{7}\, \rm M_{\odot}$, $M_{\rm BH}=3\times 10^{8}\, \rm M_{\odot}$, and $M_{\rm BH}=10^{9}\, \rm M_{\odot}$. To compare with these observations \citep[see also][]{2007ApJ...656...84G}, we use the three BH mass ranges: $1-5 \times 10^{7}\, \rm M_{\odot}$ (dark blue distribution), $1-5 \times 10^{8}\, \rm M_{\odot}$ (light blue distribution), and $0.8-2 \times 10^{9}\, \rm M_{\odot}$ (light green distribution). The normalisation here is also slightly different from the previous plot: instead of normalizing each bin by the total number of BHs in the BH mass range, we normalize by the total number of BHs in the three BH mass bins. We find good agreement for the distributions in the lower BH mass bins, i.e.  $M_{\rm BH}=3\times 10^{7}\, \rm M_{\odot}$ and  $M_{\rm BH}=3\times 10^{8}\, \rm M_{\odot}$. However, for more massive BHs, $M_{\rm BH}=10^{9}\, \rm M_{\odot}$, the number of BHs accreting at a high rate is definitely lower in the simulation than in the observations, but the slope of the distribution stays in good agreement. This may again reflect overly strong quenching accretion onto the most massive BHs, i.e. when the kinetic AGN feedback is effective. \\

\section{Galaxy sample selection}
We now turn to studying the time evolution of the star formation activity and structural properties of the galaxy population in IllustrisTNG (including both centrals and satellites), and design criteria to identify different types of galaxies.

\subsection{Defining the star-forming sequence and quiescent galaxies}
We define $M_{\star}$ in the simulations as the mass in stars, measured within a sphere with a radius twice the 3D half-stellar mass radius of the stellar body of galaxies, and instantaneous SFR as the star formation rate of gas cells within this same radius.
We start by defining the star-forming (SF) sequence (sometimes called ``Main Sequence''; MS) of galaxies in the simulations following the approach proposed by \citet{2015MNRAS.451.2933B}\footnote{A different definition of the star-forming sequence of galaxies in the IllustrisTNG simulations has been derived in \citet{2018MNRAS.477L..16T}, and is based on the median of the SFR (their Fig. 1).}. 
We compute the mean star formation rate in bins of galaxy mass in the range $10^{9}-10^{10}\, \rm M_{\odot}$ for which we do not expect many quenched galaxies, for redshifts from $z=5$ to $z=0$.
We extend the sequence to larger galaxy stellar masses, assuming that the SFR-$M_{\star}$ relation is a power law defined by:
\begin{eqnarray}
\log_{10} \rm{SFR_{\rm MS}}=\alpha + \beta \log_{10}\left( \frac{M_{\star}}{10^{10}\, \rm M_{\odot}}\right),
\end{eqnarray}
with SFR$_{\rm MS}$ in $\rm M_{\odot}/yr$ the star formation rate of galaxies on the star-forming sequence, $\alpha=1.52,0.96,0.70,0.23$, and $\beta=0.93,0.84,0.76,0.81$, for $z=4,2,1.5,0.5$ for the TNG300 simulation. The discrepancies between TNG100 and TNG300 are very small. The SF sequence of TNG100 is very slightly less steep at $z=2$, which mostly affects the galaxy mass range $M_{\star}\geqslant 10^{11} \rm M_{\odot}$. By $z=0$ they are almost indistinguishable.
We refer the reader to Table~\ref{table:table_SFR} for a comparison of the parameters of Eq. (9) for TNG300 and TNG100.

\begin{table*}
\caption{Parameters of the SF sequence at different redshifts, for TNG300 and TNG100. We find good agreement between the two simulations, and some discrepancies with the observations from the {\sc candels} observations.}
\begin{center}
\begin{tabular}{ccccccccc}
\hline
 && \multicolumn{3}{c}{$\alpha$} && \multicolumn{3}{c}{$\beta$} \\
\multicolumn{1}{c}{Redshift} &&  \multicolumn{1}{c}{TNG300} & \multicolumn{1}{c}{TNG100} & \multicolumn{1}{c}{Obs} && \multicolumn{1}{c}{TNG300} & \multicolumn{1}{c}{TNG100} & \multicolumn{1}{c}{Obs} \\

\hline         
$z=4$    && 1.52 & 1.53 &         && 0.93 & 0.93 & \\
$z=2$    && 0.96 & 0.90 & 1.15 && 0.84 & 0.82 & 0.58\\
$z=1.5$ && 0.70 & 0.70 & 0.99 && 0.76 & 0.85 & 0.68 \\
$z=0.5$ &&  0.23 & 0.16 &        && 0.81 & 0.84 & \\

\hline
\end {tabular}
\end{center}
\label{table:table_SFR}
\end{table*}

\begin{figure*}
\centering
\includegraphics[scale=0.56]{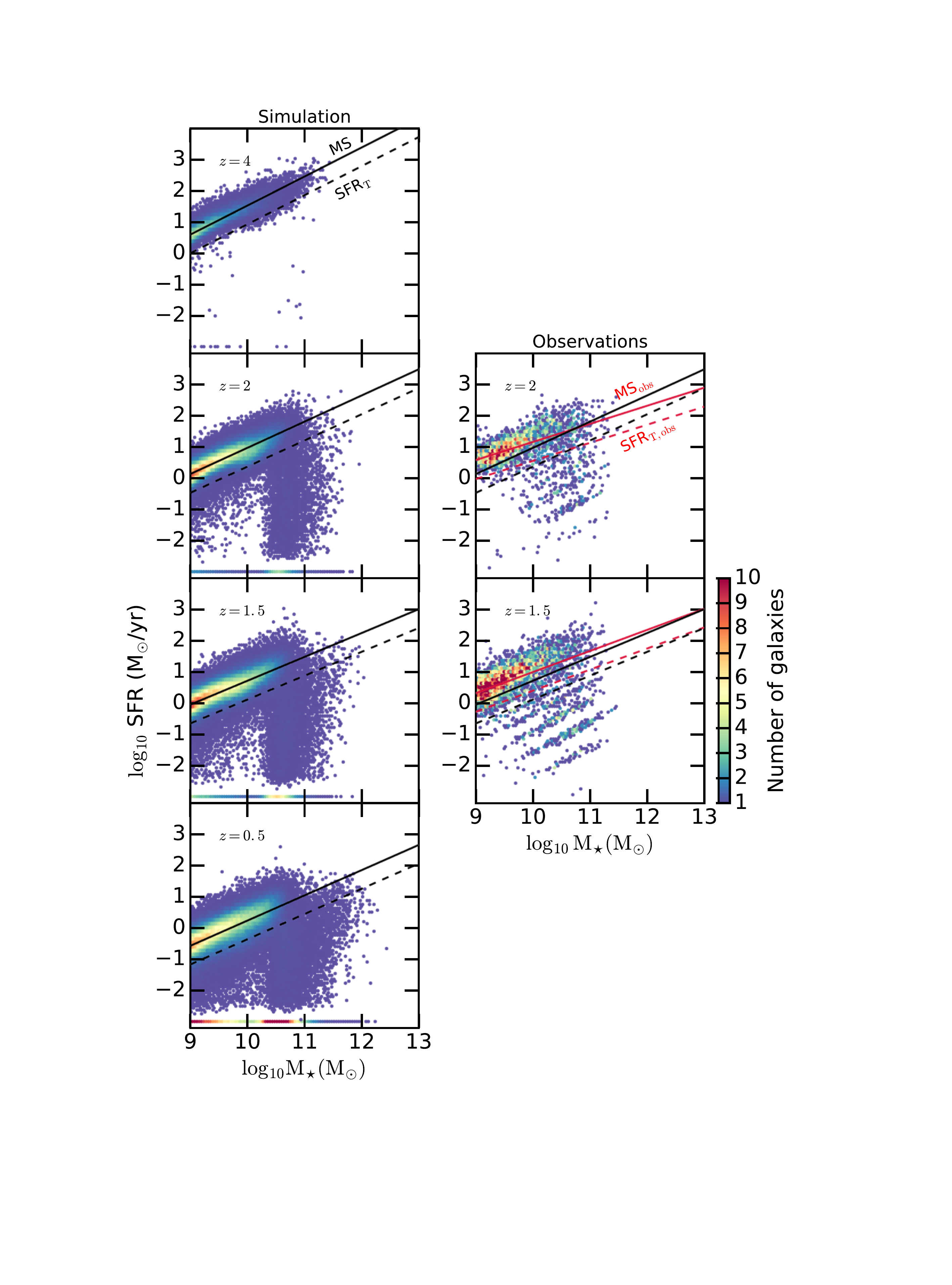}
\caption{Star formation rate (in $\rm M_{\odot}/yr$ unit) as a function of $M_{\star}$ for galaxies with mass $M_{\star}\geqslant 10^{9}\, \rm M_{\odot}$ at redshifts $z=4, 2, 1.5, 0.5$ for the simulation on the left panels (TNG300) and for the {\sc candels} observations on the right panels. The data are color coded by the number density of galaxies in the range 1-10 for the observations, and 1-100 for the simulation. For each redshift, we derive the star-forming sequence as a fit to the mean SFR in the range $M_{\star}=10^{9-10}\, \rm M_{\odot}$ (less affected by the population of quenched galaxies) as a function of stellar mass. We extrapolate the main sequence to larger galaxy mass, and show it as solid black lines (MS) for the simulation. We also show with dashed black lines ($\rm SFR_{\rm T}$) our selection criterion to identify star-forming vs. quiescent galaxies, defined as  $25\%$ of the main sequence SFR. Galaxies that are above this limit are considered as star-forming, those under the limit as quiescent galaxies.
We also repeat these black lines in the observation panels, and add the $\rm MS_{obs}$ and $\rm SFR_{\rm T,obs}$ derived from the observations as solid and dashed red lines.
Main sequences of simulated and observed galaxies are in quite good agreement (with some discrepancies in the normalization and the slope), especially for the galaxy mass range of interest $M_{\star}\geqslant 10^{10}\, \rm M_{\odot}$.}
\label{fig:sfr_mstar}
\end{figure*}

We consider a galaxy as being star-forming if its SFR is greater than $25\%$ of the SF sequence value for the same stellar mass and redshift \citep[following][]{2015MNRAS.451.2933B}. Conversely, a galaxy is considered as quiescent if its SFR is lower than $25\%$ of the SF sequence SFR. This criterion to distinguish star-forming from quiescent galaxies is established for a given population of galaxies, i.e. for a given cosmological simulation, at a given redshift. 
On the left panels of Fig.~\ref{fig:sfr_mstar} we show the SFR of galaxies as a function of their stellar mass in small hexabins color coded by the number of galaxies in each bin. The solid black line represents the fit of the SF sequence, and the dashed black line our threshold to identify galaxies as star-forming or quiescent.

The number of quenched galaxies increases with time in the simulations. Using TNG300, we find that the percentage of $M_{\star}\geqslant 10^{9} \, \rm M_{\odot}$  quenched galaxies is $3\%$ at $z=3$, and increases to $42\%$ by $z=0$. Similarly for more massive galaxies of $M_{\star}\geqslant 10^{10} \, \rm M_{\odot}$, for which we find $13\%$ and $67\%$ of quenched galaxies at $z=3$ and $z=0$ respectively, and $58\%$ and $98\%$ of $M_{\star}\geqslant 10^{11} \, \rm M_{\odot}$ quenched galaxies. 

In this paper, we carry out a close comparison between IllustrisTNG and observations from the {\sc candels} survey. We particularly focus our investigations on two narrow redshift bins allowed by the observations, i.e. $1.4<z<1.8$ (referred to as $z\sim 1.5$ hereafter) and $1.8<z<2.2$ ($z\sim 2$), which can be compared to the outputs at $z=1.5$ and $z=2$ of the simulation.
We first start by comparing galaxies in the SFR-$M_{\star}$ plane, as shown in the right panels of Fig.~\ref{fig:sfr_mstar}. The red solid lines show the main sequence relation we derive for the {\sc candels} sample\footnote{The stripes in the observations in Fig.~\ref{fig:sfr_mstar} are due to the use of parameter grids in the fitting codes to derive the candels catalog galaxy properties.}, and the red dashed lines the threshold that we use to define star-forming and quiescent galaxies, while the black lines are for the simulation. 
We fit the main sequence as in Eq. (9), with the parameters $\alpha=1.45,0.99$ and $\beta=0.58,0.68$ for the observations in the ranges $z\sim 2$ and $z\sim 1.5$ respectively, and $\alpha=0.96,0.70$ ($z=2$) and $\beta=0.84,0.76$ ($z=1.5$) for the simulation. 
We find good agreement overall  between the {\sc candels} observations and the simulation, especially in the galaxy mass range $M_{\star} \geqslant 10^{10} \, \rm M_{\odot}$. Some discrepancies are found in the normalization, where the normalization of the main sequence is higher in the observations than in the simulation. The largest discrepancy is the slope of the main sequence for $z=2$, which is shallower in the observations (and steeper in the observations at $z=1.5$) but this only very slightly affects the selection of galaxies in the range of interest $10^{10}\leqslant M_{\star}\leqslant 10^{11}\, \rm M_{\odot}$, for which we will compare the AGN fraction for the observed and simulated galaxies in the following Section 6.2.

\subsection{Time evolution of the structural properties of galaxies}
We first study the time evolution of galaxy sizes in the simulations. At early times $z\geqslant 4$, the sizes of galaxies are smaller than 5 pkpc, and sizes slightly decrease for more massive galaxies of $M_{\star}\sim 10^{11}\, \rm M_{\odot}$. 
At later times $z\leqslant 2$, we observe a turnaround in the slope of the size-mass relation. Sizes of galaxies spanning stellar mass between $M_{\star}\sim 10^{10}\, \rm M_{\odot}$ to $M_{\star}\sim 10^{12}\, \rm M_{\odot}$ increase with stellar mass.
The slope of the size-stellar mass relation strongly increases with time.
We divide the population of galaxies into star-forming and quiescent galaxies, as described above. Star-forming galaxies have larger sizes than quiescent galaxies. The effect is larger for the high resolution simulation.  
\citet{2017arXiv170705327G} provide a complete description and analysis of the time evolution of galaxy sizes, decomposed as well into star-forming and quiescent galaxies, and show that good agreement is found with observations \citep{2003MNRAS.343..978S,2011MNRAS.412L...6B,2014ApJ...788...28V}. 

In order to distinguish compact galaxies from more extended galaxies, we introduce a criterion which is redshift and simulation dependent. We use the projected stellar mass surface density as a measure of galaxy compactness, which we define as $\Sigma_{\rm e}= 0.5 M_{\star}/(\pi R_{\rm r-band, 2D}^{2})$, where $R_{\rm r-band, 2D}$ is the projected galaxy size containing half the optical light in the r-band, using projections along a random line-of-sight \citep{2017arXiv170705327G}.
Our choice of definitions is motivated by observational studies \citep[e.g.][]{2014ApJ...791...52B,2017ApJ...840...47B, 2017ApJ...846..112K}. Particularly, we prefer to use $\Sigma_{\rm e}$ over $\Sigma_{1}$ (the projected stellar mass density with 1 pkpc) in the following, as the resolutions of TNG100/TNG300 are not quite sufficient to robustly measure the latter quantity.

\begin{figure*}
\centering
\includegraphics[scale=0.56]{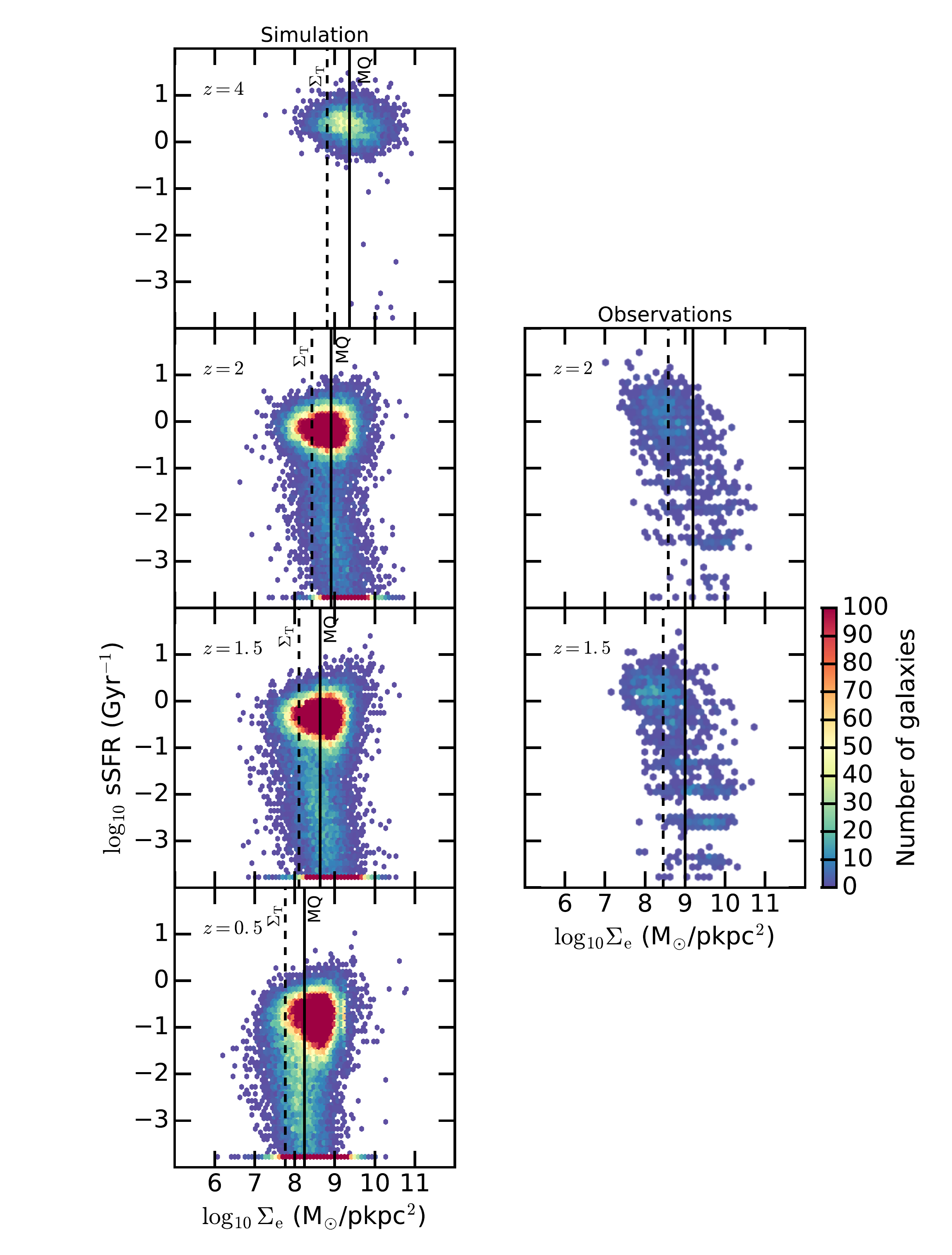}
\caption{{\it Left panels:} Redshift evolution of the galaxy population (with $M_{\star}\geqslant 10^{10}\, \rm M_{\odot}$) in TNG300 in sSFR-$\Sigma_{\rm e}$ space. The figure is color coded by the number density of galaxies in hexabins, in the range 1-100 for both the simulation and the observations. {Right panels:} sSFR-$\Sigma_{\rm e}$ plane for the observational {\sc candels} samples. For all panels, we set the minimal specific star formation rate to $\rm sSFR=2\times10^{-4}\, Gyr^{-1}$ for galaxies with a lower value. Solid black lines indicate the mean $\Sigma_{\rm e}$ value of the quiescent galaxies (MQ). The dashed black lines indicate $\Sigma_{\rm T}$, which corresponds to the 25th percentile of the quiescent galaxy population $\Sigma_{\rm e}$ distribution at a given redshift. We use this limit as a criterion to define compact vs extended galaxies: galaxies with $\Sigma_{\rm e}>\Sigma_{\rm T}$ are considered compact, and extended otherwise. }
\label{fig:second_criterion}
\end{figure*}

The specific star formation rate is defined by sSFR=SFR/$M_{\star}$. 
We show the time evolution of $M_{\star}\geqslant 10^{10}\, \rm M_{\odot}$ galaxies in the sSFR-$\Sigma_{\rm e}$ diagram at $z=4,2,1.5,0.5$ for the $300$ cMpc simulation box in the left panels of Fig.~\ref{fig:second_criterion}. The colorbar indicates the number density of galaxies in hexabins (ranging from 0 to 100 galaxies). We set sSFR$=2\times10^{-4}\, \rm Gyr^{-1}$ for galaxies with specific star formation rate below this value. 
The regions with the highest density in galaxies in the diagram are shown with reddish colors in Fig.~\ref{fig:second_criterion} and are defined by $\log_{10}  {\rm sSFR}=-1-0\, \rm Gyr^{-1}$, and $\log_{10} \Sigma_{\rm e}=8-9\, \rm M_{\odot}/pkpc^{2}$, which corresponds to the SF main sequence of galaxies, and sSFR$=2\times10^{-4}\, \rm Gyr^{-1}$ and $\log_{10} \Sigma_{\rm e}=8-9\, \rm M_{\odot}/pkpc^{2}$ which is populated by quenched galaxies. 
The $\Sigma_{\rm e}$ distribution evolves slightly to lower $\Sigma_{\rm e}$ values with time. We find similar evolution for the higher resolution IllustrisTNG simulation, i.e. with $100$ cMpc side length.
We define the mean $\Sigma_{\rm e}$ value of the quiescent population (MQ), and report it as solid black lines in Fig.~\ref{fig:second_criterion}. We find $\log_{10} \Sigma_{\rm e}=9.37,8.91,8.63,8.24 \, \rm M_{\odot}/pkpc^{2}$ for $z=4,2,1.5,0.5$ for the simulation TNG300, and very consistent values with the simulation TNG100. We refer the reader to Table~\ref{table:table_sigma} for the comparison of the mean $\Sigma_{\rm e}$ for TNG300, TNG100, and the observations from {\sc candels}.  

\begin{table}
\caption{Mean $\log_{10} \Sigma_{\rm e}$ ($\rm M_{\odot}/pkpc^{2}$) at different redshifts, for TNG300, TNG100 and the {\sc candels} galaxy population. }
\begin{center}
\begin{tabular}{ccccc}
\hline
 && \multicolumn{3}{c}{mean $\log_{10} \Sigma_{\rm e}$}  \\
\multicolumn{1}{c}{Redshift} &&  \multicolumn{1}{c}{TNG300} & \multicolumn{1}{c}{TNG100} & \multicolumn{1}{c}{Obs}  \\

\hline         
$z=4$    && 9.37 & 9.61 &  \\
$z=2$    && 8.91 & 9.06 &  9.19 \\
$z=1.5$ && 8.63 & 8.78 &  8.99  \\
$z=0.5$ &&  8.24 & 3.36 &  \\
\hline
\end {tabular}
\end{center}
\label{table:table_sigma}
\end{table}

In order to identify a broad sample of compact galaxies, we follow \citet{2017ApJ...840...47B} and define a threshold in the stellar mass surface density $\Sigma_{\rm T}$, which corresponds to the 25th percentile of the quiescent galaxies. This allows us to consider most of the peak of the quiescent population as compact galaxies. 
Galaxies with $\Sigma_{\rm e} \geqslant \Sigma_{\rm T}$ are labelled as compact, and extended otherwise.

We show the sSFR-$\Sigma_{\rm e}$ diagram for the observations in the right panels of Fig.~\ref{fig:second_criterion}. Observed galaxies populate the same region of the diagram as simulated galaxies. 
We apply our selection method as described above to identify compact and extended galaxies. 
The determination of the thresholds being sample dependent we do not expect them to be exactly the same for the simulation and the observations. We find $\log_{10} \Sigma_{\rm T} /(\rm M_{\odot}/pkpc^{-2})=8.43,8.10$ for $z=2,1.5$ in the simulation and $\log_{10} \Sigma_{\rm T} /(\rm M_{\odot}/pkpc^{-2})=8.58,8.46$ for $z\sim2,1.5$ in the observations, which represents reasonable agreement. 

\subsection{Number density of compact galaxies}
In this subsection, we discuss the number density of compact SF galaxies and compact quiescent galaxies. Many studies have derived, observationally or theoretically, the number density of these two populations of galaxies, and have found that their number density spreads over several orders of magnitude, from $n=10^{-6} \, \rm cMpc^{-3}$ to $n={\rm a \, few} \, 10^{-3}\, \rm cMpc^{-3}$ from $z=3$ to $z=0$, for $M_{\star}\geqslant 10^{10}\, \rm M_{\odot}$ \citep{2009A&A...501...15F,2010ApJ...709..644I,2011ApJ...739...24B,2013ApJ...777...18M,2013ApJ...765..104B,2013ApJ...773..112C,2014ApJ...791...52B,2014ApJ...788...28V,2015ApJ...813...23V,2015ApJ...806..158D,Charbonnier:2017psq}.
%
We do not attempt to carry out a comparison with all the observational/theoretical studies that can be found in the literature, partly because all studies use different definitions of what are compact and quiescent galaxies, but rather show that our population of compact quiescent galaxies is consistent in number density with observational findings. In addition we show the general trends of these populations' evolution with redshift.

\begin{figure}
\centering
\includegraphics[scale=0.5]{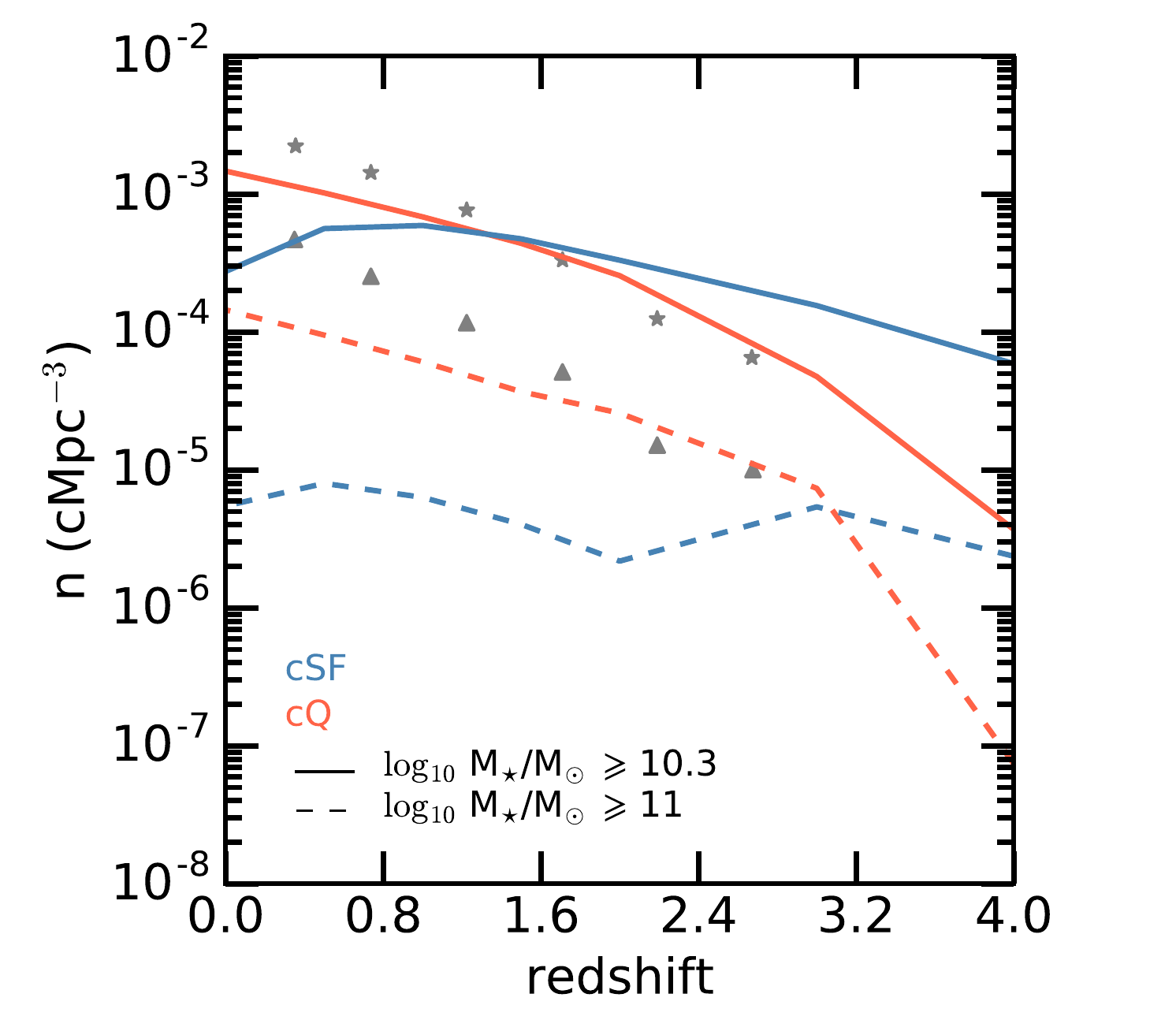}
\caption{Time evolution of the number density of SF and quiescent compact galaxies in TNG300, for two stellar mass bins ($\log_{10} {M}_{\star}/\rm M_{\odot} \geqslant 10.3$ (solid lines) and $\log_{10} {M}_{\star}/\rm M_{\odot} \geqslant 11$ (dashed lines)). The number density of SF compact galaxies with $\log_{10} {M}_{\star}/\rm M_{\odot} \geqslant 10.3$ increases from high redshift to $z\sim1$, and then decreases until $z=0$. However, the number density of quiescent galaxies increases even more strongly to $z=0$. The populations of SF compact galaxies dominate at high redshift, and the quiescent compact galaxies at low redshift.
The number density of the two populations crosses at about $z\sim1$ for $\log_{10} {M}_{\star}/\rm M_{\odot} \geqslant 10.3$ galaxies, and earlier for $\log_{10} {M}_{\star}/\rm M_{\odot} \geqslant 11$ galaxies. For comparison, we also show the number density of observed quiescent galaxies from \citet{2013ApJ...777...18M} with the star and triangle symbols to be compared with the red solid and dashed lines, respectively.
}
\label{fig:nb_density}
\end{figure}

We show in Fig.~\ref{fig:nb_density} the time evolution of the number density of compact SF (blue lines) and quiescent galaxies (orange lines) with $\log_{10} M_{\star}/\rm M_{\odot}\geqslant 10.3$ with solid lines, and high mass of $\log_{10} M_{\star}/\rm M_{\odot}\geqslant 11$ with dashed lines. We find good qualitative agreement for these mass bins with for example the results of \citet{2013ApJ...777...18M}, which is based on the UltraVista survey. We report the corresponding number density of their observed quiescent galaxies with star and triangle symbols, to be compared with the red solid and dashed lines, respectively.
The number density of SF compact galaxies in TNG300 increases from high redshift to $z\sim 1$, where it reaches a peak, and then decreases until $z=0$. However, the number density of quiescent galaxies increases even more strongly down to $z\sim 0$ \footnote{This is slightly different from the results presented in \citet{2015ApJ...813...23V}, where a decrease of the number density of compact quiescent galaxies is found with time for $z<1.8$, probably reflecting the size growth of the compact quiescent galaxies driven by minor mergers. Compact galaxies are defined with a fixed $\Sigma$ threshold in \citet{2015ApJ...813...23V}, while we use a redshift-dependent threshold that depends on the total population of quiescent galaxies. Our threshold is designed to identify a broad population of compact galaxies. Employing a more selective threshold relying on the size-mass relation leads to a similar behavior of the number density as in \citet{2015ApJ...813...23V} with a change in the slope at $z\sim 2$ and a peak at $z\sim 1.5$.}.
At $z=0$, the compact SF and quiescent populations differ by almost one order of magnitude in their number density.
The number density of the SF/quiescent populations crosses at about $z\sim 2$ for $\log_{10} {M}_{\star}/\rm M_{\odot} \geqslant 10.3$, and earlier, around $z \sim 3$, for more massive galaxies with $\log_{10} {M}_{\star}/\rm M_{\odot} \geqslant 11$. Quiescent galaxies become the predominant population of galaxies earlier for more massive galaxies. This indicates that massive galaxies become quiescent at earlier times, while less massive galaxies will quench at later times.

This time evolution of the population fractions suggests, as widely discussed in the literature, that compact SF and quiescent galaxies are not two independent galaxy populations, but instead could represent the same galaxies at a different time of their evolution. This suggests that the compact SF galaxies could be the progenitors of the compact quiescent galaxies. We further explore the paths of individual galaxies evolution in Section 7.

\section{Linking BH activity and galaxy properties}
In Section 4, we have analyzed the population of BHs in IllustrisTNG, and their evolution across cosmic times. We particularly discussed the AGN feedback modeling and its consequences for the evolution of the BH population. In Section 5, we focused our attention on the galaxy population, and derived redshift- and simulation-dependent criteria to select SF/quiescent galaxies, as well as extended/compact galaxies. We applied the same selection criteria to observations from {\sc candels}, and found sufficiently good agreement to be confident that we can now investigate the connection between the properties of BHs and the properties of their host galaxies in the simulation, and establish comparisons with these observations.

\begin{figure*}
\centering
\includegraphics[scale=0.58]{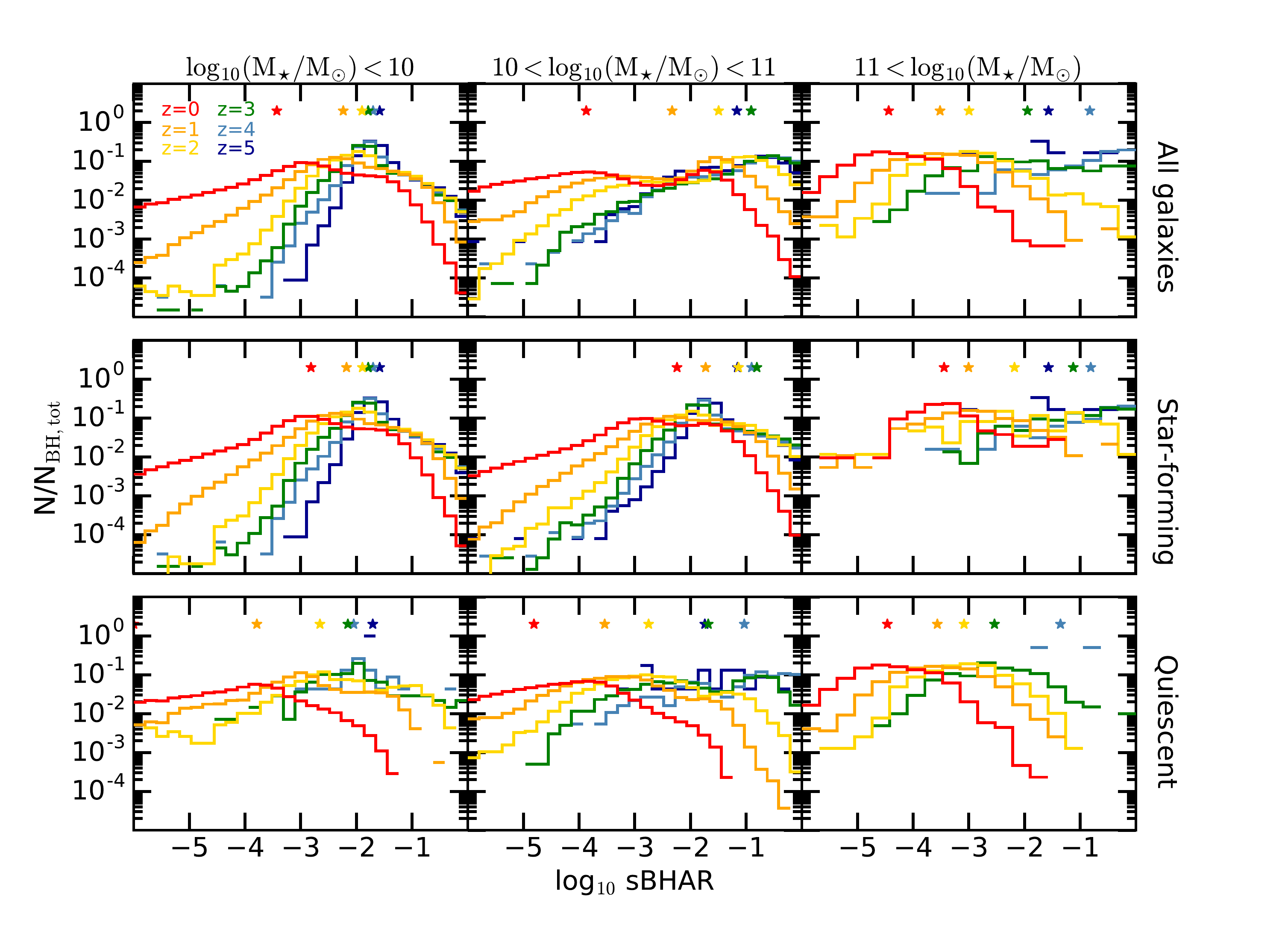}
\caption{Distribution of specific BH accretion rates (TNG300), as a function of redshift, as labelled in the first top left panel, and as a function of galaxy masses ({\it left panels}: $\log_{10}(M_{\star}/\rm M_{\odot})<10$, {\it middle panels}: $10<\log_{10}(M_{\star}/\rm M_{\odot})<11$, {\it right panels}: $11<\log_{10}(M_{\star}/\rm M_{\odot})$). We divide the galaxy population (top row), into star-forming galaxies (middle row), and quiescent (bottom row). Star symbols represent the mean $\langle \log_{10} \rm sBHAR\rangle$ of the distributions.}
\label{fig:distri_bhar}
\end{figure*}

\subsection{Galaxy star-forming properties and specific BH accretion}
In this subsection, we start by analyzing the BH accretion properties as a function of the star-formation properties of their host galaxies. Indeed, BH accretion properties are likely to depend on/correlate with galaxy properties, i.e. gas content, merger history, and star formation history. We expect the distribution of BH accretion properties to depend on whether the host galaxies are quiescent or star-forming. In the following, in order to address these potential differences we again divide the full population of galaxies into star-forming and quiescent galaxies as described above. Star-forming galaxies are defined as having a SFR higher than the 25th percentile of the main star-forming sequence; galaxies are considered quiescent otherwise.

In Fig.~\ref{fig:distri_bhar}, we show the distribution of specific BH accretion rates (sBHAR), which we define similarly as in \citet{2017arXiv170501132A}\footnote{The expression we employ here is based on the BH bolometric luminosity, while \citet{2017arXiv170501132A} use BH X-ray luminosity corrected to bolometric, but we find identical results with the two definitions.} by:
\begin{eqnarray}
{\rm sBHAR}= \frac{L_{\rm BH, bol}}{10^{38} \, {\rm erg/s}}\times \left( 0.002\times \frac{M_{\star}}{\rm M_{\odot}}\right)^{-1}.
\end{eqnarray}
This quantity becomes very similar to the Eddington ratio $f_{\rm Edd}$ (shown in Fig.~\ref{fig:edd_ratio} and Fig.~\ref{fig:edd_ratio_compheckman}) if the mass of BHs and the mass of their host galaxies correlate well. This quantity gives us insight into the rate at which a galaxy can grow its central BH relative to the mass of the host galaxy.
Here, we trace AGN activity within SF and quiescent galaxies as a function of host galaxy stellar mass and redshift. Indeed, we show the distribution of sBHAR in the full population of galaxies in the top panels of Fig.~\ref{fig:distri_bhar}, for SF galaxies in the middle panels, and finally in the quiescent galaxies in the bottom panels, for three different stellar mass ranges: $\log_{10} (M_{\star}/\rm M_{\odot})<10$ (left panels), $10<\log_{10} (M_{\star}/\rm M_{\odot})<11$ (middle panels), and  $11<\log_{10} (M_{\star}/\rm M_{\odot})$ (right panels). Different lines indicate different redshifts, as labelled in the figure. Star symbols on the top of each panel indicate the mean value $\langle \log_{10} \rm sBHAR\rangle$ of the distribution.

We first investigate the time evolution of the sBHAR distributions including all galaxies (i.e. SF and quiescent) in the top panels. The peaks of the distributions move to lower values of sBHAR with time, meaning that BHs are globally accreting at lower rates.
The effect is small for $\log_{10} M_{\star}/M_{\odot}<10$, but increases for more massive galaxies (as indicated by the position of the star symbols in the different galaxy mass bin panels). 
The distributions are broader at low redshift $z\leqslant 2$, and extend down to low values of $\log_{10} \rm sBHAR\leqslant -5$. The time evolution of the sBHAR is in good agreement with recent studies from {\sc candels} and {\sc UltraVISTA} observations \citep{2017arXiv170501132A}.

We divide the total population of galaxies into two populations, SF and quiescent galaxies. We compare our results to the results of \citet{2017arXiv170501132A} to highlight the behavior of the sBHAR distributions, and the good agreement with observations. This initial comparison motivates further dedicated comparative studies (using the same exact definitions for the different samples, etc.) with observational constraints on the sBHAR.
For the SF and quiescent galaxy populations, we find similar behaviors as for the total population, i.e. broader distributions and peaks at lower sBHAR with time.
We find good agreement between the peak of the sBHAR distributions of SF galaxies in the simulation and in observations for $M_{\star}<10^{11}\, \rm M_{\odot}$. 
For example, at $z=3,2,1,0$ the simulated sBHAR peaks at $\langle \log_{10} \rm sBHAR\rangle =-0.8,-1.1, -1.7, -2.2$ while observational constraints indicate $[-1,0], \sim-1, [-2,-1], \sim -3$ for approximately the same redshift range \citep[values estimated from Fig. 3 of][]{2017arXiv170501132A}.
The distributions peak at slightly lower sBHAR for more massive galaxies of $M_{\star}>10^{11}\, \rm M_{\odot}$ in the simulation than in the observations. Indeed for $z=3,2,1,0$ the sBHAR peak at $\langle \log_{10} \rm sBHAR\rangle =-1.1,-2.1,-3,-3.4 $ in the simulation, and ranges of $\sim -1, [-2,-1], \sim -2, \sim -3$ are found in observations.
Good agreement is found for the distributions of the quiescent galaxies, with the distributions from the simulation peaking at slightly lower values than the observations. At $z=3,2,1,0$ for $M_{\star}<10^{11}\, \rm M_{\odot}$, we find $\langle \log_{10} \rm sBHAR\rangle=-1.7,-2.7,-3.5,-4.8$, while in the observations $\sim -2, [-2,-1], [-3,-2], \sim -4$. 
For more massive quiescent galaxies of $M_{\star}>10^{11}\, \rm M_{\odot}$, we find $\langle \log_{10} \rm sBHAR\rangle=-2.5,-3,-3.5,-4.4$, while the redshift evolution trend seems weaker in observations, with $\sim -3, \sim -3, \sim -3, \sim -3.5$.

How do the sBHAR distributions of SF galaxies compare to those of the quiescent population?
The distributions indicate that BHs in quiescent galaxies accrete at lower rates than SF galaxies, at all redshifts. Similar results are found in observations \citep{2017arXiv170501132A}. Particularly at high redshift, the quiescent distributions are broader, while for SF galaxies sBHAR distributions are narrower and peak at around $\log_{10} \rm sBHAR\sim -1.5$. The largest discrepancy between the two samples appears for the most massive galaxies (right panels), where e.g. at high redshift the accretion rate of BHs is significantly sub-Eddington for quiescent galaxies, while SF galaxies host BHs accreting at high rates. This difference propagates to lower redshift.

From this analysis, it is clear that BHs in SF galaxies have a higher probability of accreting at higher rates than in quiescent galaxies. We find that the sBHAR distribution peaks are slightly redshift dependent for SF galaxies and $\log_{10} M_{\star}/M_{\odot}<11$ (left and middle panels), but strongly dependent for more massive galaxies $\log_{10} M_{\star}/M_{\odot}>11$. The same trend is found in observations, for which the turnover in the sBHAR distributions of the SF galaxies moves down slightly to lower sBHAR values in the redshift range $1<z<3$ before shifting more strongly to even lower sBHAR for $z<1$ \citep{2017arXiv170501132A}.
In the simulation, the redshift dependence is more appreciable in the quiescent galaxy population, with again an enhancement for the most massive galaxies. 

Looking at different mass bins for a given redshift allows us to investigate the galaxy stellar mass dependence of the sBHAR distributions.
The stellar mass dependence is minor for SF galaxies, in the two lower mass bins ($\log_{10}(M_{\star}/\rm M_{\odot})<10$, $10<\log_{10} (M_{\star}/\rm M_{\odot})<11$), and small in the last mass bin ($\log_{10} \rm M_{\star}/M_{\odot}>11$) with a flatter distribution at high redshift, where BHs accrete at lower rates. Similarly, the dependence is small for the quiescent sample of the simulation and for the two lower mass bins. The distributions also flatten at high redshift for intermediate galaxy masses. The last stellar mass bin $\log_{10} M_{\star}/\rm M_{\odot}>11$ exhibits the strongest variation, where the peak at $z=0$ is lower by one order of magnitude compared to the bin $10<\log_{10} M_{\star}/\rm M_{\odot}<11$ .

One notable aspect in the top middle panel (i.e. for $10<\log_{10} M_{\star}/\rm M_{\odot}<11$ and full galaxy population), is the bimodality of the sBHAR distributions for $z \leqslant 2$. Indeed the distributions have two peaks, one at high BH specific accretion rate, $\log_{10} \rm sBHAR\sim -1$, and a second peak at $\log_{10} \rm sBHAR\sim -3$. 
As pointed out previously, this bimodality indicates the transition between the SF and quiescent regimes of galaxies, where the first peak is also identified in the SF sample (even if some BHs already start to have much lower sBHAR), and the second in the quiescent sample.

\subsection{Linking BH properties to galaxy sizes: Fraction of AGN in the sSFR-$\Sigma_{\rm e}$ diagram}
We now turn to investigate connections between BH accretion, SF, and the structural properties of galaxies.
To do so we use the criterion presented in Section 4 to distinguish between extended and compact galaxies. 
In this section, we use the specific star formation - effective stellar mass surface density diagram as described previously, which is commonly used in observational papers. This diagram can be divided into four distinct quadrants: SF extended galaxies, SF compact, quiescent extended, and quiescent compact galaxies. SF galaxies occupy the top of the diagram, and quiescent galaxies the bottom quadrants. Left quadrants host extended galaxies, and right quadrants compact galaxies.

\subsubsection{sSFR-$\Sigma_{\rm e}$ diagram of the simulated galaxies color coded by BH X-ray luminosity}

\begin{figure}
\centering
\includegraphics[scale=0.49]{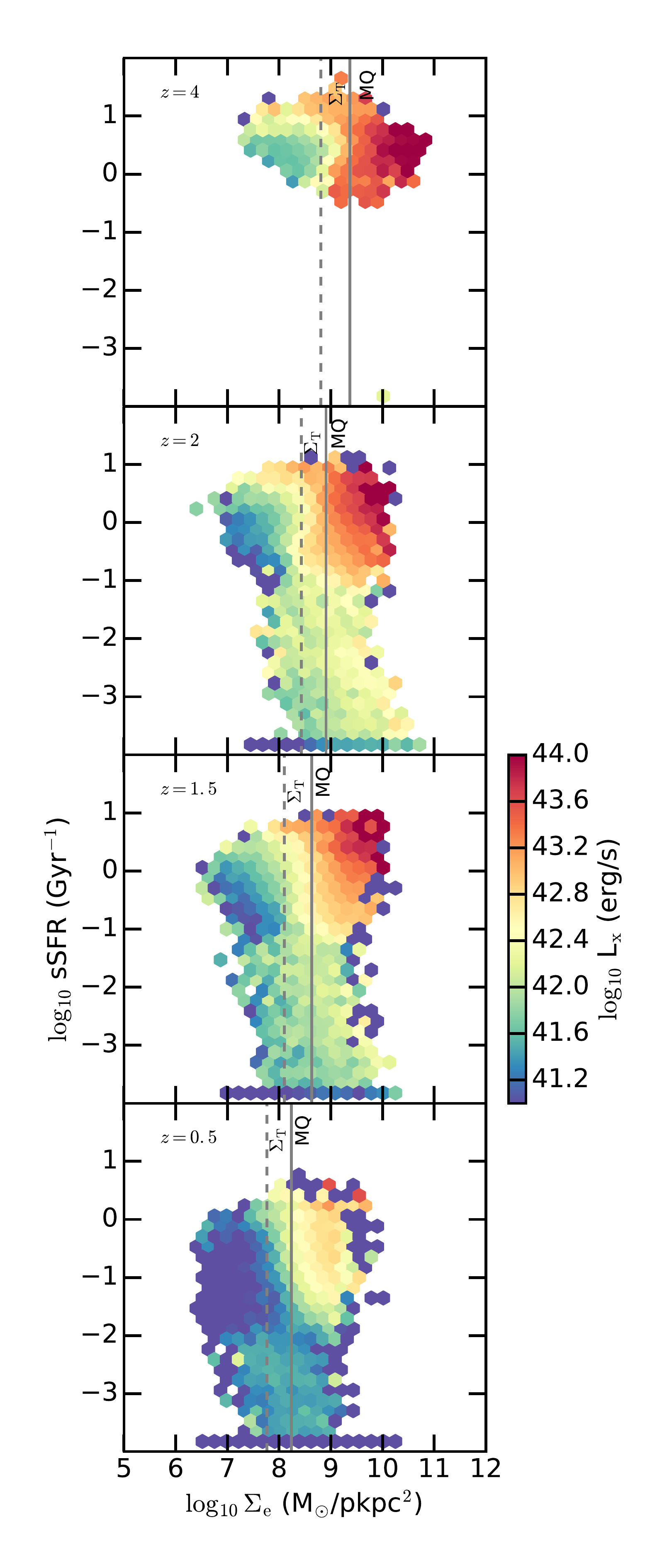}
\caption{Mean hard (2-10 keV) X-ray luminosity of AGN hosted by galaxies with $M_{\star}\geqslant 10^{9.5}\, \rm M_{\odot}$ in the sSFR-$\Sigma_{\rm e}$ diagram in TNG300, at $z=4,2,1.5,0.5$. Galaxies with $\rm sSFR<10^{-4}\, \rm Gyr^{-1}$ are set at this value. AGN in the top right quadrant of the diagrams,  i.e. AGN hosted by compact star-forming galaxies, are brighter. However galaxies below $\log_{10}\rm sSFR\sim -0.5$ for $z=2$, and $\log_{10}\rm sSFR\sim -1$ for $z\leqslant 1.5$, i.e. quiescent galaxies, have much lower mean X-ray luminosities.}
\label{fig:agn_lum}
\end{figure}

In Fig.~\ref{fig:agn_lum}, we show the sSFR-$\Sigma_{\rm e}$ diagram for both central and satellite galaxies with $M_{\star}\geqslant 10^{9.5} \, \rm M_{\odot}$ in hexabins\footnote{Only bins with more than one galaxy are displayed.} color coded by the mean BH X-ray luminosity in bins in the range $10^{41}\leqslant L_{\rm x}\leqslant 10^{44}\, \rm erg\, s^{-1}$ for $z=4,2,1.5,0.5$. The AGN X-ray luminosity is computed as described in the previous sections and here does {\rm not} contain any correction for obscuration. We set sSFR $=2\times 10^{-4} \, \rm Gyr^{-1} $ for galaxies with lower or vanishing sSFR. We report the median surface density of the quiescent population as a grey vertical line, as well as the 25th percentile of this population as a dashed vertical line; this defines our threshold to identify compact galaxies. 
For reference, all the quiescent galaxies with $M_{\star}\geqslant 10^{10}\, \rm M_{\odot}$ are found below $\log_{10}\rm sSFR\sim -0.5$ for $z=2$, and below $\log_{10}\rm sSFR\sim -1$ for $z\leqslant 1.5$, referred hereafter as the {\it quiescent limit}.
 
The mean AGN luminosity in bins is not uniform in the diagram and appears to be strongly correlated with both sSFR and $\Sigma_{\rm e}$. The star-forming galaxies and the compact galaxies have higher AGN luminosities. 
Here we show galaxies with $M_{\star}\geqslant 10^{9.5} \, \rm M_{\odot}$ to get a sense of the evolution of these galaxies as a function of their mass. The lowest mass galaxies occupy the extended star-forming part of the diagram; for comparison the left panels of Fig.~\ref{fig:agn_fraction} display galaxies with $M_{\star}\geqslant 10^{10} \, \rm M_{\odot}$. The most star-forming extended galaxies have higher X-ray luminosities, while extended galaxies with lower sSFR can host a BH with lower luminosity $L_{\rm x}\leqslant 10^{42}\, \rm erg/s$. On the top right side of the diagram, compact star-forming galaxies host BHs with the highest mean X-ray luminosity with $L_{\rm x}\geqslant 10^{43}\, \rm erg/s$ at high redshift, and lower luminosity of $L_{\rm x}\geqslant 10^{42}\, \rm erg/s$ by the end of the simulation (bottom panel).
For quiescent galaxies with lower specific star formation rates below the quiescent limit, the mean luminosity of BHs drops to lower values.

\subsubsection{AGN fraction in the sSFR-$\Sigma_{\rm e}$ diagram for the simulation}
\begin{figure*}
\centering
\includegraphics[scale=0.52]{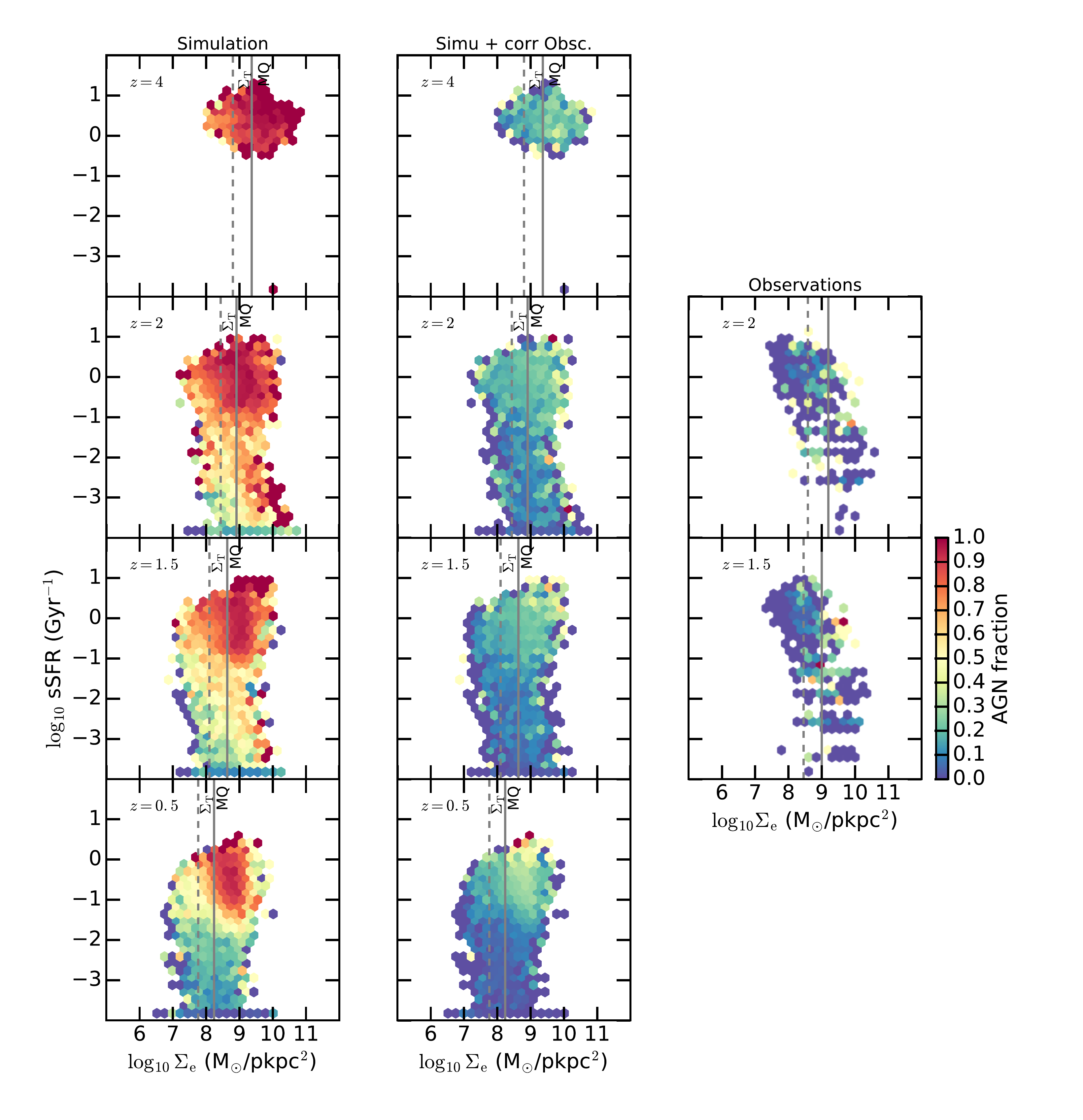}
\caption{Fraction of hard X-ray AGN (2-10 keV) in galaxies with $M_{\star}\geqslant 10^{10}\, \rm M_{\odot}$ in hexagonal bins; the associated colorbar ranges from 0 to 1. Galaxies with $\rm sSFR<10^{-4}\, \rm Gyr^{-1}$ are set at this value. {\it Left panels}: AGN fraction in TNG300. {\it Middle panels}: AGN fraction in TNG300 when we apply our correction for AGN obscuration. {\it Right panels}: AGN fraction for the {\sc candels} observations.
There is a clear enhancement in AGN activity in the top right quadrant of all the subplots, i.e. in compact star-forming galaxies. Star-forming extended galaxies (top left quadrant of all subplots) are also found to host a high fraction of AGN in the simulation, for the most star-forming galaxies, $\rm sSFR\geqslant10^{0.5}\, \rm Gyr^{-1}$. The AGN fraction is considerably lower in quiescent galaxies (bottom quadrants).}
\label{fig:agn_fraction}
\end{figure*}

To elaborate on these results, we now show the AGN fraction of $M_{\star}\geqslant 10^{10} \, \rm M_{\odot}$ galaxies in hexabins on the right panel of Fig.~\ref{fig:agn_fraction} for $z=4,2,1.5,0.5$. To compute the AGN fraction in bins we consider a BH as being an AGN if $L_{\rm x}\geqslant 10^{42} \, \rm erg/s$. On average, galaxies with higher sSFR have a higher probability of hosting an AGN, and even more so for more compact galaxies, with fractions approaching unity.
We note that extended star-forming galaxies also have on average a high AGN fraction. However, galaxies with lower sSFR have a much lower probability of hosting an X-ray-detectable AGN. With time, the AGN fraction of these compact and extended quenched galaxies decreases.
Some galaxies with $\Sigma_{e}\sim 10^{10} \, \rm M_{\odot}/pkpc^{2}$ and sSFR$=10^{-3}\, \rm Gyr^{-1}$ (red bins on the very right) at $z=2$ also have a high fraction of AGN, in contrast to other quiescent galaxies. A large fraction of those are satellite galaxies, but centrals are also present, having a high AGN activity at $z\sim2$, which also decreases at lower redshift (bottom panel at $z=0.5$).

\subsubsection{Comparison of the simulated and observed AGN fraction in the sSFR-$\Sigma_{\rm e}$ diagram}
In this section, we compare the AGN fraction that we found in the simulation to observations from {\sc candels}. 
We divide our analysis of the observations into two different redshift ranges. We first investigate the fraction of AGN in a large redshift range $1.4\leqslant z \leqslant 3$. For clarity, the figures relative to this first analysis are shown in the Appendix.  To make a closer comparison with the outputs of the simulation, we then focus our analysis on two narrower redshift bins $z\sim 1.5$ and $z\sim2$ of the {\sc candels} data. The results that we present in this section on the AGN fraction are summarized in Table~\ref{table:table_fraction}.\\

\noindent{\bf Large redshift bin $1.4\leqslant z \leqslant 3.0$} \\ 
\noindent 
Using observations from the {\sc candels} fields, \citet{2017ApJ...846..112K} derived the AGN fraction for different types of galaxies with $M_{\star}\geqslant 10^{10}\, \rm M_{\odot}$ in the redshift range $1.4\leqslant z\leqslant 3$, using the same X-ray luminosity limit of $L_{\rm x}\geqslant10^{42}\, \rm erg\, s^{-1}$ to confirm the presence of an X-ray luminous AGN. With their selection method based on dust-corrected rest-frame U-V color, they find that $40\%$ of compact star-forming galaxies host an X-ray AGN, while $8\%$ of SF extended, $11\%$ of quiescent extended, and $8\%$ of quiescent compact galaxies host an X-ray AGN. 
Instead here we use our selection method relying on SFR and $\Sigma_{\rm e}$ to identify galaxy types.
We compute the AGN fractions among the different types of galaxies in the same redshift range $1.4\leqslant z \leqslant 3$, and find that $13.4\%$ of compact star-forming galaxies, $2.7\%$ of the extended star-forming, $5.8\%$ of the extended quiescent, and $10.8\%$ of the compact quiescent galaxies host an X-ray AGN. The left panel of Fig.~\ref{fig:candels_AGNfraction} shows the sSFR-$\Sigma_{\rm e}$ diagram color coded in hexabins by the AGN fraction. 
The qualitative trend seen in the observations is in agreement with the simulation predictions, i.e. that quiescent compact galaxies have a lower probability of hosting X-ray AGN than compact star-forming galaxies.
The differences in the galaxy type selection methods lead to discrepancies in the AGN fractions, and we report this discussion in the Appendix. 
The right panel of Fig.~\ref{fig:candels_AGNfraction} shows the time evolution of the AGN fractions in narrower redshift bins of $\Delta z = 0.2$ for the different types of galaxies. We do not show the AGN fraction for the extended quiescent galaxies as the sample does not include more than 30 galaxies and is therefore subject to large uncertainties. AGN fractions in small redshift bins fluctuate strongly with time, but on average we find that the compact star-forming galaxies have a higher probability of hosting AGN than compact quiescent galaxies. 
\\

\noindent {\bf Narrower redshift bins $z\sim1.5$ and $z\sim 2.0$}  \\
\noindent To make a close comparison with the simulation, we focus our analysis on the two redshift bins described in the previous sections, i.e. $1.4\leqslant z\leqslant 1.8$ and $1.8\leqslant z\leqslant 2.2$, that we compare with the $z=1.5$ and $z=2$ outputs of the simulation. 

We show the AGN fraction for the observations on the right panels of Fig.~\ref{fig:agn_fraction}, with the same colorbar as for the simulation on the left panels.
As a consequence of the good agreement of the observed and simulated populations in the the SFR-$M_{\star}$ plane, the quiescent galaxies with $M_{\star}\geqslant 10^{10}\, \rm M_{\odot}$ in observations are also found below the quiescent limit $\log_{10} {\rm sSFR/ Gyr^{-1}} \sim -0.5$ for $z\sim 2$, and $\log_{10} {\rm sSFR/ Gyr^{-1}} \sim -1$ for $z\sim 1.5$, as for the simulation.
We also report here our thresholds to distinguish compact from extended galaxies with dark dashed lines.

From Fig.~\ref{fig:agn_fraction}, we note that the AGN fractions are very different in amplitude in the observations and simulation. Indeed, in the observations we find that $13\%\, (16\%)$ of compact star-forming galaxies host an X-ray AGN at $z\sim2 \,(1.5)$, $6\% \,(4\%)$ of extended star-forming, $10\%\, (5\%)$ of extended quiescent, and $9\% (10\%)$ of compact quiescent galaxies. These fractions are reported in Table~\ref{table:table_fraction}. However in the simulation when we consider both central and satellite galaxies we find that $93\% \,(88\%)$ of compact star-forming galaxies host an X-ray AGN, $86\% \,(73\%)$ of extended star-forming, $61\% \,(36\%)$ of extended quiescent, and $45\% \,(29\%)$ of compact quiescent galaxies for $z=2$ and $z=1.5$ respectively. In both observations and simulation, we find that the compact star-forming galaxies have the highest fraction of AGN, and the compact quiescent galaxies a lower fraction (the lowest fractions for the simulation). We note that in the simulation the galaxies with the highest sSFR also have the highest AGN fraction for both extended and more compact galaxies, which is not the case in the observations, where extended star-forming galaxies have a very low fraction of (X-ray detected) AGN. We find a very large difference between compact star-forming and compact quiescent galaxies in the simulations, which exists in the observations but with a much lower amplitude. 
This could be due to the fact that observed quiescent galaxies are not very far from the SF main sequence, as shown in Fig.~\ref{fig:sfr_mstar} where only very few galaxies are found below $\log_{10} \rm SFR/(M_{\odot}/yr) = -2$, whereas a large fraction of the simulated galaxies have very low values of the instantaneous sSFR (shown at sSFR=$2\times 10^{-4} \, \rm Gyr^{-1}$ in Fig.~\ref{fig:sfr_mstar}). Indeed, we clearly see in the simulation that for galaxies below the main sequence, the AGN fraction strongly decreases as a function of the deviation from the main sequence.\\

\noindent {\bf Correction for AGN obscuration}\\
\noindent
The raw AGN fractions from the simulation are much higher than the observed ones, for all structural galaxy types. However, the left panels of Fig.~\ref{fig:agn_fraction} show AGN fractions from the simulation with no correction for obscured AGN. The observational appearance of an active BH is, however, not only determined by its accretion and emission properties, but is also impacted by the nature of the material in the line-of-sight through which we observe its emission.
Obscuration of AGN can arise from different sources, either from the gas and dust in the very close surroundings of a BH, often modeled as gas and dust clumpy clouds in the torus around a BH, or from the gas and dust in its host galaxy.  As the fraction of AGN that are obscured is subject to large uncertainties, we test two different obscuration corrections that we apply to the simulation.

\begin{figure}
\centering
\includegraphics[scale=0.43]{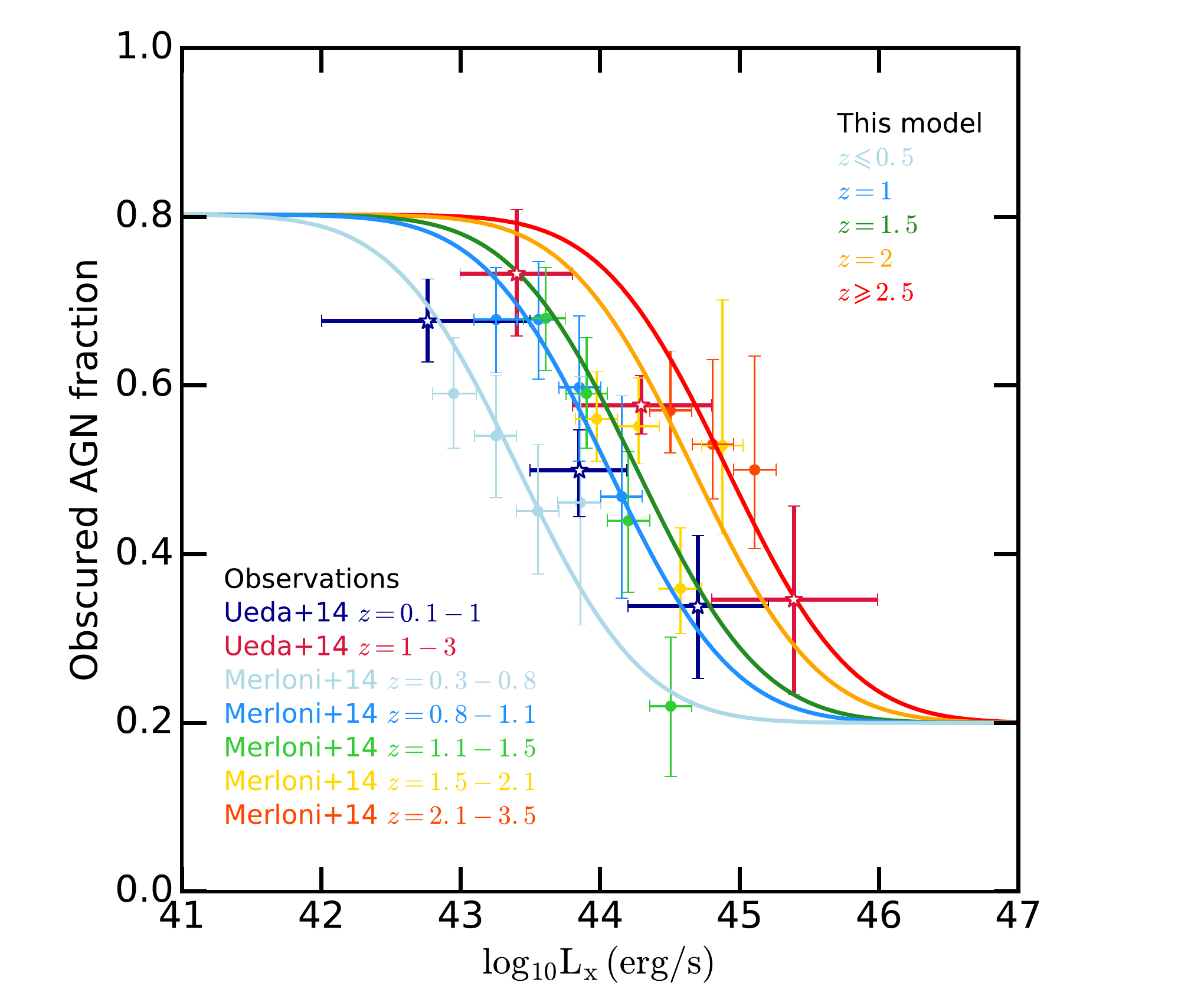}
\caption{Observational constraints (data points with error bars) for the fraction of obscured AGN from hard X-ray detected AGN samples \citep{2014ApJ...786..104U,2014MNRAS.437.3550M}. Solid lines show our redshift- and AGN X-ray luminosity-dependent model for the obscured AGN fraction. Our model assumes an anti-correlation between the fraction of obscured AGN and their hard X-ray luminosity, and the presence of more obscured AGN at higher redshifts.}
\label{fig:obscuration_model}
\end{figure}

We first try a model for the obscured fraction of AGN that is independent of the AGN X-ray luminosity as well as galaxy properties. Our first model simply assumes that a fixed fraction of either $30\%$ or $50\%$ of the AGN are obscured. Our model does not distinguish between Compton-thick AGN, i.e. AGN obscured by column density of $N_{\rm H}\geqslant 10^{24}\, \rm cm^{-2}$, and Compton-thin AGN with $N_{\rm H}<10^{24}\, \rm cm^{-2}$.
We randomly assign a null luminosity to the AGN of $30\%$ to $50\%$ of the simulated galaxy population with $M_{\star}\geqslant 10^{9}\, \rm M_{\odot}$, then we divide the population into galaxy types, and compute the fraction of AGN among those samples. We perform 100 realizations to compute the errors.
With our simple correction for obscured AGN, we find lower AGN fractions in all the different galaxy types, but the amplitude of the AGN fractions in the simulation is still higher than those in observations. Indeed, the AGN fraction among compact star-forming galaxies is for example $46-65\%$ at $z=2$ (corresponding to $50-30\%$ of obscured AGN) instead of $93\%$ with no correction. 

Evidence that the fraction of obscured AGN depends on the AGN luminosity has been accumulating over the years \citep{1982ApJ...256..410L,2003ApJ...596L..23S,Ueda2003,2005MNRAS.360..565S,2008A&A...490..905H,2010A&A...510A..56B,2013ApJ...772...26A}.
Here we use the results of \citet{2014MNRAS.437.3550M} from a large sample of AGN selected by XMM-Newton in the COSMOS field in the redshift range $0.3<z<3.5$. AGN were selected based on either optical spectral properties or the shape of their X-ray spectrum. In this study a dependence on redshift is not evident with optical selection, but is strongly revealed with X-ray selection. We show their derived obscured fraction of AGN classified from X-ray, in Fig.~\ref{fig:obscuration_model} as colored dots for different bins of redshift. We also show the fraction of obscured AGN from \citet{2014ApJ...786..104U} from the Swift/BAT, AMSS and SXDS hard X-ray band samples, averaged in the redshift range $z=0.1-1$ (blue star symbol) and $z=1-3$ (red star symbol). Both analyses show a strong anti-correlation between the absorption fraction of AGN and their luminosity.
Dependence on redshift has also been found: there are more obscured AGN toward higher redshifts, but the anti-correlation with luminosity seems to be preserved \citep{2006ApJ...653.1070B,2007A&A...463...79G,2008A&A...490..905H,2012A&A...537A..16F,2014ApJ...786..104U}.

Based on these observations, we build an empirical model for the fraction of obscured AGN that depends on both redshift and the X-ray luminosity of the BHs. Following \citet{2014ApJ...786..104U} we assume that the obscuration function has two asymptotes: faint AGN have a probability of $80\%$ to be obscured, and the brightest AGN a probability of $20\%$. The transition between these two regimes is obtained by fitting the observational data of \citet{2014MNRAS.437.3550M} for the corresponding redshift bins. Our model is shown in Fig.~\ref{fig:obscuration_model} with solid colored lines for different redshifts corresponding to the outputs of the simulation.
Again we perform 100 realizations with this second model for the obscuration to compute the AGN fraction in the different types of galaxies. 

One of these realizations is shown in the second column of Fig.~\ref{fig:agn_fraction}. We note some very minor variations of the AGN fraction in some hexabins of Fig.~\ref{fig:agn_fraction} for the different realizations, but they result in errorbars on the AGN fractions below $0.4\%$ for the compact SF and quiescent galaxies.
From this figure we see that the trend is similar to that obtained without any corrections; more AGN are found in the compact star-forming galaxies than in the compact quiescent galaxies, maintaining a good agreement with the observed trends. Moreover the amplitude of the AGN fractions on this diagram, i.e. mostly in the range $\sim 0-50\%$, are in better agreement with the AGN fractions among the observed galaxies. AGN fractions of the compact star-forming galaxies are now $20\%$ in the simulation, and $6-9\%$ for the compact quiescent galaxies.
We report the values for the AGN fractions in the 4 quadrants in Table 3, and discuss them in the following subsection.
\\

\begin{figure}
\centering
\includegraphics[scale=0.43]{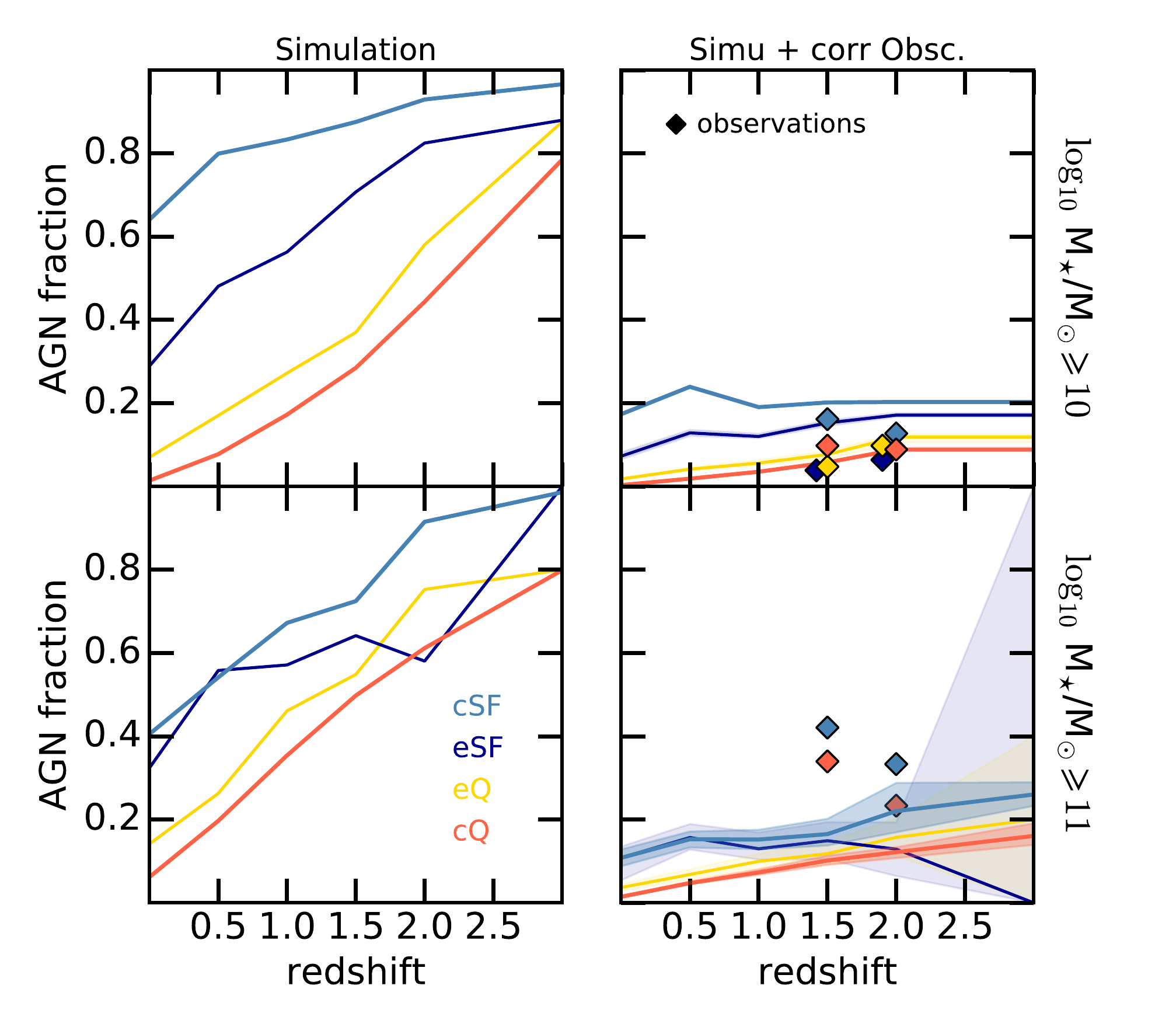}
\caption{Time evolution of the AGN fraction in each galaxy type quadrant (solid lines, as labelled in the figure). Left panels show the results from TNG300 with no correction for obscured AGN, while the right panels include our model for obscuration (we also include error bars from the different realizations when we apply the correction). On the right panels, we also include our results from the {\sc candels} observations as diamond symbols. The fractions are normalized to the total number of galaxies with $\log_{10} M_{\star}/\rm M_{\odot}\geqslant 10.$ (top panels) and $\log_{10}M_{\star}/\rm M_{\odot}\geqslant 11$ (bottom panels). We find good quantitative agreement between the AGN fractions from the observations and the simulation, especially for galaxies with $M_{\star}\geqslant 10^{10}\, \rm M_{\odot}$.}
\label{fig:fraction_quadrant}
\end{figure}

\noindent {\bf Time evolution of the AGN fractions}\\
We show in the left panels of Fig.~\ref{fig:fraction_quadrant} the time evolution of the AGN fraction in each quadrant (different colored solid lines, as labelled in the figure) of the sSFR-$\Sigma_{\rm e }$ diagram for the simulation without the correction for obscured AGN. The top panel shows the fraction for galaxies more massive than $\log_{10} M_{\star}/{\rm M}_{\odot}\geqslant 10$; the bottom panel corresponds to $\log_{10} M_{\star}/{\rm M}_{\odot}\geqslant 11$. Compact star-forming galaxies are shown in red (cSF), extended star-forming galaxies in light blue (eSF), extended quiescent galaxies in dark blue (eQ), and compact quiescent galaxies in yellow (cQ). As expected the AGN fraction decreases with time for all galaxy types; the decrease is steeper for extended and compact quiescent galaxies than for star-forming galaxies. Clearly, star-forming compact galaxies have the highest probability of hosting an AGN at all times. At high redshift $z=3$, almost all compact star-forming galaxies host an active BH; this probability does not go below $60\%$ for $M_{\star}\geqslant 10^{10}\, \rm M_{\odot}$, and $40\%$ for $M_{\star}\geqslant 10^{11}\, \rm M_{\odot}$. With time, more and more galaxies move away from the main sequence and quench. Compact quiescent galaxies have the lowest probability of hosting an AGN at any redshift, this probability drops below $10\%$ by the end of the simulation. Between these two populations we find that extended galaxies have intermediate AGN fraction, higher for extended star-forming galaxies.

\begin{table*}
\caption{Properties of the observational ({\sc candels}) and simulated (TNG300) galaxy samples. We consider galaxies with $M_{\star}\geqslant 10^{10}\, \rm M_{\odot}$. The first columns show the median of the $\log_{10} M_{\star}/\rm M_{\odot}$ of the samples, and the last columns the AGN fraction among these samples. The first row indicates our results for the simulation without any correction for the fraction of obscured AGN, the second row shows the results for the simulation when we assume that 50 or 30 $\%$ of the galaxies host an obscured AGN, and the third row shows the results when we apply a redshift and luminosity-dependent model for the obscured AGN fraction. The last row indicates our results for the observations.}
\begin{center}
\begin{tabular}{cccccccccccc}
\hline
\multicolumn{1}{c}{Redshift} & \multicolumn{1}{c}{Obs/Simu}& & \multicolumn{4}{c}{ med($\log_{10} M_{\star}/\rm M_{\odot}$)} & & \multicolumn{4}{c}{  AGN fraction ($\%$)}\\
&&& cSF & cQ & eSF & eQ && cSF & cQ & eSF & eQ \\
\hline
\hline

$z=2$ & Simulation                    && 10.25 & 10.63 & 10.13 & 10.71  && 93 & 44 & 82 & 58\\
$z=2$ & Simu corr (50-30$\%$) && &  & &                                         && 46-65 & 23-30 & 40-57 & 29-39\\
$z=2$ & Simu corr ($\rm L_{x}$-dep.)                   && &  &  &                                        && {\bf 20} & {\bf 9} & 17 & 12\\
$z\sim2$ & Observations            && 10.50 & 10.65 & 10.27 & 10.36  && {\bf 13} & {\bf 9} & 6 & 10 \\
\hline

$z=1.5$ & Simulation                     && 10.24 & 10.60 & 10.13 & 10.73   && 88 & 28 & 71 & 37 \\
$z=1.5$ & Simu corr (50-30$\%$) &&  &  &  &                                         && 44-61 & 14-20 & 36-50 & 17-25 \\
$z=1.5$ & Simu corr ($\rm L_{x}$-dep.)                   && &  &  &                                          && {\bf 20} & {\bf 6} & 15 & 8\\
$z\sim1.5$ & Observations            && 10.44 & 10.59 & 10.24 & 10.23    && {\bf 16} & {\bf 10} & 4 & 5\\

\hline
\end {tabular}
\end{center}
\label{table:table_fraction}
\end{table*}

Similarly, we show in the right panels the median of the AGN fractions for the simulation when we correct the AGN population with a luminosity- and redshift-dependent correction for obscuration. Shaded areas represent the standard deviation $\pm \sigma$ of the 100 realizations of the model. For galaxies with $M_{\star}\geqslant 10^{10}\, \rm M_{\odot}$, the errors on the AGN fractions are below $0.4\%$ for the compact star-forming and quiescent galaxies, and below $2\%$ for the extended galaxies. The corresponding error bars in the top panel of Fig.~\ref{fig:fraction_quadrant} are invisible. As a consequence of the smaller samples of $M_{\star}\geqslant 10^{11}\, \rm M_{\odot}$ galaxies, the standard deviations are higher in this case.
While the decrease with time of the AGN fractions was strong on the left panels, with the obscuration correction we now find that the decrease with time is less dramatic, but still visible, and even more so for the most massive galaxies (bottom right panel).

In Fig.~\ref{fig:fraction_quadrant} we also show the AGN fraction among the different types of galaxies in the {\sc candels} observations for redshift $z\sim 1.5$ and $z\sim 2$, with diamond symbols. We only show the results for samples with more than 30 galaxies to avoid non-robust fractions; the AGN fraction of the extended star-forming and quiescent galaxies does not appear on the bottom panel for this reason. The probability of hosting an AGN is the highest for compact star-forming galaxies, followed by compact quiescent and extended star-forming galaxies. 
We find a good agreement between the amplitude of the AGN fractions in observed and simulated galaxies for $M_{\star}\geqslant 10^{10}\, \rm M_{\odot}$.
Indeed while in the simulation $20\%$ of compact star-forming galaxies host an AGN at both $z=2$ and $z=1.5$, $13\%$ ($z\sim2$) and $16\%$ ($z\sim 1.5$) of observed galaxies do so. 
We also find good agreement for the compact quiescent galaxies;  $9\%$ of both the simulated and observed cQ galaxies host an AGN at $z\sim2$, and $6\%$ for the simulated galaxies against $10\%$ in the observations at $z\sim1.5$.
However the agreement is weaker for more massive galaxies of $M_{\star}\geqslant 10^{11}\, \rm M_{\odot}$, for which we find that the observed compact star-forming galaxies have higher AGN fractions of $33,42\%$ ($z\sim2,1.5$) compared to $22,12\%$ for the simulated galaxies. Similarly, $23,34\%$ observed compact quiescent galaxies host an AGN, against $12,10\%$ ($z=2,1.5$) in simulation. 

For the particular bins that we employ here we find that the AGN fractions of observed compact star-forming and quiescent galaxies increase slightly with time. The right panel of Fig.~\ref{fig:candels_AGNfraction} demonstrates the large time variation of the AGN fractions in observations and therefore the complexity of interpreting the results from observations as a function of redshift. In fact, using smaller bins of $\Delta z =0.2$ shows the opposite time evolution.\\

\subsubsection{Distributions of the AGN luminosity in the different galaxy types}

\begin{figure*}
\centering
\includegraphics[scale=0.55]{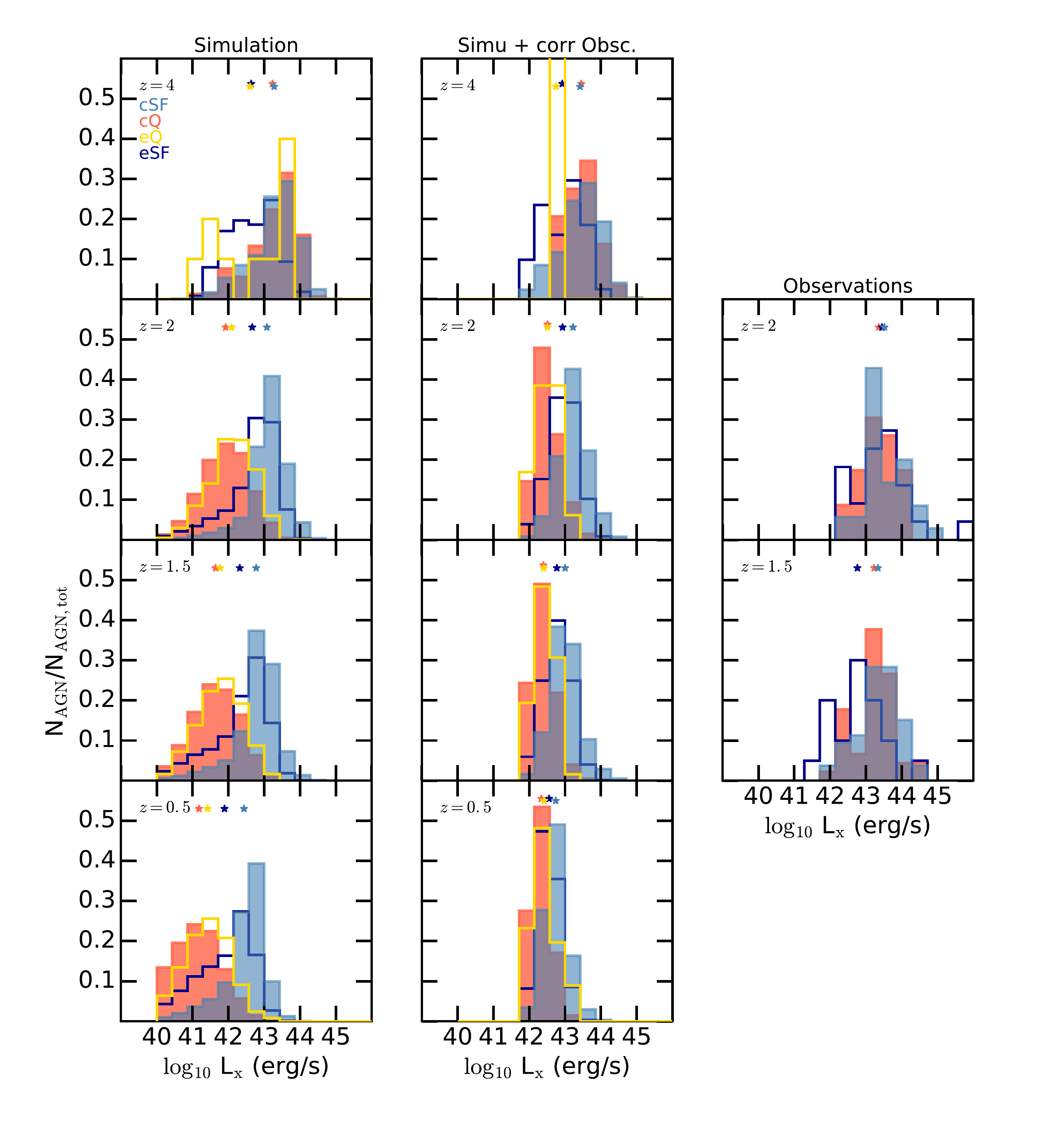}
\caption{Histogram of AGN hard (2-10 keV) X-ray luminosity, for each quadrant (light blue: extended star-forming, dark blue: extended non star-forming, orange: compact star-forming, yellow: quiescent galaxies), and for several redshifts $z=4,2,0.5$. {\it Left panels}: TNG300. {\it Middle panels}: TNG300 with our correction for the AGN obscuration fraction and a cut in luminosity $L_{\rm x}\geqslant 10^{42}\, \rm erg/s$ similar to the observational limit. {\it Right panels}: observations from {\sc candels}. The mean of the distributions are indicated with star symbols on top of each panel. While in the simulation, the mean of the AGN luminosity distributions of the cQ and cSF galaxies are separated by more than one order of magnitude (slightly less with the luminosity cut $L_{\rm x}\geqslant 10^{42}\, \rm erg/s$), they are very similar in the observations.}
\label{fig:histo_xray}
\end{figure*}

To investigate in more detail the discrepancies between the compact star-forming and compact quiescent populations, we draw the distribution of AGN hard X-ray luminosities in the 4 quadrants for the $M_{\star}\geqslant 10^{10} \, \rm M_{\odot}$ galaxies in Fig.~\ref{fig:histo_xray}. Left panels represent the simulation, middle panels represent the simulation after applying our correction for obscuration, as well as a cut of $10^{42}\, \rm erg/s$ to detect the X-ray AGN (similar to the observations), and right panels show the {\sc candels} observations.
For all panels we exclude galaxies with lower luminosities than $L_{\rm x}\leqslant 10^{40}\, \rm erg\, s^{-1}$, i.e. mostly galaxies without active BHs. Star symbols on the top of each panel represent the mean values $\langle \log_{10} L_{\rm x}\rangle$ of the distributions. 

At high redshift ($z=4$) most of the simulated galaxies (left panels) are extended or compact star-forming systems, and host BHs with high X-ray luminosities. We find that the mean luminosity for the compact star-forming galaxies is $\langle \log_{10} L_{\rm x, cSF}\rangle=43.3$ and $\langle \log_{10} L_{\rm x, cQ}\rangle=43.2$ for the quiescent compact galaxies.
With time we see the population of quiescent galaxies emerging, i.e. that the distribution of their luminosity separates from the compact star-forming population with $\langle \log_{10} L_{\rm x, cSF}\rangle=43.1, 42.4$ and $\langle\log_{10} L_{\rm x, cQ}\rangle=41.9, 40.7$ for $z=2,0.5$. Consequently the histogram of quiescent compact galaxies differs from the one for compact star-forming galaxies with mean luminosity values smaller by more than one order of magnitude than those of the compact star-forming galaxy population.
 
Clearly we find that the distributions of luminosities of the AGN populations in {\sc candels} (right panels) and IllustrisTNG are different. In observations, the distributions of compact star-forming and quiescent galaxies span a similar range of luminosity in $10^{42} \leqslant L_{\rm x}\leqslant 10^{45}\, \rm erg\, s^{-1}$, as shown in the right panels of Fig.~\ref{fig:histo_xray}. The mean luminosity of compact star-forming galaxies is $\langle\log_{10} L_{\rm x, cSF}/ \rm erg\, s^{-1}\rangle=43.4$ ($z=1.5$) and $\langle\log_{10} L_{\rm x, cSF}\rangle=43.5$ ($z=2$), and $\langle\log_{10} L_{\rm x, cQ}\rangle=43.2$ ($z=1.5$) and $\langle\log_{10} L_{\rm x, cQ}\rangle=43.4$ ($z=2$) for compact quiescent galaxies. However, in the simulation, AGN in compact star-forming galaxies have much higher luminosities spanning a range $10^{40} \leqslant L_{\rm x}\leqslant 10^{45}\, \rm erg\, s^{-1}$, whereas compact quiescent galaxies do not host AGN with luminosity higher than $L_{\rm x}= 10^{43.5}\, \rm erg\, s^{-1}$. The mean luminosity of compact star-forming galaxies is $\langle\log_{10} L_{\rm x, cSF}\rangle=42.8$ ($z=1.5$) and $\langle\log_{10} L_{\rm x, cSF}\rangle=43.1$ ($z=2$), more than one order of magnitude higher than for compact quiescent galaxies with $\langle\log_{10} L_{\rm x, cQ}\rangle=41.6$ ($z=1.5$) and $\langle\log_{10} L_{\rm x, cQ}\rangle=41.9$ ($z=2$). 
In the middle panels of Fig.~\ref{fig:histo_xray} we show the results for the simulation corrected for AGN obscuration, and we also add a cut in luminosity $L_{\rm x}\geqslant 10^{42}\, \rm erg/s$ mimicking the detectability of AGN in observations. The mean value of the cSF and cQ are closer to each other than in the left panels (i.e. when we apply no correction), but we can still distinguish two distinct luminosity distributions, which is not the case in the observations.

Our comparison highlights strong differences in the properties of AGN among the different types of galaxies in observations and the simulation. The observational picture here could be biased by the fact that we only observed quite luminous AGN with $L_{\rm x}\geqslant 10^{42}\, \rm erg\, s^{-1}$, fainter objects being even more challenging to observe and identify robustly as being powered by an accreting black hole. Again from Fig.~\ref{fig:sfr_mstar}, we know that we do not observe galaxies deviating as far below the main sequence as in the simulation, for which Fig.~\ref{fig:agn_lum} indicates a decrease in AGN luminosity with decreasing sSFR values. Faint AGN are also more likely to be obscured, which complicates their detection as well, and may bias observational results. 

Finally, in our analysis we have not corrected the X-ray luminosity for the presence of X-ray binaries. Our results should not be strongly affected by the presence of X-ray binaries as we focus on study on bright AGN with $L_{x}\geqslant 10^{42}\, \rm erg/s$.

\subsubsection{Other effects impacting the comparisons between observations and the simulation}
In the following we explore several effects that could affect the comparison between AGN fractions in the simulation and in the observations. The AGN fraction can be correlated with galaxy mass, where more massive galaxies tend to have a higher AGN fraction \citep{2010ApJ...720..368X,Aird2012}. We report the median of galaxy mass for the different samples in Table~\ref{table:table_fraction}, and discuss the differences between simulations and observations below. 

As shown in Table~\ref{table:table_fraction} (column $\rm med(\log_{10}M_{\star}/M_{\odot}$)), the compact quiescent galaxy samples have the largest galaxy stellar mass median in both observations and the simulation, and demonstrate good agreement between the two. 
The median $M_{\star}$ of the SF galaxies is lower compared to the observations. While the medians of the galaxy stellar mass in compact quiescent and star-forming galaxies are not strongly different in the observations, we find a larger offset between the two in the simulation. 
The ratios between the median of the compact SF and extended SF are slightly more similar in the simulation; this could partly explain the similarity of the AGN fraction among cSF and eSF in the simulation. 
We here refer the reader to \citet{2017ApJ...846..112K}, where they find that the AGN fraction among eSF galaxies increases when they employ a mass-matched sample with the cSF galaxies.
The biggest discrepancy that we find is that the observed sample of the extended quiescent galaxies has a low stellar mass median, while in the simulation the eQ galaxies have the largest mass median. 
Moreover, the simulated galaxy samples include more massive galaxies than the observed ones. For this reason we also compute the fraction of AGN among the simulated galaxy samples cut at the maximum stellar mass of the corresponding observational samples. We find identical AGN fractions for the compact star-forming and quiescent galaxies as before (i.e. without the mass cuts), and removing the most massive galaxies only very slightly affects the fraction of AGN among the extended galaxies.
In summary, we do not find that accounting for mass differences among the simulated and observed samples could strongly change the results presented here. The similarity between the stellar mass of the extended and compact star-forming galaxy samples could perhaps explain the higher AGN fraction in extended SF galaxies compared to observations.  

To be consistent in this paper we have derived sample-dependent thresholds to identify galaxy types, and our criteria are therefore slightly different for the observations and the simulation, even if we obtained a pretty good agreement between the two.
As a final test, we also compute the AGN fractions when we apply the same thresholds (SFR and $\Sigma_{\rm e}$) derived from the simulation to the observational samples. We find roughly the same AGN fractions for the different type galaxies, and this does not significantly affect our results. Indeed, the AGN fraction of observed compact star-forming galaxies goes from $8.9$ to $9.4\%$ at $z\sim 2$, and from $9.8$ to $9.4\%$ at $z\sim 1.5$. The AGN fraction of the compact quiescent galaxies does not change at $z\sim 2$, and goes from $16$ to $12\%$ at $z\sim 1.5$.

\section{On the global evolution of galaxies: a case study following the progenitors of two representative quiescent galaxies back in time}
\begin{figure*}
\centering
\includegraphics[width=\columnwidth]{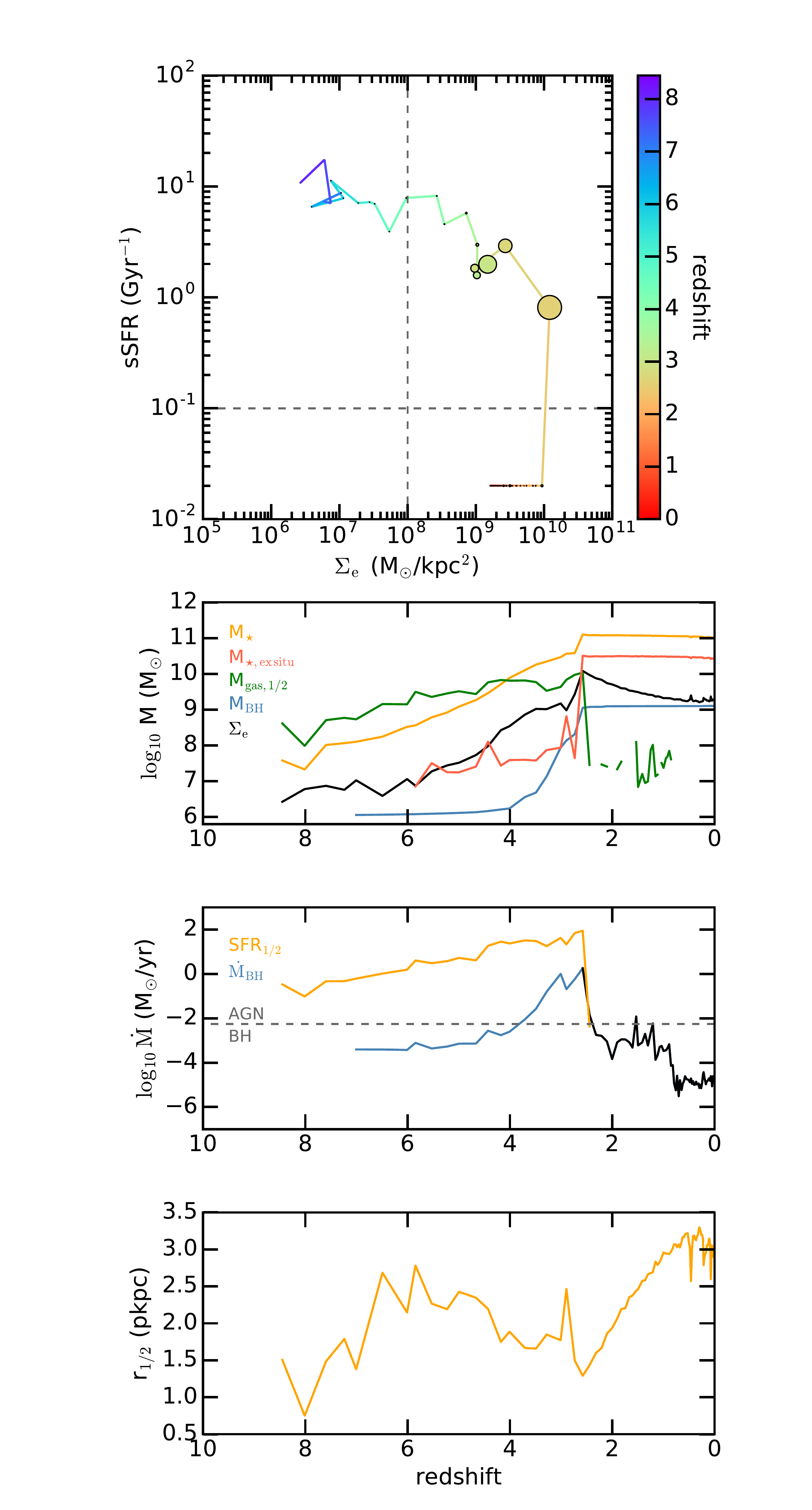}
\includegraphics[width=\columnwidth]{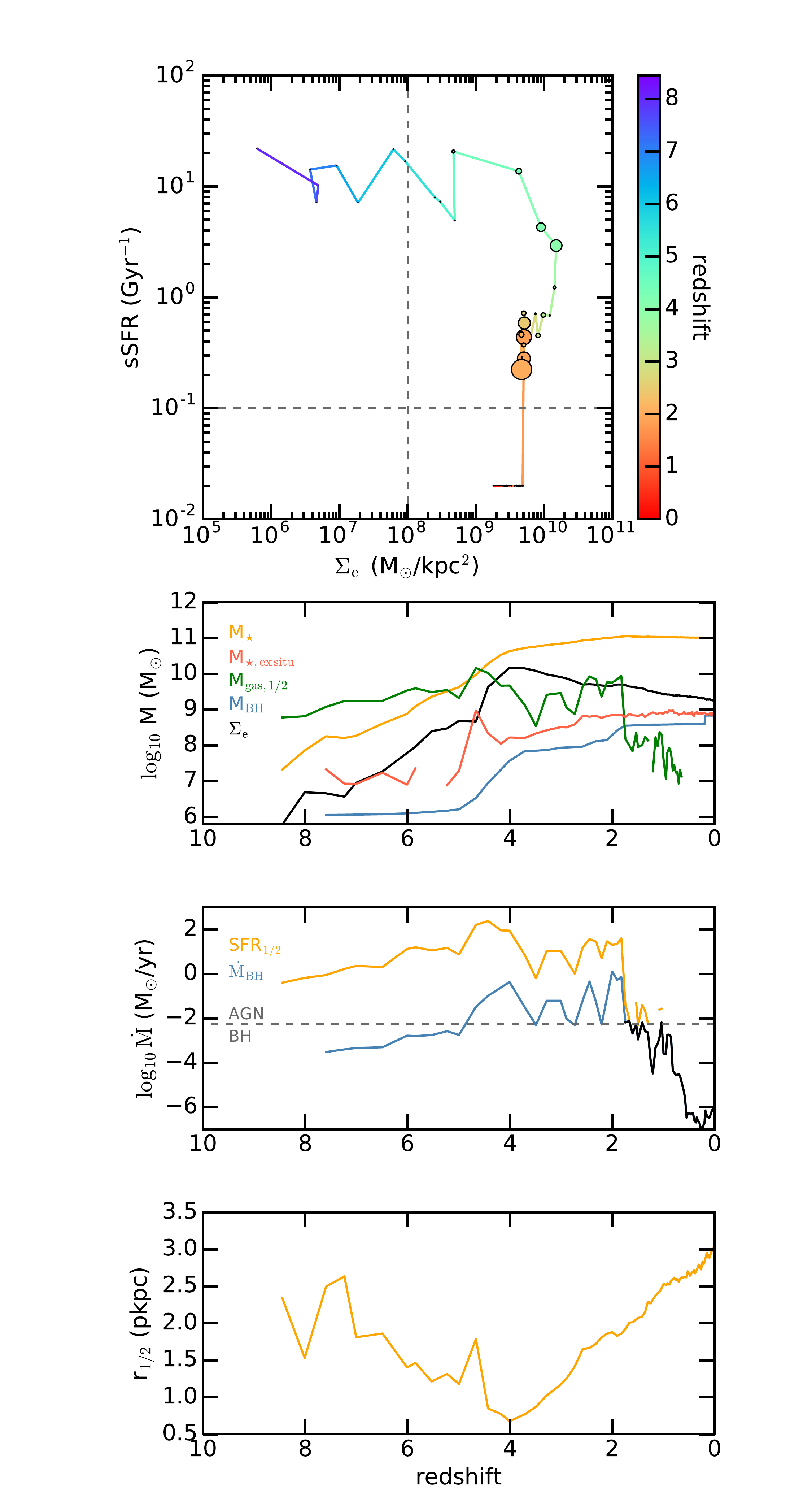}
\caption{Time evolution of two individual galaxies in TNG300. {\it Top panels}: specific star formation rate sSFR as a function of the effective stellar mass surface density $\Sigma_{\rm e}$. Circle radii are proportional to BH accretion rate, and are color coded by redshift. We set a floor of $\rm sSFR=2\times 10^{-2} Gyr^{-1}$ on the sSFR. {\it Second panels}: time evolution of several mass quantities -- the stellar mass of the galaxy ($M_{\star}$, in yellow), the ex-situ stellar mass ($M_{\star,\rm exsitu}$, in red), which is a proxy for galaxy mergers, the gas mass within the half-mass radius of the galaxy ($M_{\rm gas,1/2}$, in green), the BH mass ($M_{\rm BH}$, in blue). We also show $\Sigma_{\rm e}$ with black solid lines. {\it Third panels}: time evolution of the star formation rate enclosed in the half-mass radius of the galaxy ($\rm SFR_{1/2}$, in yellow), and of the BH accretion rate ($\dot{M}_{\rm BH}$, in blue). The horizontal dashed lines indicate the threshold above which the BH can be detected as an X-ray AGN ($L_{\rm x}>10^{42}$ erg/s). The solid black lines indicate when the kinetic AGN feedback mode is active. {\it Bottom panels}: time evolution of the half-mass radius of the galaxy. Common features: the stellar mass of the galaxy increases with time, as well as its gas content until it reaches a peak. This marks a compaction event, which occurs at a more or less constant specific star formation rate. Star formation rate and BH accretion rate also peak simultaneously at the same time, which also corresponds to a minimum in the galaxy size. After this peak of activity, which defines the end of the compaction event, the galaxy is quenched, the gas content of the galaxy drops, as well as the star formation rate, and the BH activity. Consequently, the galaxy stellar mass and BH mass also stop evolving rapidly. In these two examples, one can also identify a peak in the ex-situ stellar mass at the time of compaction, which indicates a merger has occurred.}
\label{fig:follow_quiescent}
\end{figure*}

Our goal here is to illustrate the coevolution between galaxy structural properties and BH activity through cosmic time for individual objects, in light of the statistical properties that we derived in the previous section. 
We trace back in time the evolution of the most massive and quiescent galaxies of $M_{\star}\geqslant 10^{11}\, \rm M_{\odot}$ at $z=0$. We find about $\sim 4500$ such galaxies in TNG300. A very large fraction of these galaxies have similar global evolution in the sSFR-$\Sigma_{\rm e}$ plane; to illustrate this we show the time evolution of 600 of these galaxies in Fig.~\ref{fig:global_evol}.
We do not aim at providing an exhaustive study of individual galaxy evolutionary paths. Therefore in Fig.~\ref{fig:follow_quiescent} we show only two representative examples of individual galaxies with $M_{\star}\geqslant 10^{11}\, \rm M_{\odot}$.

The top panels show the specific star formation rate as a function of the effective stellar mass surface density, i.e. the sSFR-$\Sigma_{\rm e}$ diagram. Circle radii give an indication of the level of BH activity. These circles and the lines are color coded by redshift. 
The next panels down represent the time evolution of several mass quantities related to the galaxy content: in yellow the stellar mass of the galaxy, in red the ex-situ stellar mass \citep[from the merger tree catalogs,]{2015MNRAS.449...49R}, which is a proxy for galaxy mergers, in green the gas mass within the half-mass radius of the galaxy, in blue the BH mass. We also add in black the evolution of the effective stellar mass surface density $\Sigma_{\rm e}$. The third panels down show the time evolution of the star formation rate (in yellow) enclosed in the half-mass radius of the galaxy, and the BH accretion rate is shown in blue. The horizontal dashed lines indicate the threshold above which the BH can be detected as an X-ray AGN (with an X-ray luminosity of $L_{\rm x}\geqslant 10^{42}$ erg/s). The black solid lines indicate where the kinetic feedback is active. Finally the bottom panels show the time evolution of the stellar half-mass 3D radius of the galaxy.

We see that galaxy stellar mass increases continuously with time, as well as the gas content, until a steep increase leading to a peak, at about $z\sim 2.5$ for the first galaxy, and about $z\sim 4.5$ for the second. This represents the compaction of the galaxy. Compaction is defined here as the reduction of the galaxy size, which eventually reaches a minimum. The compaction of the galaxies happens at more or less constant specific star formation rate, and can be appreciated on the top panels.
Star formation rate and BH accretion then peak simultaneously, which also corresponds to a minimum in the galaxy size, and roughly coincides with a peak in the evolution of $\Sigma_{\rm e}$.
After this peak of activity, which defines the end of the compaction event, the galaxy experiences a full quenching (galaxy 1), i.e. the suppression of SFR is sustained in time down to $z=0$, or a quenching attempt (galaxy 2), i.e. a suppression of the SFR followed by new peaks of star formation. 
The gas content of the galaxy drops, as well as the star formation rate, and the BH activity. Consequently, the galaxy stellar mass and BH mass also cease to evolve rapidly.
After this stage, the galaxy radius starts increasing again.

For the second galaxy, the SFR and BH activity are not quenched forever, but instead it experiences an oscillation of activity, where the BH can still be detected as an AGN, until another quenching event, that will lead to the final full quenching of the galaxy, at about $z\sim 2$.
For galaxy 1, we see that the quenching of the galaxy corresponds exactly to the time when AGN kinetic feedback becomes active, rather than the onset of the thermal feedback. However, for galaxy 2, the first quenching attempt of the galaxy does not correspond to the time AGN kinetic feedback is active, which tends to illustrate that galaxy quenching is a complex process, that may involve several mechanisms, such as gas consumption, SN feedback and thermal AGN feedback.
In these two examples, one can also identify a peak in the ex-situ stellar mass at time of compaction, which indicates that in some cases mergers play a role in galaxy compaction.

For the purpose of this paper, we do not detail the mechanisms for compaction in IllustrisTNG.
We have purposely focussed our attention on massive galaxies of $M_{\star}\geqslant 10^{10}-10^{11}\, \rm M_{\odot}$, to address a comparison with observational results. 
We have seen that those galaxies generally first experience a compaction event, and a quenching event. We also find in the simulations that some lower mass galaxies, with $M_{\star}=10^{9}-10^{10}\, \rm M_{\odot}$, seem to be able to quench without experiencing a strong compaction event, but we do not explore these in detail here. We are, particularly at $z\sim 2$, able to distinguish between these two channels of quenching, one after a compaction event and therefore producing compact galaxies (for $\Sigma_{\rm e}\sim 10^{8.5}-10^{10}\, \rm M_{\odot}/pkpc^{2}$), the second for lower mass galaxies that are also more extended (for $\Sigma_{\rm e}\sim 10^{7}-10^{8}\, \rm M_{\odot}/pkpc^{2}$) \citep[see also Fig. 19 in][]{2014MNRAS.445..175G}.

\section{Discussion and Conclusions}
We have analyzed the new generation of magneto-hydrodynamic cosmological simulations IllustrisTNG, with a particular interest in the coevolution between BH activity and the star formation and structural properties of their host galaxies through cosmic time. We have carried out a close comparison with observations from the {\sc candels} survey to investigate to what extent the simulated populations of active BHs and their host galaxies were able to reproduce and explain key observational findings such as AGN fractions as a function of a galaxy's location in the SFR-$\Sigma_{\rm e}$ diagram, where $\Sigma_{\rm e}$ is used as an indicator of galaxy compactness.\\

In the following we summarize our results in two sections, the first one on the evolution of the active BH population in IllustrisTNG, and the second on the coevolution of these BHs and the properties of their host galaxies.\\

\noindent {\bf Population of BHs}
\begin{itemize}
\item The hard X-ray luminosity function (Fig.~\ref{fig:xlf}, Fig.~\ref{fig:LF_churazov}, and Fig.~\ref{fig:xlf_eps}) is in good agreement with observations at very high redshift ($z=5$), but significantly overproduces AGN with $\log_{10} L_{\rm x}/ (\rm erg \, s^{-1})=42-45$ at early times. By $z=2-1$, we still see an overproduction of AGN with $\log_{10} L_{\rm x}/ (\rm erg \, s^{-1})=42-43$. The simulation does not seem to produce enough bright AGN compared to observational constraints \citep{Hop_bol_2007,2015ApJ...802...89B}. These results vary quantitatively but not qualitatively with BH luminosity estimation method (Fig.~\ref{fig:LF_churazov}).

\item The large volume of the simulation allows us to statistically study the time evolution of the fraction of bright AGN ($L_{\rm bol}\geqslant 10^{44}\, \rm erg/s$) hosted by massive galaxies with $M_{\star}\geqslant 10^{11}\, \rm M_{\odot}$ (Fig.~\ref{fig:occupation_fraction_massivegal}). Except for a lack of these AGN at low redshift $z\lesssim 1$, we find very good agreement with observations \citep{2007ApJ...669..776K,2016MNRAS.457..629C,2017ApJ...842...21M}.

\item We use the AGN occupation fraction as a diagnostic for the impact of AGN feedback modeling in the simulation (Fig.~\ref{fig:bh_occ_fraction}). 
The AGN occupation fraction is close to unity for galaxies with $M_{\star}\leqslant 10^{10}\, \rm M_{\odot}$, but then drops rapidly for more massive galaxies with mean BH mass around $M_{\rm BH}\sim 10^{8}\, \rm M_{\odot}$, i.e. when BHs transition from the quasar mode thermal feedback to the very efficient kinetic mode of AGN feedback.

\item The distribution of the Eddington ratios (Fig.~\ref{fig:edd_ratio_compheckman}) is in good agreement at $z=0$ for $M_{\rm BH}<10^{9}\, \rm M_{\odot}$ with the observational constraints of \citet{2004ApJ...613..109H}. However, the Eddington ratios of more massive BHs of $M_{\rm BH}\gtrsim 10^{9}\, \rm M_{\odot}$ are about one order of magnitude below the observational constraint, which suggests that the simulated massive BHs accrete at lower rates than the observed ones due to the efficient kinetic feedback. The Eddington ratios of the observations predict that most of these massive BHs are in the radiatively inefficient regime. Therefore, while the simulation may be overly efficient at clearing gas around these BHs, the mode of accretion/feedback may agree with the observational data.
We observe a bimodal Eddington ratio distribution for the BH mass range $M_{\rm BH}=10^{8}-10^{9} \, \rm M_{\odot}$, which reflects the transition between the quasar and kinetic modes of AGN feedback.
Moreover the time evolution of the mean Eddington ratio distributions (Fig.~\ref{fig:mean_fedd}) is in good agreement with the observational constraints of \citet{2012ApJ...746..169S}. 

\end{itemize}

In the following we further discuss some of our conclusions.
The simulations show reasonably good agreement with observational constraints, although we find that the faint-end of the AGN population is overproduced, and the very bright-end underproduced. The presence of too many faint AGN compared to observational constraints has also been found in other state-of-the-art cosmological simulations \citep{2015MNRAS.452..575S,2016MNRAS.460.2979V}, while other simulations show a smaller overproduction \citep{Hirschmann2012}. Applying a cut in BH or host galaxy mass reduces the number of low luminosity AGN \citep[see also][]{2012MNRAS.420.2662D,2015MNRAS.452..575S,2016MNRAS.460.2979V}, which suggests that poorly resolved or insufficiently efficient SN feedback could be responsible for the overproduction of faint AGNs. 
The faint end of the X-ray luminosity function is affected by the bolometric estimation method used \citep[see also][]{Hirschmann2012}. The number of faint AGN decreases when accounting for both radiatively efficient and inefficient AGN, especially for $z\leqslant 1$, but does not affect qualitatively our findings. 


The simulations seem to predict a lower number density of very bright AGN than observed at all redshifts, which is due to the very efficient kinetic mode of the AGN feedback implemented here.
AGN accretion rates can vary by orders of magnitude on potentially short timescales from days to Myrs. 
The bright end of the XLF could be affected by Eddington bias due to additional sources of scatter in the AGN luminosity, such as short time variability of the accretion onto BHs that is not captured by the simulations.

We have derived all the BH properties presented here for the large simulated volume TNG300 and the higher resolution but lower volume simulation TNG100 to investigate resolution effects.
The two simulations show similar behaviours. Some discrepancies are observed in the BH - stellar mass relation where TNG100 shows a very tight correlation, and the larger volume of TNG300 allows for more scatter around the correlation \citep[see][]{2017arXiv171004659W}. As discussed in \citet{2017arXiv171004659W}, the higher/lower resolution of TNG100/TNG300 affects the evolution of BH mass and BH accretion for fixed halo mass. Consequently the overproduction of faint AGN is more severe in TNG100.

Finally, we have found that some of the diagnostics presented in our study keep the imprint of the AGN feedback modeling. In the simulations, the dividing point between the thermal and the kinetic modes of the AGN feedback depends on BH mass. We found a bimodality in the Eddington ratio distributions for the BH mass range for which a substantial number of BHs enter the kinetic mode of the AGN feedback ($\log_{10}M_{\rm BH}=8-9 \,\rm M_{\odot}$). We also found a definite drop in the AGN occupation fraction above the BH mass at which the kinetic mode is more effective. 
Further development/effort to constrain these two quantities (Eddington ratio distribution, AGN occupation fraction) from observations \citep[such as][]{2017arXiv170501132A}, as a function of e.g., galaxy stellar mass, BH mass or luminosity, and following dedicated comparisons with the simulations, will be crucial to test the current implementation of AGN feedback, and/or develop new ones.\\

\noindent {\bf Correlation between galaxy structural properties and BH activity}\\
In the second part of the paper we have investigated the connections between the population of active BHs in IllustrisTNG and their host galaxy properties, and carefully compared them with observations of galaxies and AGN from the {\sc candels} survey \citep{2017ApJ...846..112K}.

\begin{itemize}

\item We find good agreement between the populations of simulated and observed galaxies in terms of the SFR-$M_{\star}$ sequence (Fig.~\ref{fig:sfr_mstar}), sSFR-$\Sigma_{\rm e}$ (Fig.~\ref{fig:second_criterion}), and relatively good agreement for the mean relation between sizes and galaxy stellar mass \citep[see][for a detailed discussion]{2017arXiv170705327G}. 

\item We have established redshift- and mass-dependent criteria to classify galaxies into different types, i.e. star-forming vs quiescent galaxies, and compact vs extended galaxies, based on galaxy SFR and stellar mass surface density $\Sigma_{\rm e}$. 
Star-forming galaxies are defined here as galaxies with SFR above the 25th percentile of the star-forming sequence, and compact galaxies as galaxies with $\Sigma_{\rm e}$ higher than the 25th percentile of the quiescent population distribution.
Our sample-dependent method allowed us to apply these thresholds to both observations and simulations (Fig.~\ref{fig:sfr_mstar}, Fig.~\ref{fig:second_criterion}).
The quantity SFR is a simulation-friendly quantity, whereas selections based on colors has historically been more common in observational studies \citep{2017ApJ...846..112K} that compute the AGN fraction among different galaxy types. While SFR is less well constrained in observations than colors, there are large uncertainties on predicting colors from simulations \citep[see][for more discussion on colors in simulations]{2018MNRAS.475..624N}. Here we have decided to work with the SFR, which is also easier to interpret in a physical sense.

\item The number density of compact star-forming simulated galaxies increases with time before reaching a peak (at about $z\sim 1$) and decreasing, while the number of quiescent galaxies increases all the way to $z=0$ with a steeper slope (Fig.~\ref{fig:nb_density}). The number densities of these two populations intersect at higher redshift for more massive galaxies. Number density functions are consistent with star-forming compact galaxies being the progenitors of quiescent compact galaxies. We note here that direct comparisons with observations or previous theoretical works are not straightforward as all studies use different definitions of star-forming/quiescent and extended/compact galaxies.

\item We found that the distributions of specific BH accretion (Fig.~\ref{fig:distri_bhar}, sBHAR) become broader with time and the peak of the distributions moves to lower sBHAR, in good agreement with the recent observational study of \citet{2017arXiv170501132A}. The distributions of quiescent galaxies start deviating from the star-forming ones at $z\sim2$ with lower sBHAR. Star-forming galaxies have a higher probability of hosting accreting BHs. Quiescent galaxies host BHs with lower sBHAR than SF galaxies in the same stellar mass range and redshift. The peak of the sBHAR distributions are generally strongly dependent on redshift, and weakly dependent on the host galaxy stellar mass. 

\item AGN hard X-ray luminosity (Fig.~\ref{fig:agn_lum}) and AGN fraction (Fig.~\ref{fig:agn_fraction}) are both strongly correlated with a host galaxy's location in the sSFR-$\Sigma_{\rm e}$ diagram. Star-forming compact galaxies have the highest probability of hosting an AGN of $L_{\rm x}\geqslant 10^{42}\, \rm erg/s$, 
while quiescent compact galaxies have the lowest probability. More compact and star-forming galaxies have a higher probability of hosting a more luminous AGN (up to $L_{\rm x}\sim 10^{44}\, \rm erg/s$) in the simulation.
The simulation is able to reproduce the qualitative trends found in several observational studies for AGN fractions in different galaxy types \citep{2013ApJ...763L...6T,2014ApJ...791...52B,2017ApJ...846..112K}.

\item To compare the AGN fractions with observations from the {\sc candels} survey, we have built an empirical model for obscuration that depends both on redshift and AGN luminosity, following observational results from \citet{2014ApJ...786..104U} and \citet{2014MNRAS.437.3550M}. Our model assumes an anti-correlation between the AGN obscured fraction and AGN luminosity, and that there are more obscured AGN towards higher redshifts. 
With this correction for obscured AGN we find good quantitative agreement of the AGN fractions with the observations from {\sc candels}: simulated compact star-forming galaxies have a probability of $20\%$ to host an X-ray AGN at $z=2,1.5$, while observed galaxies have a probability of $13-16\%$ to do so. Quiescent compact galaxies have a probability of $6-9\%$ to host an AGN in the simulation at $z=2,1.5$, and $9-10\%$ in the {\sc candels} observations. 
We find twice as many hard X-ray AGN in compact star-forming galaxies than in their quiescent counterparts. This reveals that BHs have a very active phase while in the compact star-forming phase. 
In the simulation, these BHs enter the efficient kinetic mode of the AGN feedback. The kinetic mode of the AGN feedback is responsible for reducing and quenching star formation over long timescales in these simulations.

\item  We have investigated several discrepancies between the simulated and observed samples of galaxies, and the distributions of AGN luminosity appear to be the most significantly discrepant (Fig.~\ref{fig:histo_xray}). In the simulation, AGN luminosity distributions of star-forming and quiescent compact galaxies are similar at high redshift (e.g., $z=4$), but then strongly deviate from one another by more than one order of magnitude for their mean values. However, compact star-forming and quiescent galaxies have quite similar AGN luminosity distributions in the {\sc candels} observations. As a consequence, the AGN fractions derived for different galaxy types, especially for compact quiescent, are more affected by the luminosity threshold used to identify AGN in the simulation than in observations.

\item We have traced back in time $z=0$ quiescent galaxies of $M_{\star}\geqslant 10^{11}\, \rm M_{\odot}$ to understand their evolution (Fig.~\ref{fig:follow_quiescent}, Fig.~\ref{fig:global_evol}). A large fraction of these galaxies first compactify and reach a smaller radius, which corresponds to a peak of SFR and BH activity. Galaxies then quench, and can no longer sustain BH accretion and star formation activity. Interestingly, we note that for some galaxies several episodes of quenching are needed to fully suppress their activity. These examples nicely illustrate the role of the AGN kinetic feedback mode in quenching galaxies in IllustrisTNG.

\end{itemize}

Quantitatively, we find that the simulation nicely reproduces AGN fractions found in the {\sc candels} survey (Fig.\ref{fig:fraction_quadrant}) for compact SF and quiescent $M_{\star} \geqslant 10^{10}\, \rm M_{odot}$ galaxies. Of course, our results are subject to the uncertainties of the BH XLF, and also to large uncertainties in the obscured AGN fraction.

Obscuration is a major uncertainty for studies aiming at deriving AGN fractions, especially when compared to observations. First of all, the obscured fraction of AGN is likely to be not only a function of redshift, and BH luminosity, as modeled here, but also certainly of galaxy properties.
Correlations between the obscuration of AGN and the properties of their host galaxies are however not clear yet. For example, \citet{2014MNRAS.437.3550M} found no correlation between obscuration and galaxy stellar mass or SFR.
However, we do not expect the column density around AGN to be independent of galaxy structural type, i.e. whether galaxies are more compact or extended. In fact, mechanisms responsible for compactifying galaxies could also result in an increase of the central column density. Compact star-forming galaxies could therefore potentially host many more obscured AGN \citep[see][for more details on the population of obscured AGN]{2017MNRAS.466L.103C}. We also refer here to \citet{2014A&A...562A..67G} for a particular case at $z=4.7$ of a compact star-forming galaxy hosting a heavily obscured, Compton-thick AGN. In their study, the star formation rate, gas depletion timescale, and the compactness of the galaxy suggest that it could be one of the progenitors of high-redshift compact quiescent galaxies.

Uncertainties in the fraction of AGN obscured by Compton-thick gas resulting in $N_{\rm H}\geqslant 10^{24}\, \rm cm^{-2}$ are large, especially because they are difficult to observe in energies $\leqslant 10 \, \rm keV$. Recent studies argue that Compton-thick AGN most likely arise from the nuclear region around the BHs \citep{2017MNRAS.465.4348B} and should therefore probably be only weakly dependent on galactic-scale properties (with the exception of some massive galaxies being able to have Compton-thick column density at galactic-scale). The fraction of these Compton-thick AGN can be potentially important. Results from \citet{2007A&A...463...79G} suggest that the ratio between Compton-thin, Compton-thick and unobscured AGN could be 4:4:1 (respectively) for an AGN X-ray luminosity of $L_{\rm x}<10^{43}\, \rm erg/s$, and 1:1:1 at higher X-ray luminosity $L_{\rm x}\geqslant 10^{44}\, \rm erg/s$.
In our analysis, we have used the observational constraints of \citet{2014MNRAS.437.3550M}, which mostly consider Compton-thin AGN, although they cannot rule out contamination by Compton-thick AGN. 
We have not further corrected for the presence of Compton-thick AGN in our simulated samples. Indeed it is not clear what would be the fraction of heavily obscured Compton-thick AGN missed by the {\sc candels} observations that we use, and whether we would totally miss these AGN, or if their apparent X-ray luminosity would be only shifted to lower luminosity.

Further modeling of obscuration as a function of galaxy properties from observational samples and comparison to semi-analytical models and hydrodynamic simulations such as those presented here are needed to better constrain the fraction of obscured and heavily obscured AGN. 
As the quantitative agreement of the AGN fractions in the different galaxy type samples between the {\sc candels} observations and the IllustrisTNG simulation strongly depends on the obscuration fraction correction, our results should be updated upon arrival of any new observational constraints.

Finally, our approach in this paper was to develop sample-based thresholds to identify galaxies with types, and to apply them to both the simulation and the observations. The differences that we found between the distributions of hard X-ray luminosity for AGN hosted by compact SF and quiescent galaxies between the simulation and the observations is very interesting. It motivates future works where observational and simulated samples could instead be matched in several key quantities such as galaxy stellar mass, SFR, and BH luminosity.

\section*{Acknowledgements}
MH thanks Roberto Gilli and James Aird for very useful discussion on the obscured AGN fraction, Vicente Rodriguez-Gomez for discussion on the merger trees of IllustrisTNG, and Fabio Vito for discussion on the observational constraints on the BH luminosity function.
We thank Nick Carrerio and Ian Fisk, as well as the entire computational core of the Flatiron Institute, for smoothly running the Flatiron Rusty cluster. The Flatiron Institute is supported by the Simons Foundation.
RSS is grateful for the generous support of the Downsbrough family.
M Hirschmann acknowledges financial support from the European Research Council (ERC) via an Advanced Grant under grant agreement no. 321323 NEOGAL.

\bibliography{biblio_complete}

\appendix
\section{AGN luminosity function and Eddington ratio}
\begin{figure*}
\centering
\includegraphics[scale=0.5]{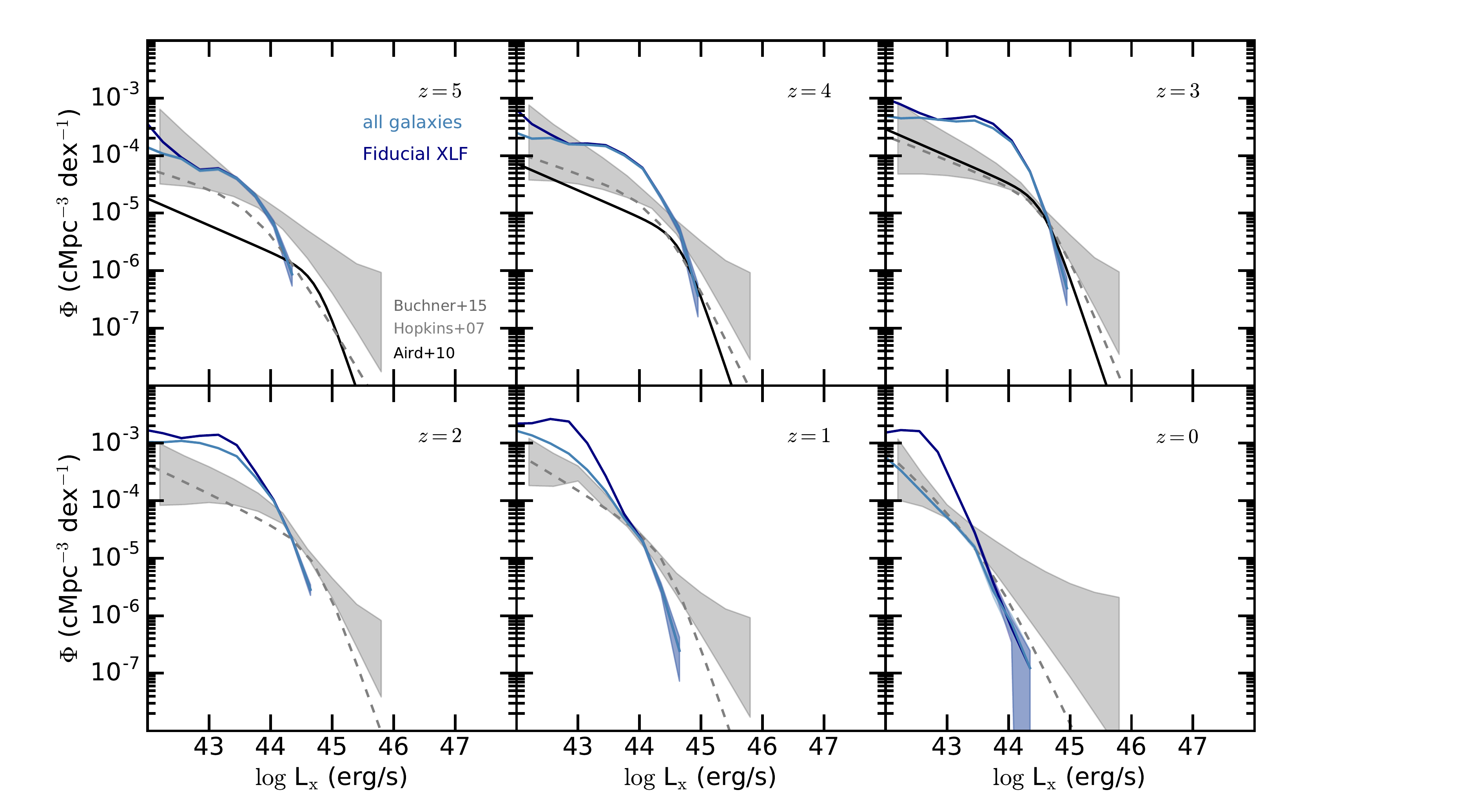}
\caption{AGN luminosity function for TNG300 (blue lines) where we introduce a distinction between radiatively efficient and inefficient AGN when we compute their bolometric luminosity \citep{Churazov2005}. The main effect is to slightly lower the number of faint AGN, specially for $z\leqslant 1$. Black and grey lines show observational constraints. For comparison, our fiducial XLF from Fig.~\ref{fig:xlf} is shown in dark blue color.} 
\label{fig:LF_churazov}
\end{figure*}

Fig.~\ref{fig:LF_churazov} shows the luminosity function of the AGN (blue lines) as in Fig.~\ref{fig:xlf}, but we now compute the AGN luminosity following \citet{Churazov2005}. The bolometric luminosity of an AGN now depends on its Eddington ratio.
As discussed in Section 4.1.2, the main effect of this modification is to lower the number of faint AGN, especially for redshift $z\leqslant 1$.

\begin{figure*}
\includegraphics[scale=0.5]{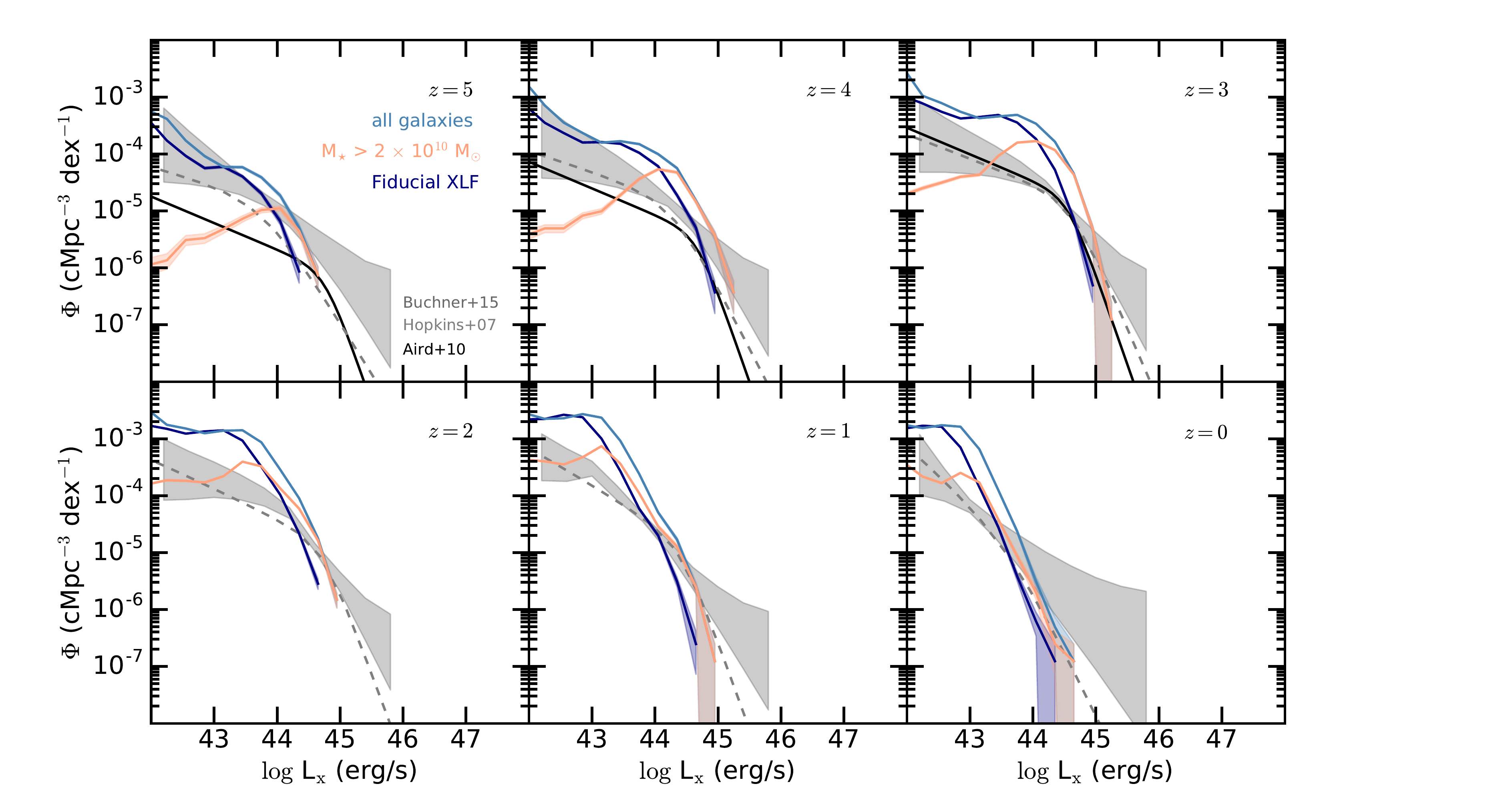}
\caption{BH hard (2-10 keV) X-ray luminosity function (TNG300), similar to Fig.~\ref{fig:xlf}, but we adopt a different value of the radiative efficiency $\epsilon_{\rm r}=0.2$. This leads to larger differences with observations at the faint end of the XLF, and better agreement at the bright end of the XLF. For comparison, our fiducial XLF from Fig.~\ref{fig:xlf} is shown in dark blue color.}
\label{fig:xlf_eps}
\end{figure*}

In Fig.~\ref{fig:xlf_eps}, we show a variation of Fig.~\ref{fig:xlf} for which we now consider a higher value of the radiative efficiency $\epsilon_{\rm r}=0.2$. This leads to larger differences with observations at the faint end of the XLF, and a better agreement at the bright end of the XLF.

\section{Time evolution of the fraction of bright AGN in massive galaxies}
\begin{figure*}
\centering
\includegraphics[scale=0.55]{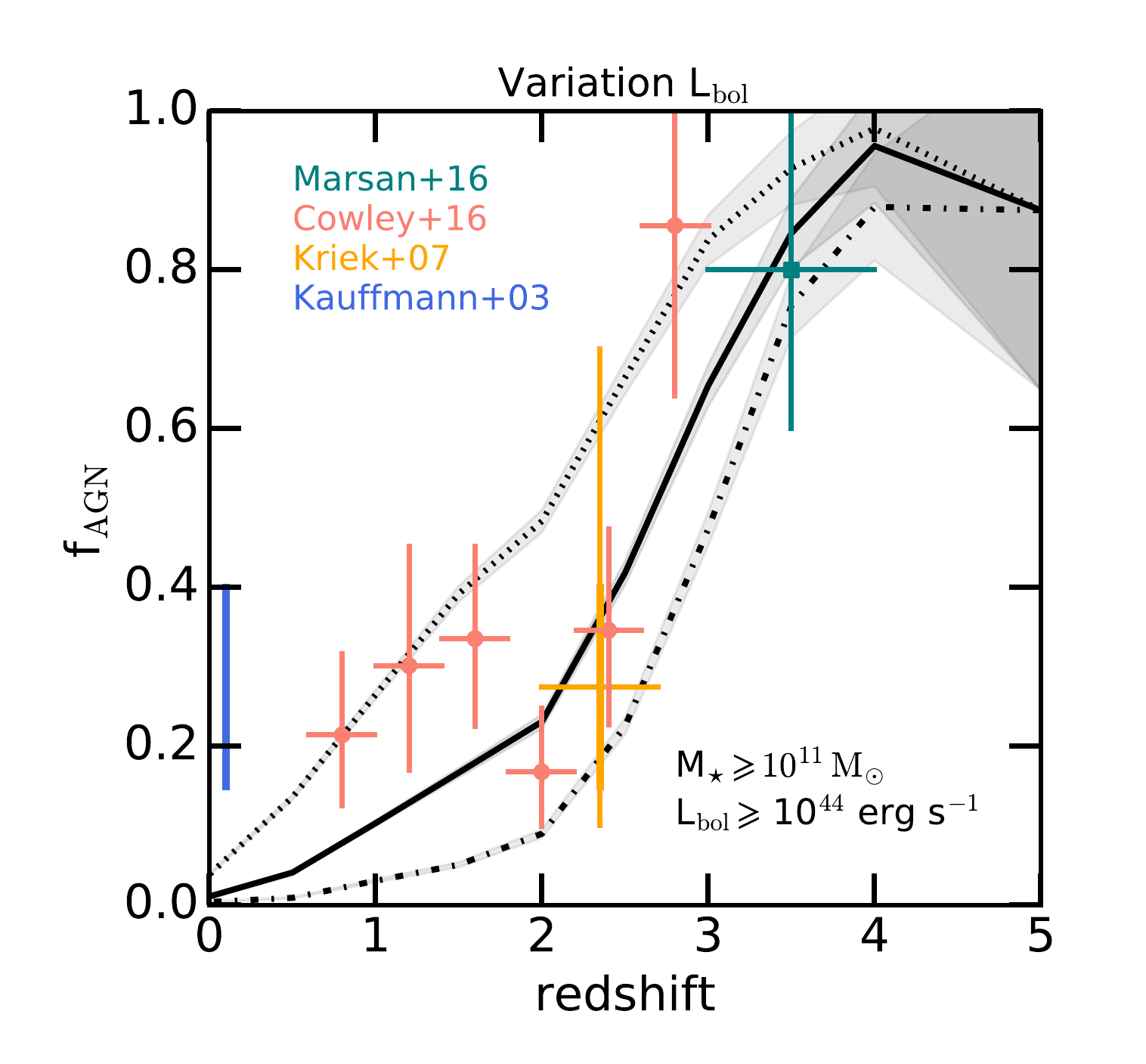}
\includegraphics[scale=0.55]{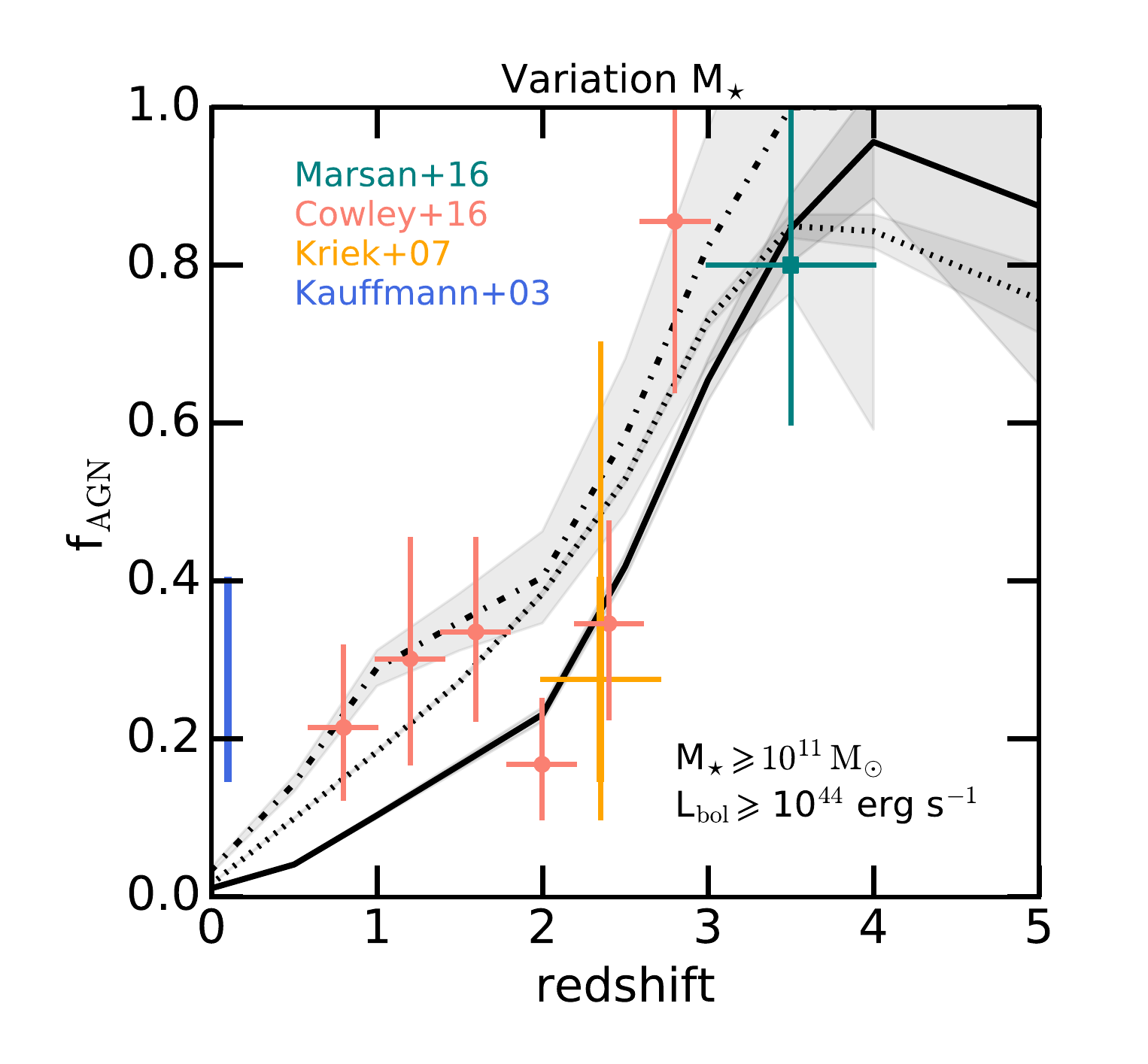}
\caption{Fraction of AGN in massive galaxies in TNG300. {\it Left panel:} Variation of the bolometric luminosity cut, retaining the $M_{\star}\geqslant 10^{11}\, \rm M_{\odot}$ cut. The black solid line shows the AGN fraction with $L_{\rm bol}\geqslant 10^{44}\, \rm erg/s$ AGN. The dashed dotted line shows the fraction for AGN with higher luminosities  $L_{\rm bol}\geqslant 10^{44.5}\, \rm erg/s$, and the dotted line shows the fraction for $L_{\rm bol}\geqslant 10^{43.5}\, \rm erg/s$ AGN. {\it Right panel:} Variation of the stellar mass cut, retaining the cut in luminosity $L_{\rm bol}\geqslant 10^{44}\, \rm erg/s$. The black solid line shows the AGN fraction for our initial $M_{\star}\geqslant 10^{11}\, \rm M_{\odot}$ cut, the dotted line uses a $M_{\star}\geqslant 10^{10.5}\, \rm M_{\odot}$ cut, and the dashed dotted line uses a $M_{\star}\geqslant 10^{11.5}\, \rm M_{\odot}$ cut.}
\label{fig:occupation_fraction_massivegal_variation}
\end{figure*}

The large volume of the simulation TNG300 allows us to investigate the fraction of bright AGN ($L_{\rm bol}\geqslant 10^{44} \rm erg/s$) found in massive galaxies ($M_{\star}\geqslant 10^{11}\, \rm M_{\odot}$), and to compare with observations. The observations described in Section 4.2 are all consistent with these cuts in bolometric luminosity and galaxy mass, and compared with TNG300 and TNG100 in Fig.~\ref{fig:occupation_fraction_massivegal}. In order to give a sense of how the results vary in the simulation, we slightly vary the $L_{\rm bol}$ cut (left panel) and the galaxy stellar mass $M_{\star}$ (right panel) in Fig.~\ref{fig:occupation_fraction_massivegal_variation}.
The solid black line in both panels is our reference ($L_{\rm bol}\geqslant 10^{44} \rm erg/s$, and $M_{\star}\geqslant 10^{11}\, \rm M_{\odot}$), and is the one to be compared with the observational constraints.
We include Poisson error bars as shaded regions; they are a function of both the number of massive galaxies, and the number of AGN among these galaxies.

In the left panel, we also show the fraction of AGN that are slightly less luminous, i.e. $L_{\rm bol}\geqslant 10^{43.5} \rm erg/s$, with the dotted line.
As we relax the luminosity cut, and therefore allow for the detection of more AGN in galaxies, it is not surprising that the fraction of AGN becomes higher than the reference line. Similarly, when we are more restrictive and only allow for the detection of brighter AGN, the fraction of these AGN is lower than the reference. In other words, the fraction of AGN among massive galaxies of $M_{\star}\geqslant 10^{11}\, \rm M_{\odot}$ increases when we consider slightly fainter AGN.
In the right panel of Fig.~\ref{fig:occupation_fraction_massivegal_variation} we slightly vary the mass threshold for massive galaxies. We consider less massive galaxies with $M_{\star}\geqslant 10^{10.5}\, \rm M_{\odot}$ for the dotted line, and we are slightly more conservative with the dashed dotted line and only include galaxies with $M_{\star}\geqslant 10^{11.5}\, \rm M_{\odot}$.
The fraction of AGN in massive galaxies is higher when we decrease or increase the cut in galaxy mass, which is not trivial to understand. This behavior can be connected to Fig.~\ref{fig:bh_occ_fraction} where we have identified a saddle point in the AGN occupation fraction at $M_{\star}\sim 10^{11}\, \rm M_{\odot}$. This saddle point is a direct outcome of the kinetic AGN feedback mode modeling. Including lower mass galaxies increases the probability of detecting more AGN, and looking at only more massive galaxies also increases the probability of finding AGN because from Fig.~\ref{fig:bh_occ_fraction} galaxies more massive than $M_{\star}\sim 10^{11}\, \rm M_{\odot}$ have a higher probability of hosting an AGN.


\section{AGN fractions in the {\sc candels} observations in the redshift range $1.4\leqslant z \leqslant 3$}

\begin{figure*}
\centering
\includegraphics[scale=0.60]{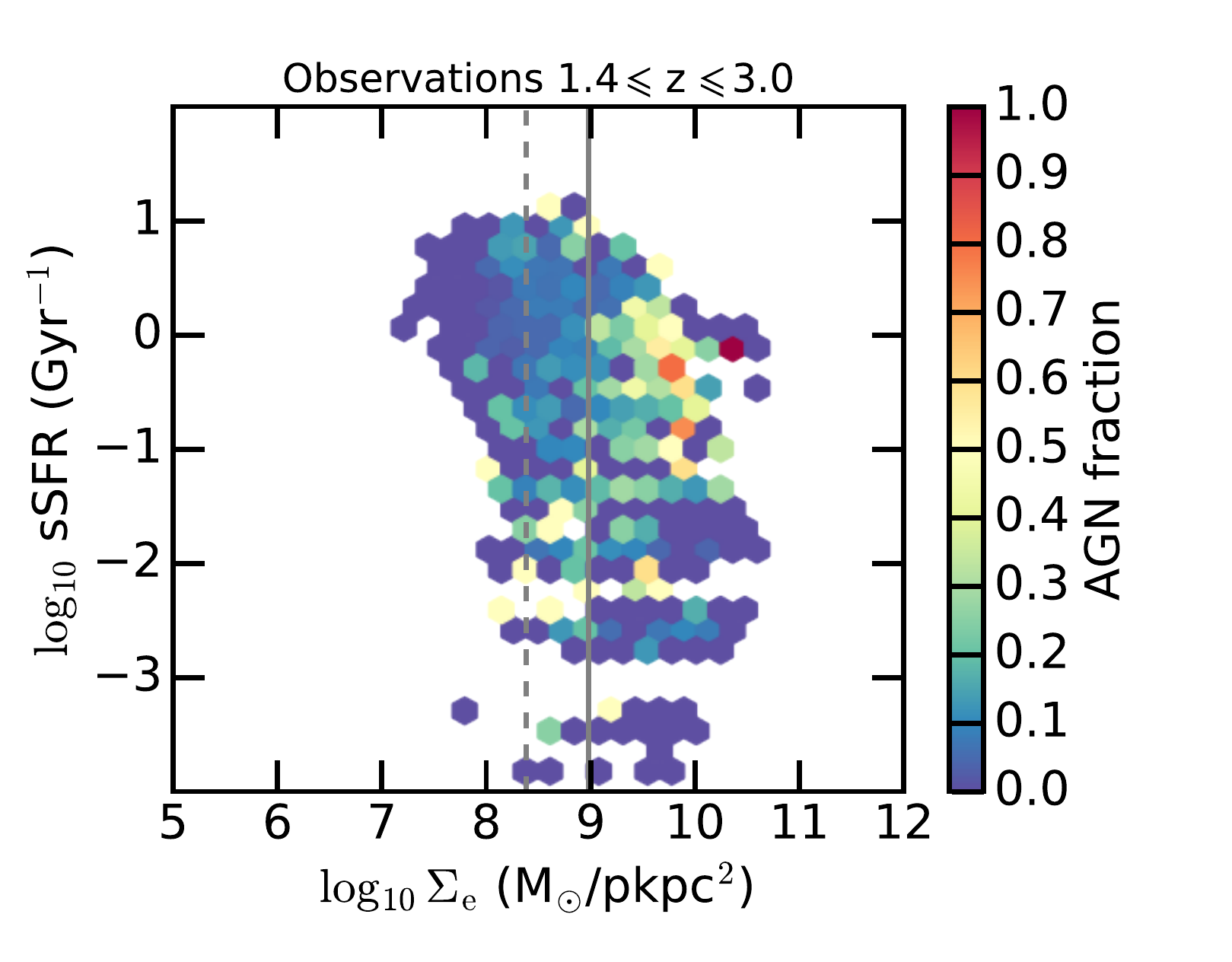}
\includegraphics[scale=0.47]{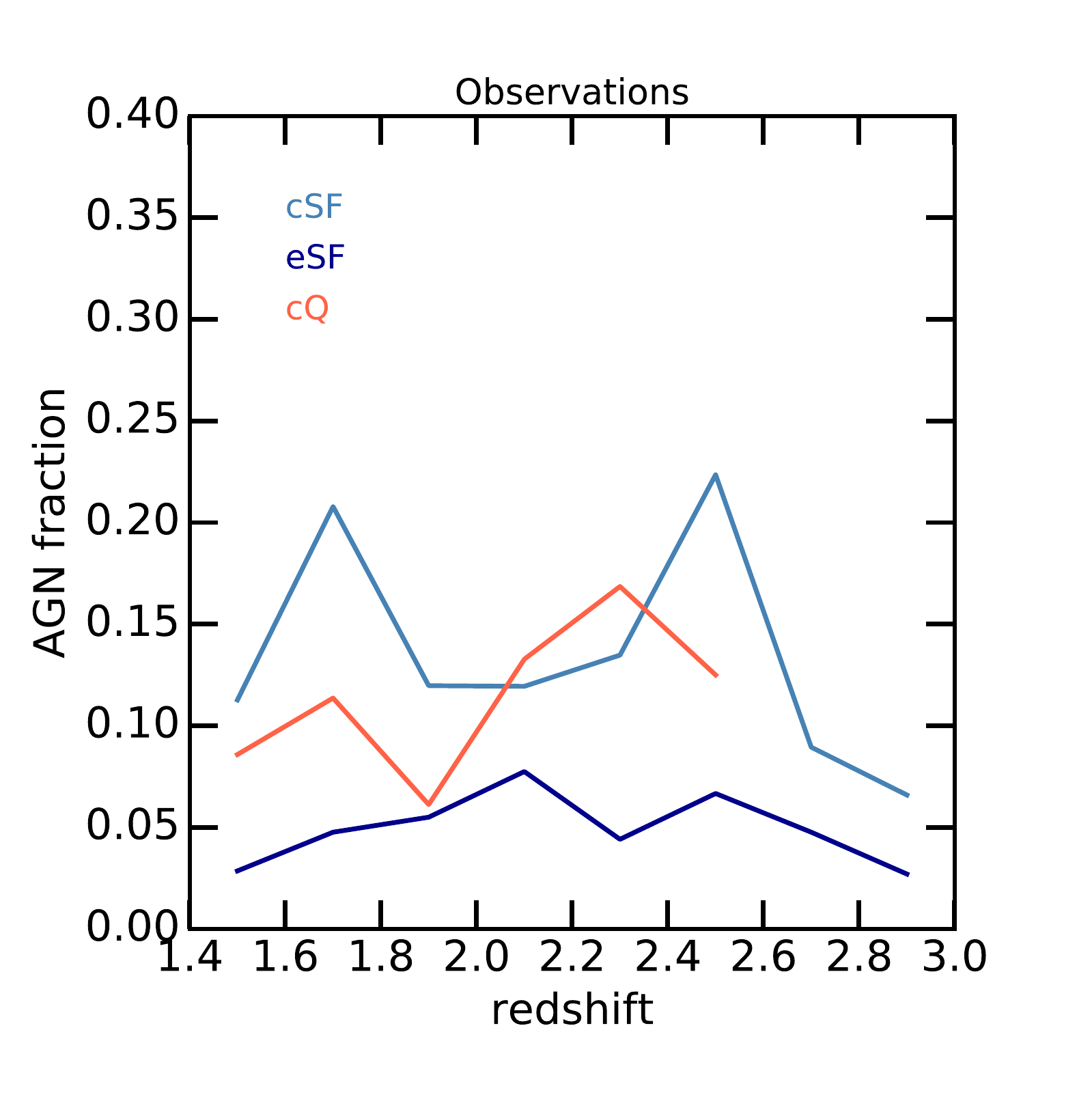}
\caption{{\it Left panel}: sSFR-$\Sigma_{\rm e}$ diagram color coded with the AGN fraction in hexabins for the observations from the {\sc candels} survey for $1.4\leqslant z\leqslant 3$ with $M_{\star}\geqslant 10^{10}\, \rm M_{\odot}$. Compact galaxies (right of the vertical dashed line) around the main sequence (horizontal dashed line) have the highest probability of hosting an AGN. {Right panel: Time evolution of the AGN fraction for compact star-forming (light blue), extended star-forming (dark blue), and compact quiescent galaxies (red solid line). We only show fractions for samples with more than 30 galaxies, and do not show the evolution of the extended quiescent galaxy population for this reason. Compact star-forming galaxies have the highest fraction of AGN, followed by compact quiescent galaxies, and extended star-forming galaxies.}}
\label{fig:candels_AGNfraction}
\end{figure*}

In this section, we use our selection method to classify galaxies by their structural type in the {\sc candels} observations in the redshift range $1.4\leqslant z\leqslant 3$ (see section 6.2.3), and compare the observed AGN fraction that we obtained to the results of \citet{2017ApJ...846..112K}.
Galaxies are considered compact if their stellar mass surface density $\Sigma_{\rm e}$ is higher than the percentile $25\%$ (grey vertical dashed line, $\log_{10} \Sigma_{\rm T}=8.38 \, \rm M_{\odot}/pkpc^{2}$) of the median $\log_{10} \Sigma_{\rm e}=8.98 \, \rm M_{\odot}/pkpc^{2}$ of quiescent galaxies (grey vertical solid line). 
Fig.~\ref{fig:candels_AGNfraction} (left panel) shows the AGN fractions among the different galaxy types.
We find indeed that $13.4\%$ of the compact star-forming galaxies host an AGN, $2.7\%$ for the extended star-forming, $5.8\%$ for the extended quiescent, and $10.8\%$ for the compact quiescent galaxies. 
The differences in the galaxy type selection methods lead to changes in the AGN fraction compared to \citet{2017ApJ...846..112K}. \citet{2017ApJ...846..112K} uses a mass-dependent threshold to distinguish compact from extended galaxies, based on the separation from the best fit of the $\Sigma_{\rm e}-M_{\star}$ relation for quiescent galaxies \citep{2017ApJ...840...47B}, i.e. $\log_{10} \Sigma_{\rm e}=-0.52[\log_{10}\left(M_{\star}/\rm M_{\odot} \right)]+9.91$, which corresponds to $\log_{10} \Sigma_{\rm e}=10.17 \, \rm M_{\odot}/pkpc^{2}$ for $M_{\star}=10^{10}\, \rm M_{\odot}$ and $\log_{10} \Sigma_{\rm e}=9.65 \, \rm M_{\odot}/pkpc^{2}$ for $M_{\star}=10^{11}\, \rm M_{\odot}$, which leads to a threshold to define compact galaxies in the range $\Sigma_{\rm T}=9.2-9.9\, \rm M_{\odot}/pkpc^{2}$ depending on galaxy mass. Employing a higher threshold $\Sigma_{\rm e}$ as in \citet{2017ApJ...846..112K} increases the AGN fraction among the compact star-forming galaxies, and increases the fraction in extended galaxies as well, as found in \citet{2017ApJ...846..112K}. This choice increases also the fraction among extended quiescent and decreases it among compact quiescent galaxies, again as found in \citet{2017ApJ...846..112K}. A lower SFR threshold to distinguish between star-forming and quiescent galaxies would also increase the AGN fraction in compact star-forming galaxies, and lower it for compact quiescent galaxies. 
\citet{2017ApJ...846..112K} use a diagnostic based on galaxy colors to distinguish star-forming and quiescent galaxies. Here we prefer to use a diagnostic based on SFR, that we can directly compare to simulations, where the large uncertainty in dust properties (specially at high redshifts) makes the computation of galaxy colors difficult.
Right panel of Fig.~\ref{fig:candels_AGNfraction} shows the time evolution of the observed AGN fractions in the different types of galaxies.

\section{Global evolution of massive galaxies}
\begin{figure}
\centering
\includegraphics[scale=0.5]{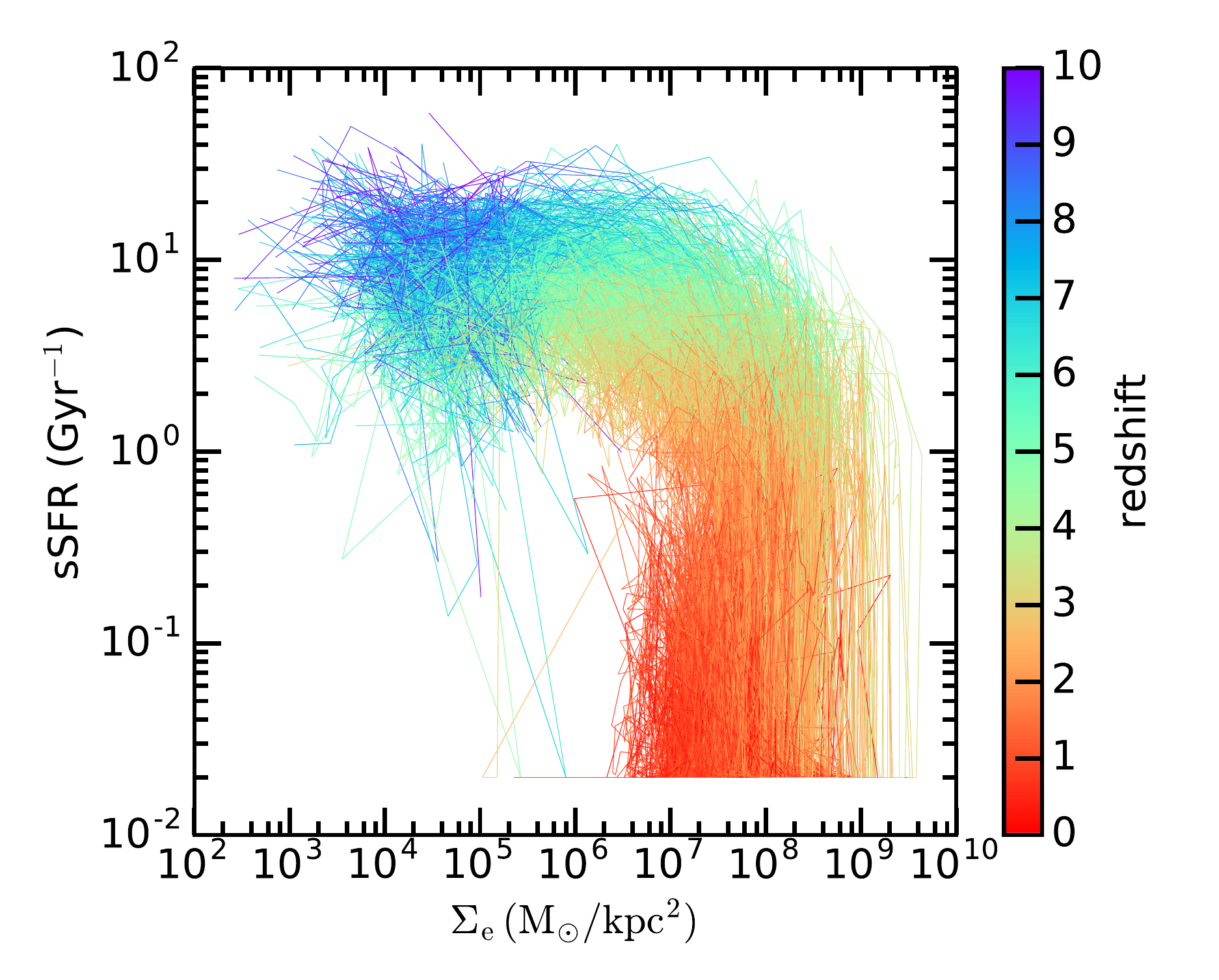}
\caption{Evolutionary paths of 600 random quiescent galaxies with $M_{\star}\geqslant 10^{11}\, \rm M_{\odot}$ at $z=0$, in the sSFR-$\Sigma_{\rm e}$ plane. The individual evolution are color coded by redshift.}
\label{fig:global_evol}
\end{figure}

In Fig.~\ref{fig:follow_quiescent}, we have presented the evolution of two individual galaxies. The global time evolution of these galaxies is representative of a large fraction of galaxies with $M_{\star}\geqslant 10^{11}\, M_{\odot}$, which are quiescent at $z=0$. 
To illustrate this large fraction of galaxies with a similar evolution in the plane sSFR-$\Sigma_{\rm e}$, we show in Fig.~\ref{fig:global_evol} the time evolution of 600 randomly-chosen individual galaxies. 
We color code their evolution with redshift. 
 
\label{lastpage}
\end{document}